\documentclass[12pt,document,nofootinbib,superscriptaddress,onecolumn,preprintnumbers,balancelastpage,table]{revtex4-1}
\pdfoutput=1
\usepackage[table,dvipsnames]{xcolor}
\usepackage{colortbl}
\usepackage{latexsym}
\usepackage{graphicx}
\usepackage{amssymb}
\usepackage{epsfig,amsmath,graphics}
\usepackage{listings}
\usepackage{pstricks}
\usepackage{booktabs}
\usepackage{color}
\usepackage{amsmath}
\usepackage{dcolumn}
\usepackage{slashed}
\usepackage{cancel}
\usepackage{enumitem}
\setlist{nolistsep}
\usepackage{comment,latexsym} 
\usepackage{bm}
\usepackage{verbatim}
\usepackage{tabularx}
\usepackage{dcolumn}
\usepackage{multirow}
\definecolor{Mahogany}{rgb}{0.62,0.24,0.15}
\definecolor{colorLink}{rgb}{0.7,0,0}
\definecolor{colorCite}{rgb}{0,.7,0}
\definecolor{colorURL}{rgb}{0,0,0.7}
\definecolor{colorTC}{rgb}{.2,.7,.2}
\definecolor{colorMD}{rgb}{0.7,.4,.5}
\definecolor{colorNT}{rgb}{0.0,.7,.2}
\definecolor{colorDP}{rgb}{.7,.7,.2}
\definecolor{colorSE}{rgb}{.2,.2,.7}
\usepackage[colorlinks=true,linktocpage=true,linkcolor=black,citecolor=colorCite,urlcolor=colorURL]{hyperref}
\usepackage{setspace}
\usepackage{xspace}
\usepackage{multirow}
\usepackage{titlesec}
\usepackage{bm}
\usepackage{epstopdf}
\usepackage[normalem]{ulem}
\AtBeginDocument{
\heavyrulewidth=.08em
\lightrulewidth=.05em
\cmidrulewidth=.03em
\belowrulesep=.65ex
\belowbottomsep=0pt
\aboverulesep=.4ex
\abovetopsep=0pt
\cmidrulesep=\doublerulesep
\cmidrulekern=.5em
\defaultaddspace=.5em
}

\def\be{\begin{equation}}
\def\ee{\end{equation}}
\newcommand{\beq}{\begin{equation}}
\newcommand{\eeq}{\end{equation}}
\newcommand{\eref}[1]{Eq.~(\ref{#1})}

\newcommand{\lsim}{\!\mathrel{\hbox{\rlap{\lower.55ex \hbox{$\sim$}} \kern-.34em \raise.4ex \hbox{$<$}}}}
\newcommand{\gsim}{\!\mathrel{\hbox{\rlap{\lower.55ex \hbox{$\sim$}} \kern-.34em \raise.4ex \hbox{$>$}}}}

\newcommand{\go}{\widetilde{g}}
\newcommand{\sq}{\widetilde{q}}

\newcommand{\MHT}{\leavevmode\cancel{H}_T}
\newcommand{\ZpJ}{$Z(\rightarrow \nu\,\overline{\nu})$ + jets}
\newcommand{\Nlino}{\widetilde{\chi}}
\newcommand{\MCMS}{M_{T2}^{\text{CMS}}}

\newcommand{\typea}{$\MHT$~\!-{\it type}~}
\newcommand{\typeb}{{\it E~scale-type}~}
\newcommand{\typec}{{\it E~struc-type}~}
\newcommand{\typeac}{$\MHT$~\!-/{\it~E~scale-type}~}
\newcommand{\typeab}{$\MHT$~\!-/{\it~E~struc-type}~}
\newcommand{\typebc}{{\it E~scale-/E~struc-type}~}
\newcommand{\typehybrid}{{\it Hybrid-type}~}

\expandafter\def\expandafter\normalsize\expandafter{%
    \normalsize
    \setlength\abovedisplayskip{8pt}
    \setlength\belowdisplayskip{8pt}
    \setlength\abovedisplayshortskip{8pt}
    \setlength\belowdisplayshortskip{8pt}
}

\newcolumntype{C}{c<{\kern\tabcolsep}@{}}

\titleformat{\section}{\center\normalfont\fontsize{14}{15}\bfseries}{\thesection.}{1em}{}
\titleformat{\subsubsection}{\center\normalfont\fontsize{12}{15}}{\thesubsubsection.}{1em}{}

\begin{document}

\begin{flushright}
FERMILAB-PUB-16-122-E\\
NSF-KITP-16-063
\end{flushright}

\vskip 50 pt

\title{Dissecting Jets and Missing Energy Searches\\[-7pt]
Using \textbf{\emph{n}}-body Extended Simplified Models
 } 

\author{Timothy Cohen}
\affiliation{Institute of Theoretical Science, University of Oregon, Eugene, OR 97403, USA}

\author{Matthew J.~Dolan}
\affiliation{
ARC Centre of Excellence for Particle Physics at the Terascale, \\
\vspace{-8pt}
School of Physics, University of Melbourne, 3010, Australia}

\author{Sonia El Hedri}
\affiliation{Institut fur Physik (THEP) Johannes Gutenberg-Universitat, D-55099, Mainz, Germany}

\author{James Hirschauer}
\affiliation{
Fermi National Accelerator Laboratory, Batavia, IL 60510, USA
}

\author{Nhan Tran}
\affiliation{
Fermi National Accelerator Laboratory, Batavia, IL 60510, USA
}

\author{Andrew Whitbeck}
\affiliation{
Fermi National Accelerator Laboratory, Batavia, IL 60510, USA
}

\begin{abstract}
\vskip 1 pt
\begin{center}
{\bf Abstract}
\end{center}
\vskip -30 pt
\quad
\begin{spacing}{1.05}
\noindent Simplified Models are a useful way to characterize new physics scenarios for the LHC.  Particle decays are often represented using non-renormalizable operators that involve the minimal number of fields required by symmetries. Generalizing to a wider class of decay operators allows one to model a variety of final states. This approach, which we dub the $n$-body extension of Simplified Models, provides a unifying treatment of the signal phase space resulting from a variety of signals. In this paper, we present the first application of this framework in the context of multijet plus missing energy searches. The main result of this work is a global performance study with the goal of identifying which set of observables yields the best discriminating power against the largest Standard Model backgrounds for a wide range of signal jet multiplicities.  Our analysis compares combinations of one, two and three variables, placing emphasis on the enhanced sensitivity gain resulting from non-trivial correlations. Utilizing boosted decision trees, we compare and classify the performance of missing energy, energy scale and energy structure observables.   We demonstrate that including an observable from each of these three classes is required to achieve optimal performance.  This work additionally serves to establish the utility of $n$-body extended Simplified Models as a diagnostic for unpacking the relative merits of different search strategies, thereby motivating their application to new physics signatures beyond jets and missing energy.
\end{spacing}
\end{abstract}


\maketitle
\newpage
\begin{spacing}{1.3}
\pagebreak
%
%
\tableofcontents

\section{Introduction}
\label{sec:Intro}
Hadron colliders provide some of the most important experimental inputs in high energy physics.  At the microscopic level the colliding particles are quarks and gluons, implying that the production cross section is highest for states that either carry color or have a large interaction strength with quarks. In many beyond the Standard Model scenarios these new physics states then decay back to colored Standard Model particles, along with some dark sector objects that escape detection.  The resulting experimental signature is multiple high $p_T$ jets and missing transverse energy, $\MHT$. Searches for new physics that are characterized by this final state have very high priority at the Large Hadron Collider (LHC)~\cite{Aad:2015iea,Aad:2015baa,Chatrchyan:2014lfa,CMS-PAS-SUS-13-020}.  Hence, many ideas for distinguishing signal from background have been proposed~\cite{Lester:1999tx,Barr:2003rg,Rogan:2010kb,Randall:2008rw}.  The framework introduced in this article has been developed in order to quantitatively compare and contrast these different approaches.  

It is particularly interesting to understand how the observables respond as a function of the final state parton multiplicity, which can vary between new physics models.  The canonical jets + $\MHT$ searches at the LHC are currently framed in terms of Simplified Models~\cite{Alves:2011wf}, the majority of which have been extracted from the Minimal Supersymmetric Standard Model (MSSM)~\cite{Martin:1997ns}.  In particular, these searches have been optimized for signals that are motivated by supersymmetric (SUSY) models, involving both gluinos $\go$ and squarks $\sq$ (fermionic color octets and scalar color triplets respectively) which decay to a stable neutral particle, the neutralino $\Nlino$, some number of light flavor $q$, bottom $b$, and top $t$ quarks, along with the option of additional weak gauge bosons $W^{\pm}$, and $Z^0$. The simplest and best studied decay modes are $\go \rightarrow q\,\overline{q}\,\Nlino$ and $\sq \rightarrow q\,\Nlino$~\cite{Alwall:2008va, Alwall:2008ag}.  In typical $R$-parity conserving models, these particles are produced in pairs and the parton level final state involves some number of colored objects and $\MHT$. Table~\ref{tab:SimpModTopologies} provides a detailed  picture a variety of the possible final states, including those which are currently being searched for, ignoring possible flavor-tagging.

\begin{table}[h!]
\renewcommand{\arraystretch}{1.6}
\setlength{\tabcolsep}{7pt}
\setlength{\arrayrulewidth}{.5mm}
\begin{tabular}{@{\extracolsep{\fill}} c  @{\extracolsep{\fill}} |  @{\extracolsep{\fill}} c @{\extracolsep{\fill}}|  @{\extracolsep{\fill}} c @{\extracolsep{\fill}}}
\hline
\,\,{\sc Production} \,\,\, & \,\,\, {\sc Decay Channel} \,\,\, & \,\,\, {\sc Final State} \,\, \\
\hline
\cellcolor[gray]{.9}$\sq\,\, \sq$ &\cellcolor[gray]{.9} $\sq \rightarrow q\,\Nlino$ & \cellcolor[gray]{.9}$2 \text{ partons} + \MHT$ \\
\multirow{2}{*}{$\sq\,\, \go$} & $\go \rightarrow q\,\overline{q}\,\Nlino$ & \multirow{2}{*}{$3 \text{ partons} + \MHT$}\\[-5pt]
& $\sq \rightarrow q\,\Nlino$ & \\
\cellcolor[gray]{.9} $\go\,\, \go$ & \cellcolor[gray]{.9} $\go \rightarrow q\,\overline{q}\,\Nlino$ & \cellcolor[gray]{.9} $4\text{ partons} + \MHT$\\
\multirow{2}{*}{$\sq\,\, \go$} & $\go \rightarrow q\,\overline{q}\,Z^0\,\Nlino$ & \multirow{2}{*}{$5\text{ partons} + \MHT$}\\[-5pt]
& $\sq \rightarrow q\,\Nlino$ & \\
\cellcolor[gray]{.9} $\widetilde{t}\,\, \widetilde{t}$ & \cellcolor[gray]{.9} $\widetilde{t} \rightarrow t\,\Nlino$ & \cellcolor[gray]{.9} $6\text{ partons} + \MHT$\\
\multirow{2}{*}{$\sq\,\, \go$} & $\go \rightarrow t\,\overline{t}\,\Nlino$ & \multirow{2}{*}{$7\text{ partons} + \MHT$}\\[-5pt]
& $\sq \rightarrow q\,\Nlino$ & \\
\cellcolor[gray]{.9} $\go\,\, \go$ & \cellcolor[gray]{.9} $\go \rightarrow q\,\overline{q}\,Z^0\,\Nlino$ & \cellcolor[gray]{.9} $8\text{ partons} + \MHT$\\
\toprule
\end{tabular}
\caption{This table shows different SUSY production modes and decay channels associated with differing numbers of final state jets, where we have ignored  flavor when counting partons. We assume that the $Z$-bosons and top quarks decay hadronically. The actual number of reconstructed partons may differ from that in this table, see Appendix~\ref{App: UV} for further details. }
\label{tab:SimpModTopologies}
\end{table}

While this suite of signal topologies covers a wide range of possible final states (not all of which have associated public results from the LHC collaborations), the relative optimizations are complicated by the fact that the different production modes do not yield the same cross section, and the presence of intermediate on-shell states can lead to additional features in the signal distributions.  It is therefore difficult to contrast the variety of approaches for digging new physics out of jets + $\MHT$.  In order to minimize the differences between the ways of generating these various final states, we are introducing a novel variation of the Simplified Models paradigm which we will refer to as the ``$n$-body extension."  As outlined in detail in the next section, we will be performing our analysis as a function of the final state parton multiplicity, which we achieve by varying the number of partons that result from the \emph{direct} decays of ``gluinos."
This allows to us compare observables in the same regions of phase-space, without needing to correct for the relative effects inherent to different Simplified Models,
and yields a concrete and transparent assessment of the performance of a wide variety of variables and their combinations.

In order to achieve a fair comparison between the observables considered below, we use boosted decision trees (BDTs) to optimize the different search cuts applied, thereby achieving maximum signal-to-background discrimination. While BDTs are by now a widely used technique experimentally, they are not as familiar to theorists -- we provide a brief technical introduction to them in Appendix~\ref{app:bdts}. We are advocating for the use of multivariate tools here, not as an in-practice analysis strategy, but instead as a guiding principle to evaluate the relative importance of different multivariable approaches.  In particular, BDTs permit the straightforward analysis of both correlated and uncorrelated variables, which in turn allows for the identification of powerful combinations.

In this study, we focus on the observables that have been used by ATLAS and CMS in their multijet plus missing energy analyses. Since our multivariate approach is not meant to be taken as a new search strategy, we neglect the possible signal and background uncertainties. Therefore, our results represent how the different variables would perform in an ideal optimistic case.  We compare the discrimination power of different sets of observables to a nearly optimal benchmark analysis, in which we combine all considered variables using a single BDT.  This `aggregate' result is of course unrealistic and should be interpreted as an approximate upper limit on the possible performance.  

Studying single variables alone leads to a classification scheme into three classes -- missing energy-, energy scale-, or energy structure-type -- along with a few hybrids which exhibit characteristics of more than one of the relevant behaviors.  In general, we find that most of the information about the final states studied can be captured using multivariable analyses including at least one of each type. For most of the parameter space, combinations of standard variables such as the number of jets, the sum of jet mass (or $H_T$), and $\MHT$ lead to near optimal performance.  For the simple topologies studied here, more sophisticated variables are usually strongly correlated with one of these well-known variables.  This provides justification for the canonical approaches already in place, and helps guide modifications that can be used when designing future searches.

The rest of this paper is organized as follows.  Section~\ref{sec:nBodySimpMod} provides a detailed definition of $n$-body extended Simplified Models, along with some comments on their theoretical consistency.  Section~\ref{Sec: toolkit} details our approach, the variables that we consider, and the signal and background simulations. Finally, Section~\ref{Sec: results} shows the results of our analysis for both compressed and non-compressed gluino-neutralino topologies. We conclude in Section~\ref{sec:Conc}.  A number of appendices are given which provide some additional details and justify some of the approximations taken in the main text.

\section{The \emph{n}-body Extended Gluino-Neutralino Simplified Model}
\label{sec:nBodySimpMod}

Simplified Models are a convenient way to organize signals relevant for new physics searches at the LHC.  The philosophy is to identify models involving the minimal number of new particles and couplings in order to populate regions of signature space.\footnote{For other another approach  to new physics searches that attempts to minimize theory bias, see~\cite{Dube:2008kf, Craig:2012pu}.} Altering the masses of the relevant states leads to kinematic differences which motivate multiple signal regions that can be designed in order to provide sensitivity across parameter space.  This approach has taken hold at both ATLAS and CMS, and most of the new physics results are now cast in terms of Simplified Models.  

Given a Simplified Model, there are many ways one can extend it.  For example, one can add additional states and couplings which could lead to new production modes, new branching ratios, and/or new kinematic features.  One of the key ideas in this paper is a new augmentation of the Simplified Model theory space, which we will refer to as the ``$n$-body" extension.  The starting point for $n$-body extended Simplified Models is a set of states and a Lagrangian.  Take the now canonical ``Gluino-Neutralino" Simplified Model, which will be the example used throughout this work. The full beyond the Standard Model new particle content is a color octet Majorana fermion $\widetilde{g}$ (the gluino) and a singlet Majorana fermion $\Nlino$ (the neutralino).  The Lagrangian is given by 
\begin{align}
\mathcal{L} = \frac{i}{2}\,\overline{\widetilde{g}} \,\gamma^\mu\, D_\mu \,\widetilde{g} - \frac{1}{2}\,  m_{\widetilde{g}}\,\overline{\widetilde{g}}\, \widetilde{g}  + \frac{i}{2}\,\overline{\Nlino} \, \gamma^\mu\, D_\mu \,\Nlino - \frac{1}{2}\,  m_{\Nlino}\,\overline{\Nlino}\, \Nlino  + \mathcal{L}_\text{decay}\, ,
\end{align}
where $\mathcal{L}_\text{decay}$ is the Lagrangian that specifies the decays of the gluino, and $D_\mu$ is the appropriate covariant derivative.\footnote{As is standard practice, we are ignoring gauge eigenstate mixing and are not tracking electroweak symmetry breaking spurions since these effects have no impact on the observables of interest here.}  

\begin{figure}[h!]
\centering
\includegraphics[width=.80 \textwidth]{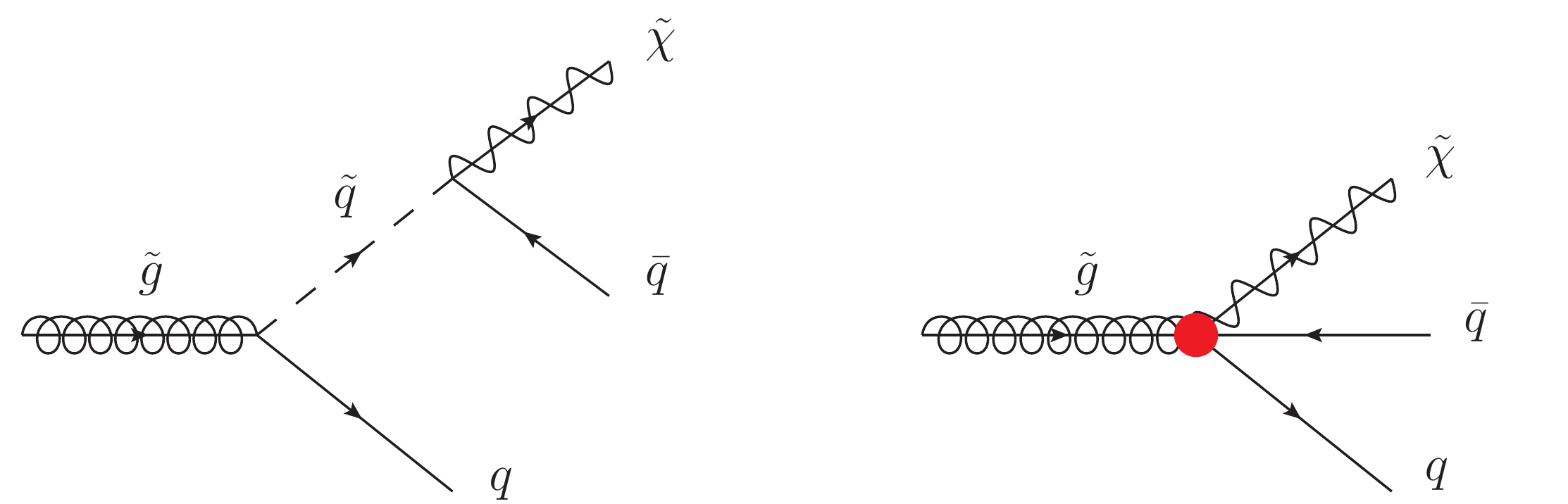}
\caption{Sketch of integrating out a squark to yield a three-body decay of the gluino.}
\label{fig:gluino3bodySketch} 
\end{figure}

We will assume $R$-parity is conserved in this study.  This parity along with gauge invariance implies that the gluino must decay via a non-renormalizable operator to some Standard Model states and a neutralino.  For example, the gluino could decay via an off-shell squark, see Fig.~\ref{fig:gluino3bodySketch}.  This yields the standard $\widetilde{g} \rightarrow q\,\overline{q}\,\Nlino$ decay channel, and the decay Lagrangian is given by the four-fermion operator
\begin{align}
 \mathcal{L}^{(2)}_\text{decay} = \frac{y^2}{\Lambda^2}\,\overline{q}\,\widetilde{g}\,q \, \overline{\Nlino} + \text{h.c.}\,,
 \label{eq:2body}
\end{align}
where $1/\Lambda$ is the suppression scale for this operator, $y$ is a dimensionless coupling and the superscript refers to the number of final state colored partons that will result from the decay.
It is straightforward to  complete this operator in the ultraviolet (UV) by introducing a squark, as illustrated in Fig.~\ref{fig:gluino3bodySketch}; the scale $\Lambda$ is proportional to the squark mass.  The quarks  $q$ in \eref{eq:2body} can be  light flavor quarks,  bottoms, or  tops; all three cases have been searched for at the LHC in the jets + $\MHT$~channel (with $b$-decay/tags when appropriate) and in channels with leptons for the decays involving tops. 

The simple extension proposed in this paper, which will be of use below when exploring the variables utilized for jets + $\MHT$~searches, is to introduce a larger set of possibilities that allow us to vary the number $n$ of partons in the final state:  
\begin{align}
\mathcal{L}^{(1)}_\text{decay} &= \frac{y^2 g_s}{16\,\pi^2\,\Lambda}\,G_{\mu\nu}\,\overline{\widetilde{g}}\,\bar{\sigma}^{\mu\nu} \, \Nlino + \text{h.c.}\, , \\
\label{Eq:l1}
\mathcal{L}^{(2)}_\text{decay} &= \frac{y^2}{\Lambda^2}\,\overline{q}\,\widetilde{g}\,q \, \overline{\Nlino}+ \text{h.c.}\, , \\
\mathcal{L}^{(3)}_\text{decay} &= \frac{y^2 g_s}{16\,\pi^2\,\Lambda^4}\,\overline{q}\,q \, G_{\mu\nu}\,\overline{\widetilde{g}}\,\bar{\sigma}^{\mu\nu}\,\Nlino + \text{h.c.}\, , \\
\mathcal{L}^{(4)}_\text{decay} &= \frac{y^4}{\Lambda^5}\,\overline{q}\,q\,\overline{q}\,\widetilde{g}\,q \, \overline{\Nlino} + \text{h.c.}\, , \\
\mathcal{L}^{(5)}_\text{decay} &= \frac{y^4 g_s}{16\,\pi^2\,\Lambda^7}\,\overline{q}\,q \,\overline{q}\,q\,G_{\mu\nu}\,\overline{\widetilde{g}}\,\bar{\sigma}^{\mu\nu}\, \Nlino + \text{h.c.}\, , 
\label{Eq:l5}
\end{align}
where $G^{\mu\nu}$ is the $SU(3)$ field strength with associated gauge coupling $g_s$, $\Lambda$ is a dimensionful scale, and $y$ is a dimensionless coupling, see Appendix~\ref{App: UV} for a detailed discussion of these operators.  Note that in practice we will ignore the angular correlations predicted by the Lorentz structure of these specific operators by assuming a flat matrix element and allowing relativistic phase space to predict the final state distribution in the parent rest frame.

Note that here we have chosen the $n$-body extension at each higher point which adds the maximum number of quarks.  It is entirely plausible that all the partons could be gluons, and the operator would just include more powers of $G^{\mu\nu}$.
However, such operators would either require the existence of lower order operators -- these would provide the dominant gluino decay modes -- or involve loop interactions. In the latter case, the additional loop suppression factors would severely restrict the parameter space available for prompt decays. This issue is discussed in more detail in Appendix~\ref{App: UV}.
It is also possible that the $n$-body extension could involve Standard Model states besides quarks and gluons.  We will leave exploring the implications of these additional directions in model space to future work.  

Considering these effective operators allows us to focus our study on final states only, while not including information about possible intermediate particles in the gluino decay chain. Most of our results, however, are expected to also apply to topologies including intermediate on-shell particles, this comparison is made more concrete in Appendix~\ref{sec:OnShellIntermediateStates}. In particular, although some of our decay Lagrangians might not admit reasonable UV completions (in the sense that many coincidences would be required such that the decay mode of interest would actually dominate), or would lead to long-lived gluinos, they can be used to study generic configurations where the gluino cascade decays through one or more intermediate on-shell particles. Possible UV completions and gluino lifetimes for the operators considered in this study are discussed in Appendix~\ref{App: UV}.

For the sake of specificity, it is worth pausing to define some crucial notation.  A ``parton" is either a quark or gluon that is produced at the matrix element level before the parton shower has been applied to a given event.  We will distinguish between partons that result from the direct decays of the gluino, the ``decay partons," from those that come from higher order initial state radiation (ISR) of gluons and/or quarks off the hard collision process as implemented using the MLM merging procedure~\cite{Alwall:2007fs}, the ``matrix element partons:"
\begin{align}
n &\equiv \text{ number of total decay partons.}\notag\\
m &\equiv \text{ maximum number of matrix element partons.} \notag
\end{align}
This terminology will be used extensively in the discussion below.

Finally, we conclude this section by contrasting $n$-body extended Simplified Models with On-Shell Effective Theories (OSETs)~\cite{ArkaniHamed:2007fw}.  OSETs are a class of non-Lagrangian based parameterizations for new physics scenarios.  They are characterized by the masses of the on-shell particles that are involved in the process of interest, along with additional parameters that determine the size and shape of the production cross section.  This framework was introduced as a suggested shift in the approach to interpreting new physics searches at the then-upcoming LHC program.  The goal was to move away from relying on ``full" models as the jumping off point for designing searches, the classic example being scans in the $M_0$ versus $M_{1/2}$ plane of the Constrained MSSM.  Instead, OSETs were invented to provide a signature based signal injection.   These ideas ultimately lead to the invention of Simplified Models and their subsequent adoption by both CMS and ATLAS.  Additionally, the MARMOSET Monte Carlo program was developed as an approach to solve the ``LHC Inverse" problem in a systematic way~\cite{ArkaniHamed:2007fw}.  In a similar spirit, many related tools have been released that facilitate the straightforward reinterpretation of LHC results~\cite{Drees:2013wra, Papucci:2014rja, Kraml:2014sna}.

One of the key points of this approach was to recognize that the leading kinematic aspects of the production cross section are simply due to the behavior of the parton distribution functions in most cases of threshold production.  This implies that one could reproduce the majority of observable features across a wide variety of models using only a few parameters.  Since the parton luminosities fall with a very high power of the momentum fraction, a Taylor expansion of the cross section is essentially truncated at leading order -- a few exceptions were identified, \emph{e.g.} $p$-wave production, and the required modifications for their inclusion into the OSET paradigm was provided~\cite{ArkaniHamed:2007fw}.  For our purposes, their work demonstrates that even though we have chosen to use ``gluinos" as our parent particles, the implications of our results are expected to hold in a much wider variety of theories that are dominated by non-zero $s$-wave production.  This is part of the justification for the comparison between the distributions provided for our 2-parton results  with stop pair production (that subsequently decay yielding $t\,\overline{t}\,\Nlino\,\Nlino$) given in Appendix~\ref{sec:OnShellIntermediateStates}.  In our view, one advantage of using Simplified Models is that they are well-defined Lagrangian based theories, which implies that one can analyze the feasibility of UV completions (see Appendix~\ref{App: UV}), and an investigation of higher order perturbative corrections can be performed straightforwardly.  Additionally, modern simulation tools can be utilized which allows us to include the impact of merging matrix elements involving different numbers of ISR partons, which are important for the modeling of signals from compressed spectra as discussed below in Sec.~\ref{sec:Degenerate}.

The OSET framework obviously must additionally involve some mechanism that decays the parent particles.  The approach taken in~\cite{ArkaniHamed:2007fw} was to ignore spin correlations and decay all unstable new physics states using a flat matrix element integrated against the standard phase space distribution.  As already discussed, this is the approach taken in this work as well.  While this assumption does not provide a good approximation for all possible models of interest, it is broadly applicable to a wide variety of scenarios and is an obvious choice for the kind of study we are performing here.  Certainly exceptions can be found, \emph{e.g.} if there is an on-shell intermediate state and the invariant mass of the resulting decay products is used as a discriminator, see~\cite{ArkaniHamed:2007fw} for additional examples along with simple modifications to move beyond the assumption of phase space only decays.  Furthermore, there are cases where the angular dependence of the decay products becomes important, see \emph{e.g.}~\cite{Kilic:2007zk, OSCELOT} for a discussion.  However, much of this information is washed out once one boosts the decay products to the lab frame and integrates over the possible orientations of the intermediate particles.  This accounts for the lack of sensitivity to these effects and further justifies the assumptions made in this work.  Clearly OSETs and our $n$-body extended Simplified Models are complementary approaches, and the work of~\cite{ArkaniHamed:2007fw} gives many of the detailed arguments for the broad applicability of the choices made here.


\section{Dissection Toolkit}
\label{Sec: toolkit}

The $n$-body extended Gluino-Neutralino Simplified Models can be used to systematically explore a range of jets + $\MHT$ final states.
Search strategies for gluinos at CMS and ATLAS predominantly employ inclusive cuts in a phase space of some number of observables, which vary from analysis to analysis~\cite{Chatrchyan:2013wxa, Chatrchyan:2014lfa, Khachatryan:2015pwa, Khachatryan:2015vra, Khachatryan:2016kdk, Aad:2013wta, Aad:2014wea, Aad:2015iea, ATLAS-CONF-2015-062, Khachatryan:2016xvy, Chatrchyan:2013mys}.
Our goal is to understand the performance sensitivity of these observables for various injected signals, including the impact of correlations that are taken advantage of through different variable combinations.  Developing quantitative intuition for which observables can be best used to distinguish between a given signal and background will lead to a better understanding of how to maximize coverage for a given space of signals. 

There is an additional practical matter due to the fact that systematic errors are present for every aspect of an analysis, be it from theoretical uncertainties, \emph{e.g.} due to working at finite order in perturbation theory, or from experimental issues that arise from a variety of sources as one goes from hits in a detector to reconstructed physics objects, \emph{e.g.} jet energy scale uncertainties, isolation requirements, and so forth.  When working with real data, any observable that one wants to include in an analysis must be validated, requiring suitable control regions along with the ability to make reliable computations in order to extrapolate into the signal region.  This implies there will always be a trade off between including more information, and maintaining a reasonable level of systematic control. Hence, in practice the number of observables is limited.
It is outside the scope of this study to quantify such systematic effects. However, when performing a multi-variable analysis it will always be desirable to optimize the sensitivity of those variables.
Given that there are several analyses which use different sets of observables, if/when a putative signal is discovered, understanding the correlations between given observables will be necessary to properly characterize the new physics. In this article we will generally consider combinations of up to three variables; the reason for this choice will become apparent as we go through the results.

To develop intuition for a broad set of variables, we will characterize sensitivity using curves which detail the signal efficiency versus background rejection power for 
a given cut in observable space, often referred to as receiver operating characteristic (ROC) curves.  
We first start by comparing single observables against each other; however, it is important to also consider multiple observables together since in practice a multi-dimensional space must be probed to achieve maximum sensitivity.
Our quantitative analysis relies on BDTs, which are preferred for their convenience and flexibility.  It is expected that while the absolute performance of the BDTs is better than that of the coarsely binned multidimensional templates often used in experimental searches.  The ROC curves are then an interesting metric for comparison of variables.
The effect of binning is not explored in greater detail as it is luminosity and background estimation method dependent and our goal is  to derive results independent of these effects.
The details of the BDT implementation are given in Appendix~\ref{app:bdts}.

\subsection{Observables}
\label{sec:Observables}
Many searches have been designed to access new physics in jets + $\MHT$.  We choose to study the following suite of observables, which incorporates the predominant variables used on the 8~TeV LHC data.
\begin{itemize}

\item $H_T$ is defined as the scalar sum of the $p_T$ of all jets in the event whose $p_{T}>30$ GeV and $|\eta|<$ 2.5. This variable is particularly powerful for topologies like ours, where a new heavy particle decays to multiple objects.

\item $\vec{\leavevmode\cancel{H}}_T$ is defined as the negative vector sum of the transverse momenta of all jets in the event whose $p_{T}>30$ GeV and $|\eta|<$ 5.0.  Then $\MHT = \Big|\,\vec{\leavevmode\cancel{H}}_T\Big|$ is the scalar missing energy.  For signal events, non-zero $\MHT$ dominantly results from neutralinos in the final state.

\item $N_j$ is defined as the number of jets in the event whose $p_{T}>30$ GeV and $|\eta|<$ 2.5.  Jets are clustered using the anti-$k_T$ algorithm with a cone size $R =0.4$.

\item $\MHT$/$\sqrt{H_T}$~\cite{BruceKnuteson:1999pna,Nachman:2013bia}, where $H_T$ and $\MHT$ are defined above. This variable discriminates against events where the $\MHT$ comes from jet mismeasurement.  

\item $ M_J \equiv \sum m_J$~\cite{Hook:2012fd}, where $m_J$ is the mass of a given anti-$k_T(R=1.0)$  jet with $m_J > 50$ GeV. This variable is predicted to be particularly useful for large multiplicity signals where multiple objects with moderate $p_T$ can be clustered into hard fat jets \cite{Hook:2012fd, Cohen:2012yc, Hedri:2013pvl, Cohen:2014epa, Aad:2015lea,CMS-PAS-SUS-15-007}.

\item $m_\text{eff} \equiv \sum p_T^{j} + \MHT$~is the effective mass variable often utilized in jets + $\MHT$ searches performed by ATLAS, see~\cite{ATLAS-CONF-2015-062} for a recent example.

\item Razor~\cite{Rogan:2010kb}, is a two-dimensional variable $\big \{m_R, R^2\big\}$, used to identify final states with two visible objects $j_1$ and $j_2$ and $\MHT$. In a so-called "Razor frame``, obtained by applying the longitudinal boost
\begin{align}
\beta^{R*}_L = \frac{p_z^{j_1} + p_z^{j_2}}{E_{j_1} + E_{j_2}}
\end{align}
from the lab frame, $M_R$ is defined as
\begin{align}
M_R = \sqrt{\big(E_{j_1} + E_{j_2}\big)^2 - \Big(p_z^{j_1} + p_z^{j_2}\Big)^2}.
\end{align}
The dimensionless variable $R$ is defined as 
\begin{align}
R = \dfrac{M_T^R}{M_R},
\end{align}
where $M_T^R = \dfrac{1}{\sqrt{2}}\sqrt{\MHT \Big(p_T^{j_1} + p_T^{j_2}\Big) - \vec{\leavevmode\cancel{H}}_T\cdot\Big(\vec{p}_T^{\,j_1} + \vec{p}_T^{\,j_2}\Big)}$.  For cases where an event contains more than two jets, jets are combined into two pseudojets by finding the combination of jets that minimizes the sum of the squared masses, $p_{\text{pseudo},1}^2+p_{\text{pseudo},2}^2$.  

\item $M_{T2}$~\cite{Lester:1999tx,Barr:2003rg}, is used to identify signals where a pair-produced particle decays semi-visibly. It is defined as
\begin{align}
M_{T2} = \min_{\vec{\slashed{p}}_{T}^{\,(1)}+\vec{\slashed{p}}_{T}^{\,(2)}=\vec{\leavevmode\cancel{H}}_T}\left[ \max\left\{m_T\Big(\vec{p}_{T}^{\,\,(1)}, \vec{\slashed{p}}_{T}^{\,(1)}\Big), m_T\Big(\vec{p}_{T}^{\,\,(2)}, \vec{\slashed{p}}_{T}^{\,(2)}\Big)\right\}\right],
\end{align}
where $\vec{p}_T^{\,\,(1),(2)}$ are the transverse momenta of the visible objects, $\vec{\slashed{p}}_T^{\,(1),(2)}$ are the possible transverse momentum assignments for the two invisible objects of an assumed mass, and $m_T$ is the usual transverse mass variable.  In the case where an event contains more than two jets, the jets are combined into two pseudojets using the same criteria as for the Razor.  

\quad\, We will utilize two versions of this variable below, denoted $M_{T2}$ and $\MCMS$.  For both, the input neutralino test mass is taken to be zero.  One makes use of the measured mass of the input pseudojets when performing the min/max calculation inherent in $M_{T2}$.  The other,  $\MCMS$,  explicitly assumes that the input jets are massless and is used by the CMS collaboration in its searches. Note that the pseudojets can have significant masses -- this will lead to significant differences in performance, and in fact we will find that these two implementations of the $M_{T2}$ have quite different behaviors as discussed in the next section.

\item $\alpha_T$~\cite{Randall:2008rw,Chatrchyan:2013mys} is used to reject multijet events in which $\MHT$ arises as a result of jet mismeasurement. It is defined as
\begin{align}
\alpha_T = \frac{1}{2}\times \frac{1 - (\Delta H_T/H_T)}{\sqrt{1 - (\MHT/H_T)^2}},
\end{align}
where $H_T$ and $\MHT$ are defined above. A di-pseudojet system is constructed using the Razor definitions above.\footnote{We also studied an alternative algorithm for defining the di-pseudojet system in which the $H_T$ of the di-jet system was minimized.  It was found that the different algorithms had little effect on performance.}  Then $\Delta H_T$ is the scalar $p_T$ difference between the two jets.
\end{itemize}

The number of $b$ jets, $n_b$, is also often used which greatly changes the background composition and suppresses light flavor jet backgrounds.  
As our study focuses primarily on kinematic properties of the event, we do not consider $n_b$ directly; however, we do consider observable performance for different backgrounds separately 
which effectively separates performance in $n_b$ bins, \textit{e.g.} searches with multiple $b$-tags will be dominated by $t\,\overline{t}$ backgrounds.


\subsection{Simulation}
We simulate both signal and background events using \texttt{MadGraph5 v1.5.14}~\cite{Alwall:2014hca} using CTEQ6L1 parton distribution functions~\cite{Pumplin:2002vw}, interfaced with \texttt{Pythia v6.4}~\cite{Sjostrand:2006za} for parton showering and hadronization. Basic detector simulation is performed in \texttt{Delphes}~\cite{deFavereau:2013fsa}, with the default implementation of the CMS detector. For the matched signal and background samples, \texttt{MadGraph} and \texttt{Pythia} are interfaced for MLM matching~\cite{Mangano:2002ea} with the $k_T$ shower scheme~\cite{Alwall:2008qv}. All our samples are generated at a center-of-mass energy $\sqrt{s}=13$~TeV.

\subsubsection*{\textbf{\textit{Signal Simulation}} }
We generate the following samples:
\begin{align}
	p\; p &\rightarrow \widetilde{g}\,\widetilde{g} + m \text{ jets}\notag\\
	\widetilde{g}\, \widetilde{g} &\rightarrow n \text{ partons} + 2\,\Nlino
\end{align}
 for $n = 2, \ldots 8$ denotes the total number of ``decay partons" while $m$ gives the number of ``matrix element partons."     When $n$ is even, we require each gluino to decay to $\frac{n}{2}\text{ partons} + \Nlino$ with $100$\% branching ratio.  For $n$ odd, we require each gluino to decay to $\frac{n \pm 1}{2}\text{ partons} + \MHT$ with $50$\% branching ratio for each decay mode and keep only the events where the gluinos decay asymmetrically.  Note that while this procedure is artificial in the case of identical particle production, it is possible to have odd numbers of final state partons in associated production, \emph{i.e}, if the production channel involves multiple states, see Table~\ref{tab:SimpModTopologies} above for examples. 
 The quark-gluon content of the final states is determined using the operators shown in Eqs.~(\ref{Eq:l1}) - (\ref{Eq:l5}).  In practice, we decay the gluinos in \texttt{Pythia} by simply specifying the final states, allowing the program to choose an appropriate color connection and decay the parent using phase space integrated against a flat decay matrix element. 
    
In the first part of our study, we set the neutralino mass to be $m_{\Nlino} = 1$ GeV in order for the gluino decay not to be constrained by phase space; this will be referred to as the ``uncompressed" signal phase space. 
We do not require any jets from ISR ($m = 0$) and consider the following gluino masses
    \begin{align}
    	m_{\tilde{g}} &= \left\{500, 1000, 1500, 2000  \right\}\,\mathrm{GeV}.
    \end{align}

The second part of our study considers ``compressed" spectra where the LSP mass is 5\% less than the gluino mass. 
Due to the limited phase space for the decays, the gluino decay products are now expected to be soft. 
We therefore consider topologies where the gluinos are boosted against one or two ISR jets by generating matched events with $m = 0,1,2$ with the matching scale set to 100 GeV. 
Specifically, we consider the following gluino and neutralino masses: 
\begin{align}
    	\big\{m_{\tilde{g}}, m_{\Nlino}\big\} = \Big(\big\{500, 475\big\}, \big\{1000, 950\big\}, \big\{1500, 1425  \big\}\Big) \text{ GeV}
\end{align}
We will only show results for $\big\{m_{\tilde{g}}, m_{\Nlino}\big\} =  \big\{1000, 950\big\} \text{ TeV}$ below, but will comment on the behavior of other masses.

\subsubsection*{\textbf{\textit{Background Simulation}}}

We generated matched samples of \ZpJ, $t\,\overline{t}$ + jets, and QCD multijet events.
We accommodate up to four partons in the final state, which determines the maximum number of jets for a given process. 
In order to efficiently populate the tails of the background distributions, we split each background into bins of the variable $S_T^*$, which is defined as the scalar sum of the $p_T$ of all generator level particles, \emph{i.e.}, at parton level.  Following the procedure detailed in~\cite{Avetisyan:2013onh}, we modify \texttt{MadGraph} to implement a cut on $S_T^*$ at generator level and require each bin to satisfy
\begin{align}
	\sigma_i = \sigma\big(\mathtt{htmax}_i > S_T^* > \mathtt{htmin}_i\big) \gtrsim 0.9 \times \sigma\big(S_T^* > \mathtt{htmin}_i\big)
\end{align}
where $\mathtt{htmax}_i$, $\mathtt{htmin}_i$ are the edges of the $i$-th bin. The final overflow bin has to satisfy $N_\text{overflow}/10 >  \sigma_\text{overflow}\times \mathcal{L}$, where $N_\text{overflow}$ is the total number of events to be generated in the overflow bin, $\sigma_\text{overflow}$ is the cross section in this bin, and $\mathcal{L}$ is the luminosity. 
Table~\ref{Tab:back} shows the various background categories, the number of events generated per $S_T^*$ bin and how many $S_T^*$ bins were generated.  
These events are then showered and passed through \texttt{Delphes} independently, before being weighted by the cross section in each bin and combined. 
\begin{table}[h!]
	\renewcommand{\arraystretch}{1.6}
\setlength{\tabcolsep}{7pt}
\setlength{\arrayrulewidth}{.5mm}
	\begin{tabular}{c|c|c|c|c}
		\hline
		Background & $m$ partons & num. of  $S_T^*$ bins  & $N_\text{MC}$ per bin & matching scale\\
		\hline
		\ZpJ & 1\,-\,3 & 7 &$\sim 4 \times 10^6$ & 60 GeV\\
		$t\,\overline{t}$ + jets & 0\,-\,2 & 7 & $\sim 7 \times 10^6$ & 60 GeV\\
		QCD & 2\,-\,4 & 10 & $\sim 1 \times 10^7$ & 60 GeV\\
		\hline
	\end{tabular}
	\caption{\label{Tab:back} Parameter choices made for the background simulations are given here: the number of matrix element partons $m$ included, the number of $S_T^*$ bins, the number of events $N_\text{MC}$ generated in each bin, and the MLM matching scale. }
\end{table}



\section{Results of Gluino-Neutralino Study}
\label{Sec: results}
Motivated primarily by trigger thresholds from LHC Run 1 analyses, we preselect our signal and background samples requiring 

\noindent \emph{Preselection:}
\begin{itemize}
\item $H_T>$ 500~GeV, 
\item $\MHT >$ 200~GeV, 
\item two or more jets above 30~GeV, 
\item no isolated leptons above 20~GeV,
\item ${\rm min}(\Delta\phi) > 0.4$,
\end{itemize}
where ${\rm min}(\Delta\phi)$ is the minimum azimuthal angle between the missing energy and the two highest $p_T$ jets.
This is a standard cut which is used to reduce the fake missing energy from mismeasured jets.\footnote{Most signals have a large value of $\min(\Delta \phi)$, which is why it is often employed in preselection.  However, this is not always the case, \emph{e.g.} searching for SUSY spectra with heavy squarks and light gluinos~\cite{Fan:2011jc}, or models that yield semi-visible jets~\cite{Cohen:2015toa} requires a modified approach.}
The motivation for these preselection thresholds are driven primarily by the current trigger thresholds of CMS and ATLAS analyses which are restricted to have $\MHT$ in the trigger.  Note this is not always the case and alternate triggers are sometimes utilized to reduce the QCD rate, although we leave this for future studies to explore.

We then evaluate the performance of the variables described in Sec.~\ref{sec:Observables} by computing the background rejection rate for different values of the signal acceptance. In the following, we consider the signal topologies described in Sec.~\ref{sec:nBodySimpMod} as well as the \ZpJ, $t\,\overline{t}$, and QCD multijet backgrounds.  Although not shown explicitly, we find that the results for $W$ + jets and  \ZpJ~backgrounds are qualitatively the same.  This is not surprising as the topologies are very similar.  To reduce complexity and number of results, we show only the results for \ZpJ~backgrounds.

\begin{figure}[h!]
\includegraphics[height=0.1\textwidth]{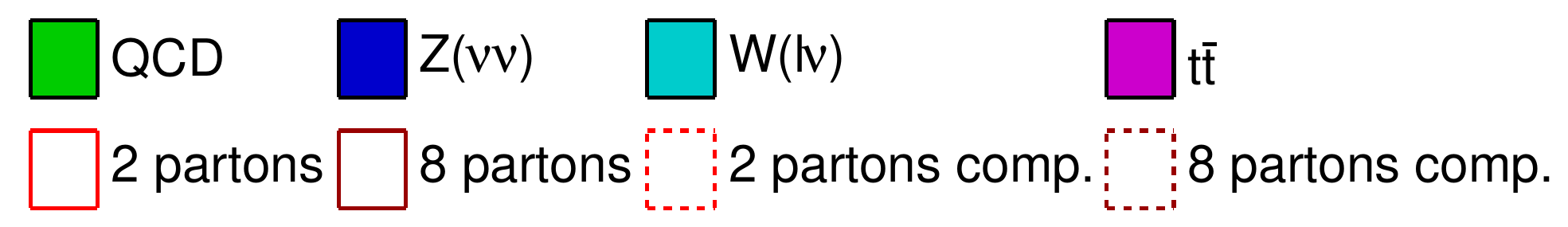}\\[20pt]
\includegraphics[width=0.49\textwidth]{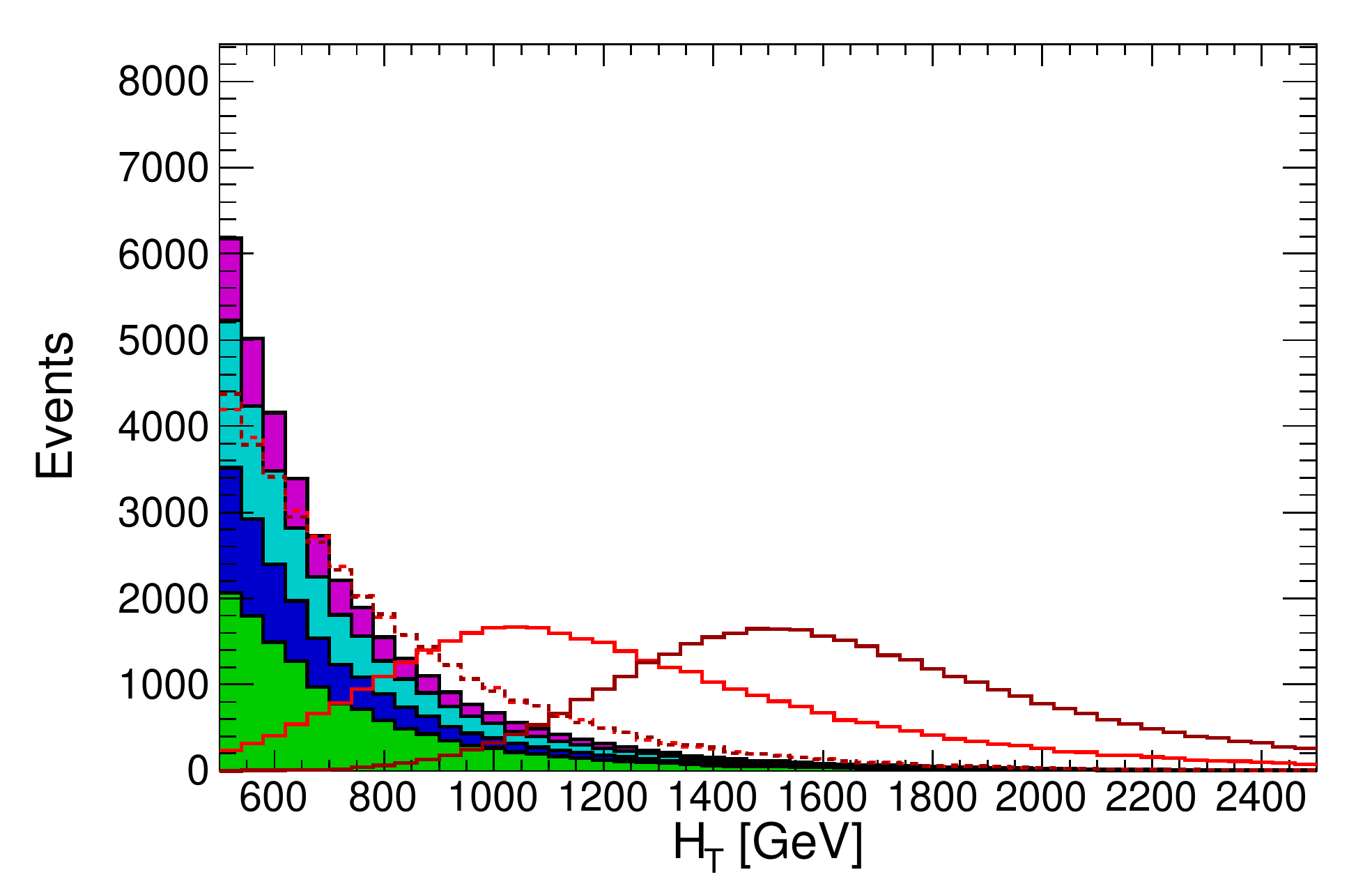}
\includegraphics[width=0.49\textwidth]{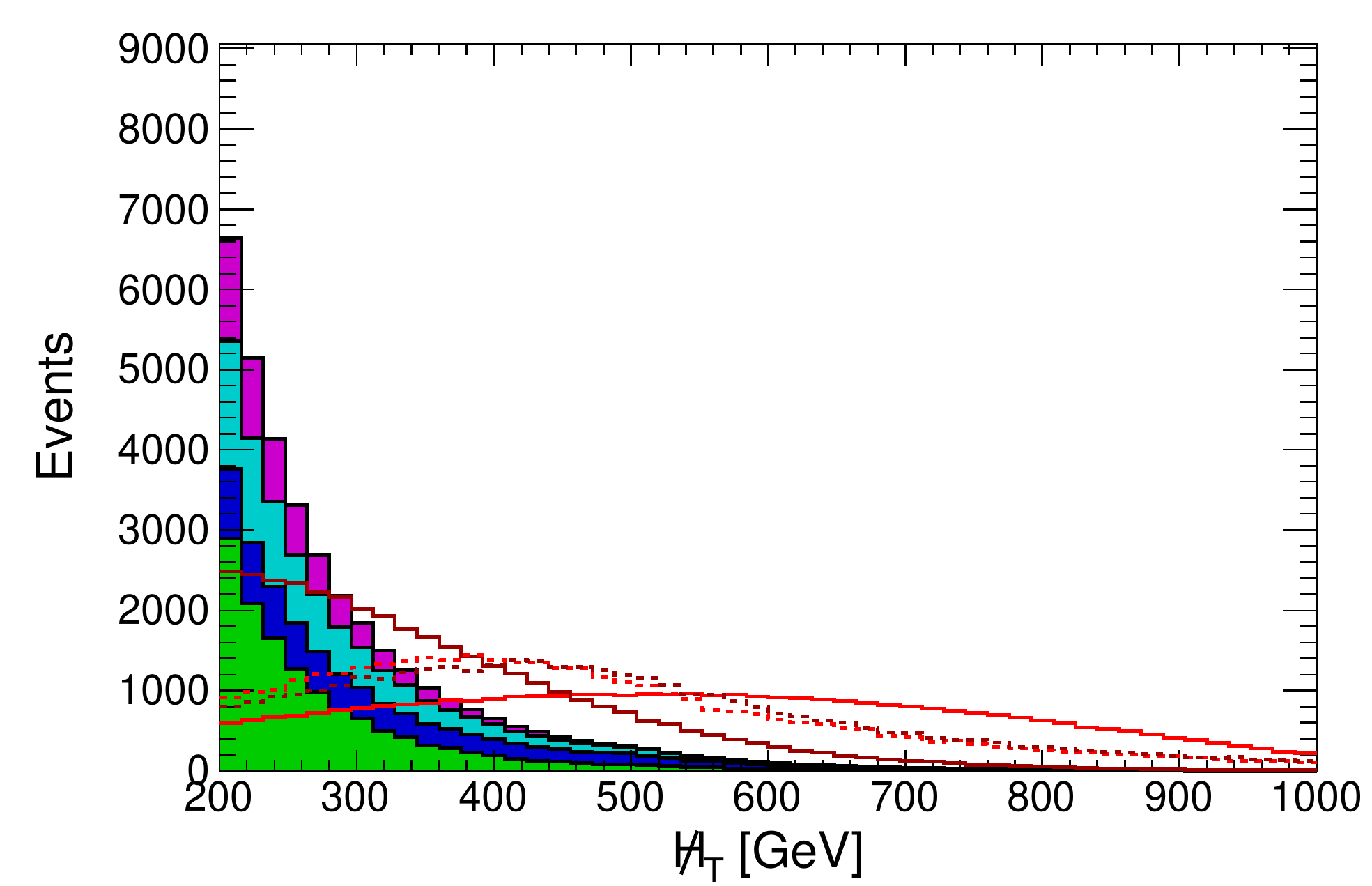}\\
\includegraphics[width=0.49\textwidth]{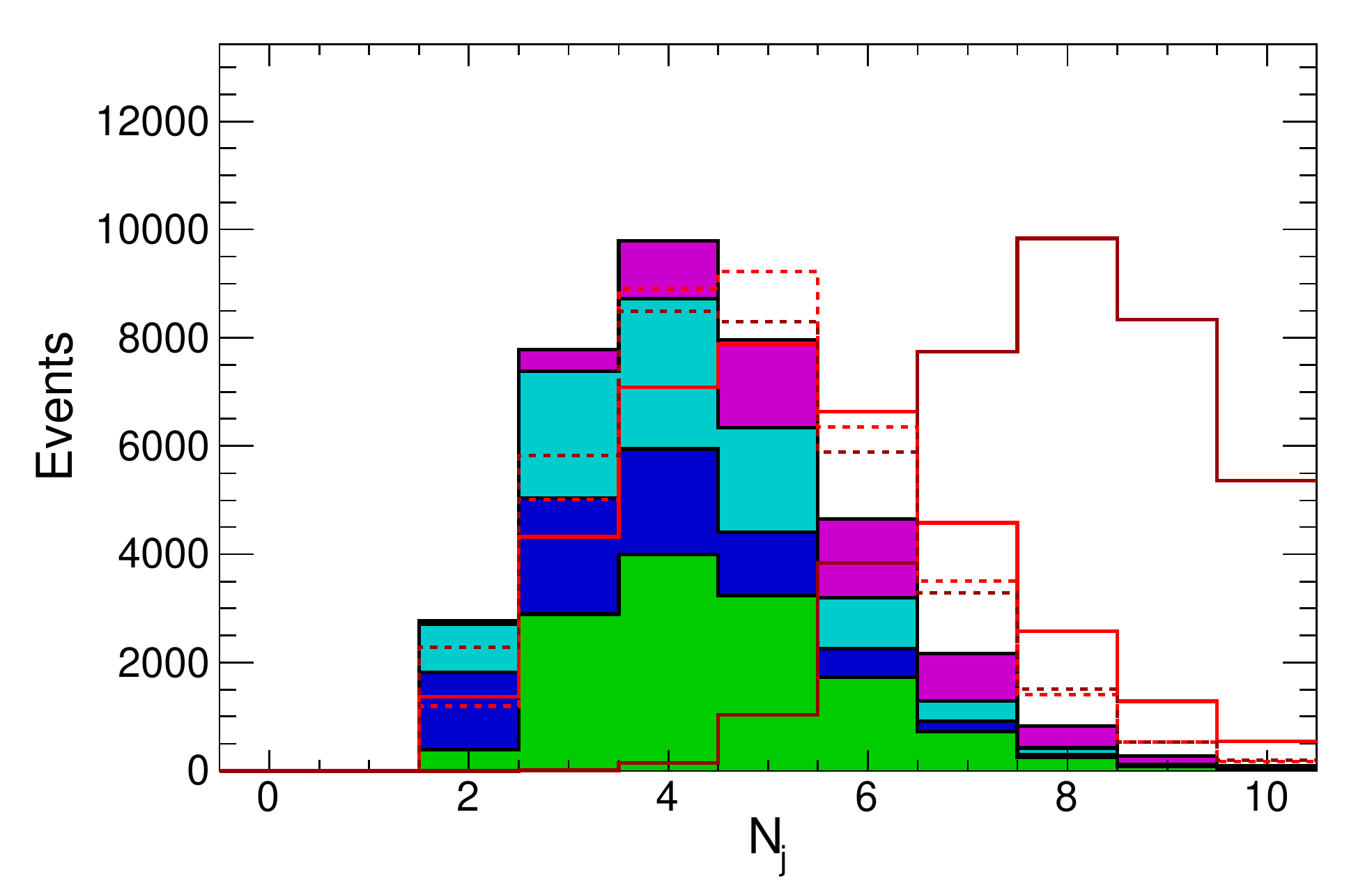}
\includegraphics[width=0.49\textwidth]{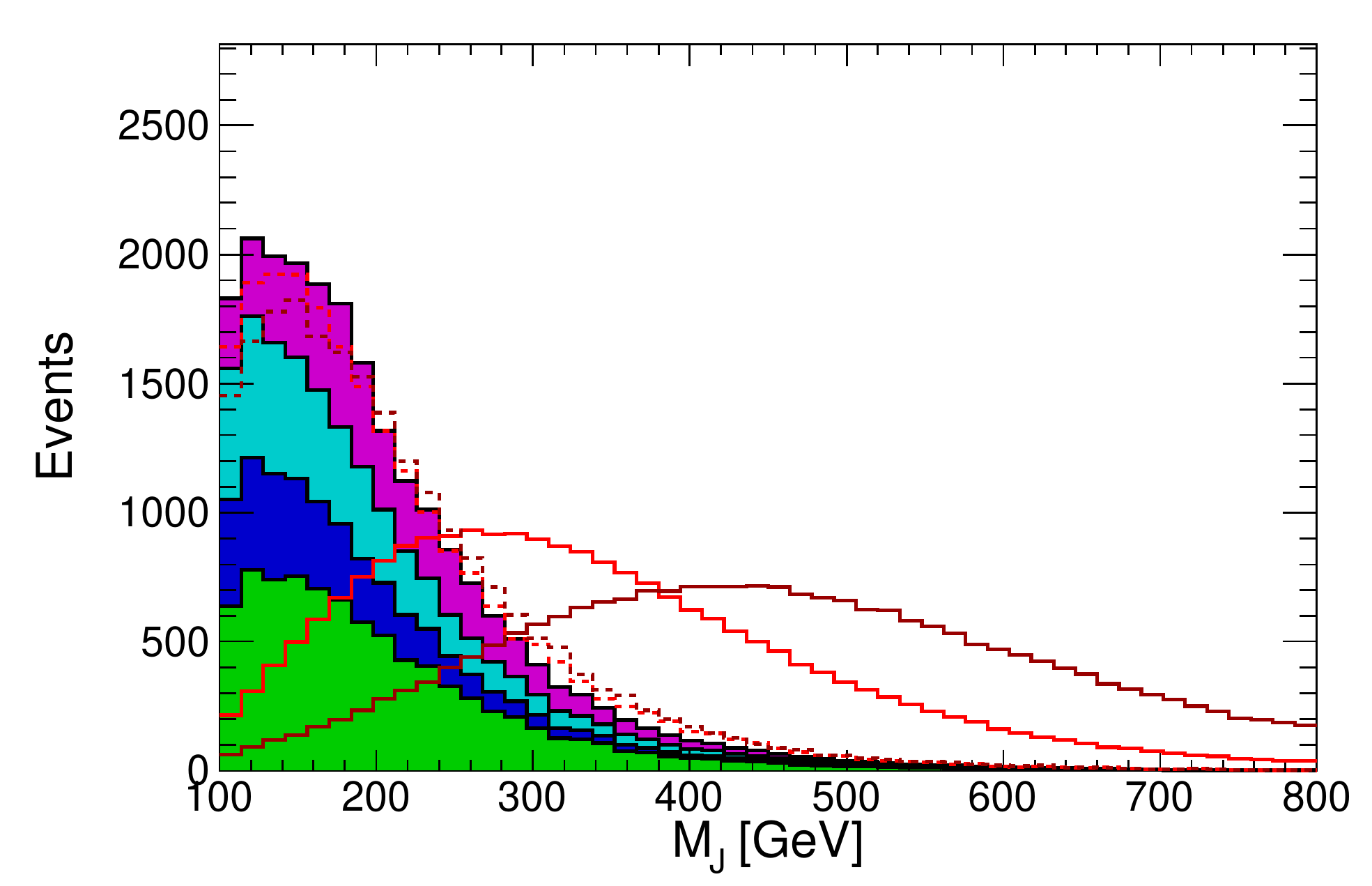}
\caption{\label{fig:1dstacks} 1D distributions of selected observables.  Backgrounds are normalized to 10 fb$^{-1}$.  Signals are normalized to the same yields as the sum of all backgrounds for shape comparison.  The gluino mass 
for all of the signal models is 1 TeV.  The solid (dashed) signal histograms are for uncompressed (compressed) spectra.  }
\end{figure}

\begin{figure}[h!]
\includegraphics[width=0.49\textwidth]{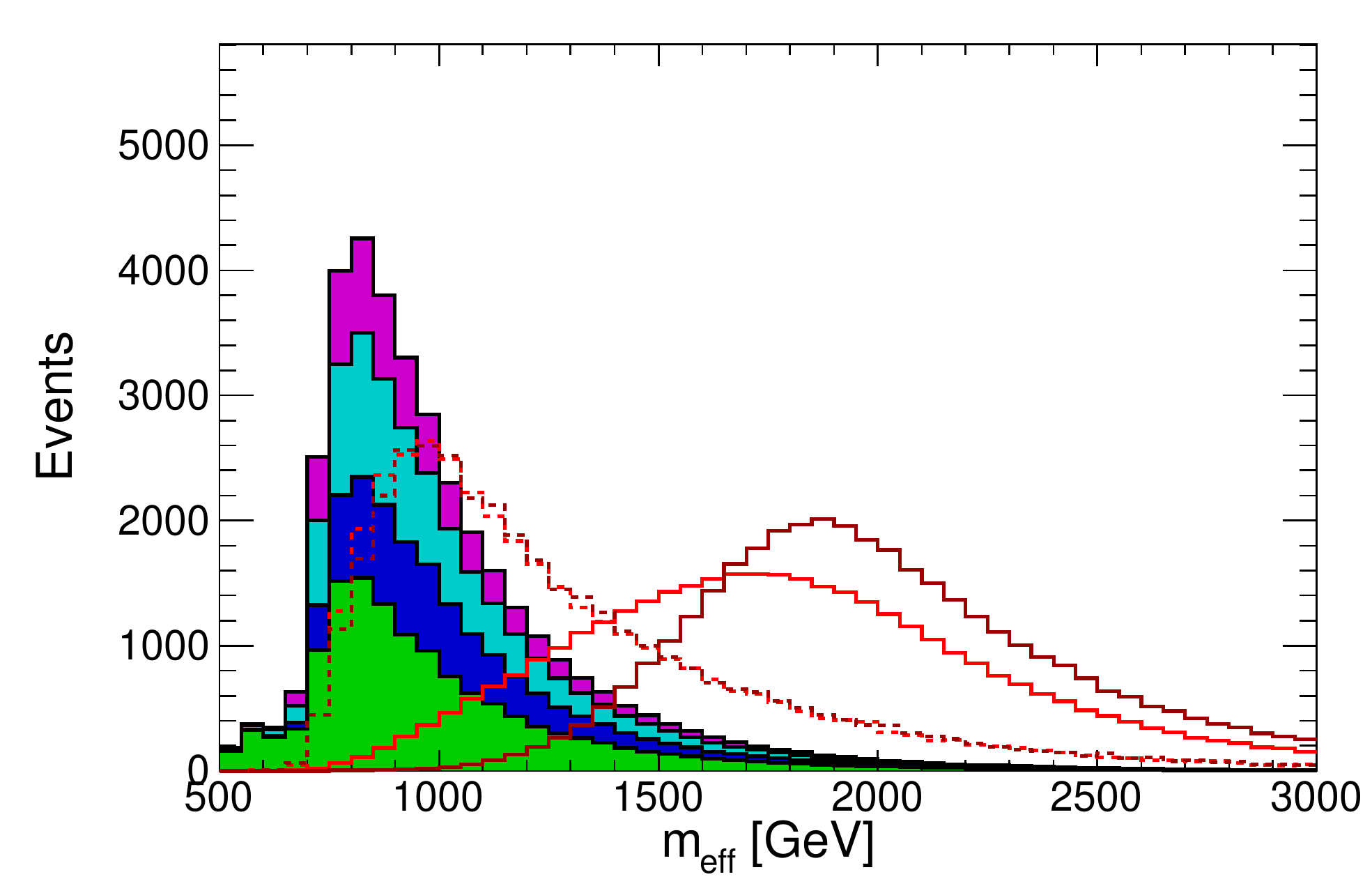} 
\includegraphics[width=0.49\textwidth]{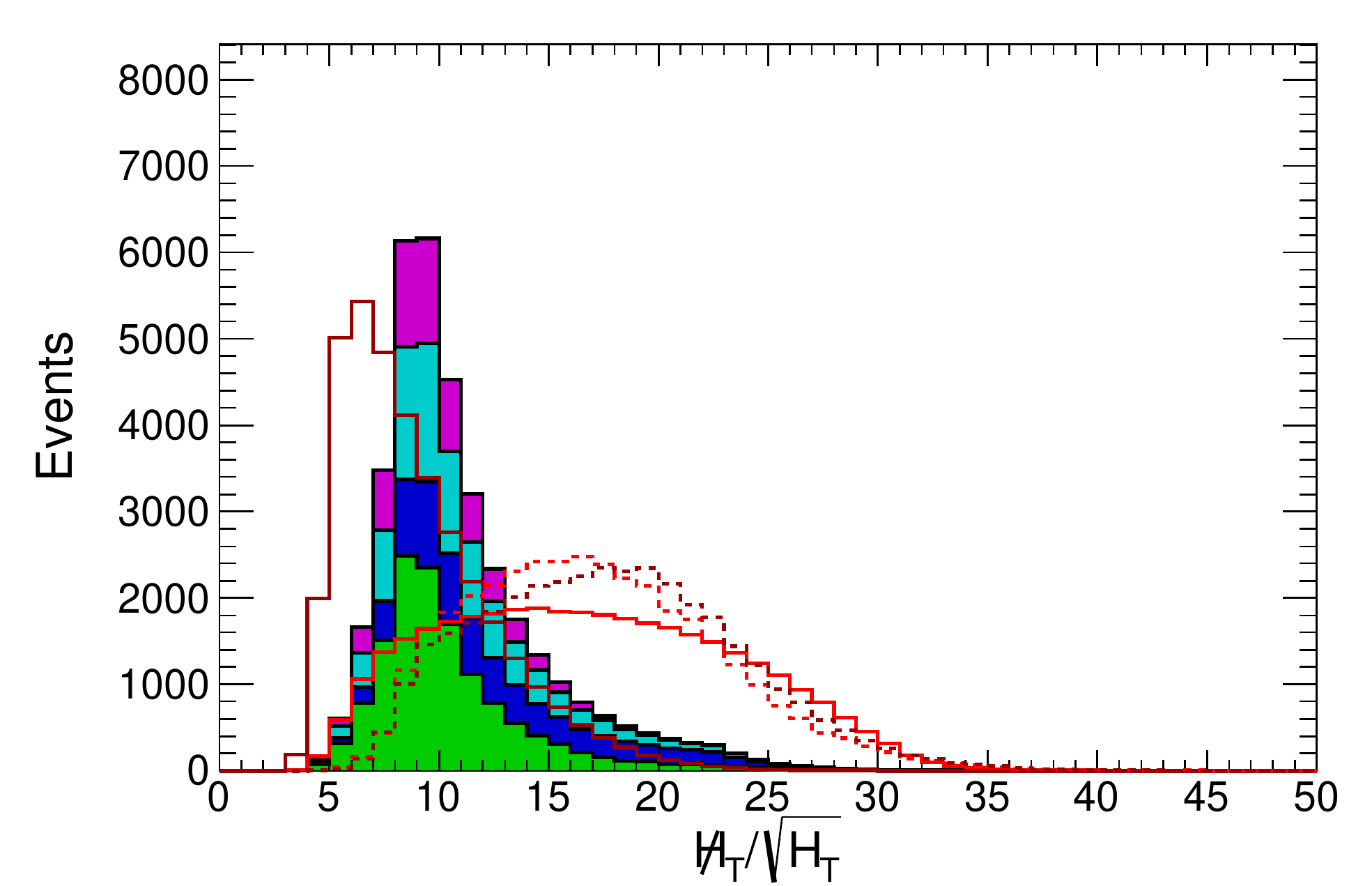}\\
\includegraphics[width=0.49\textwidth]{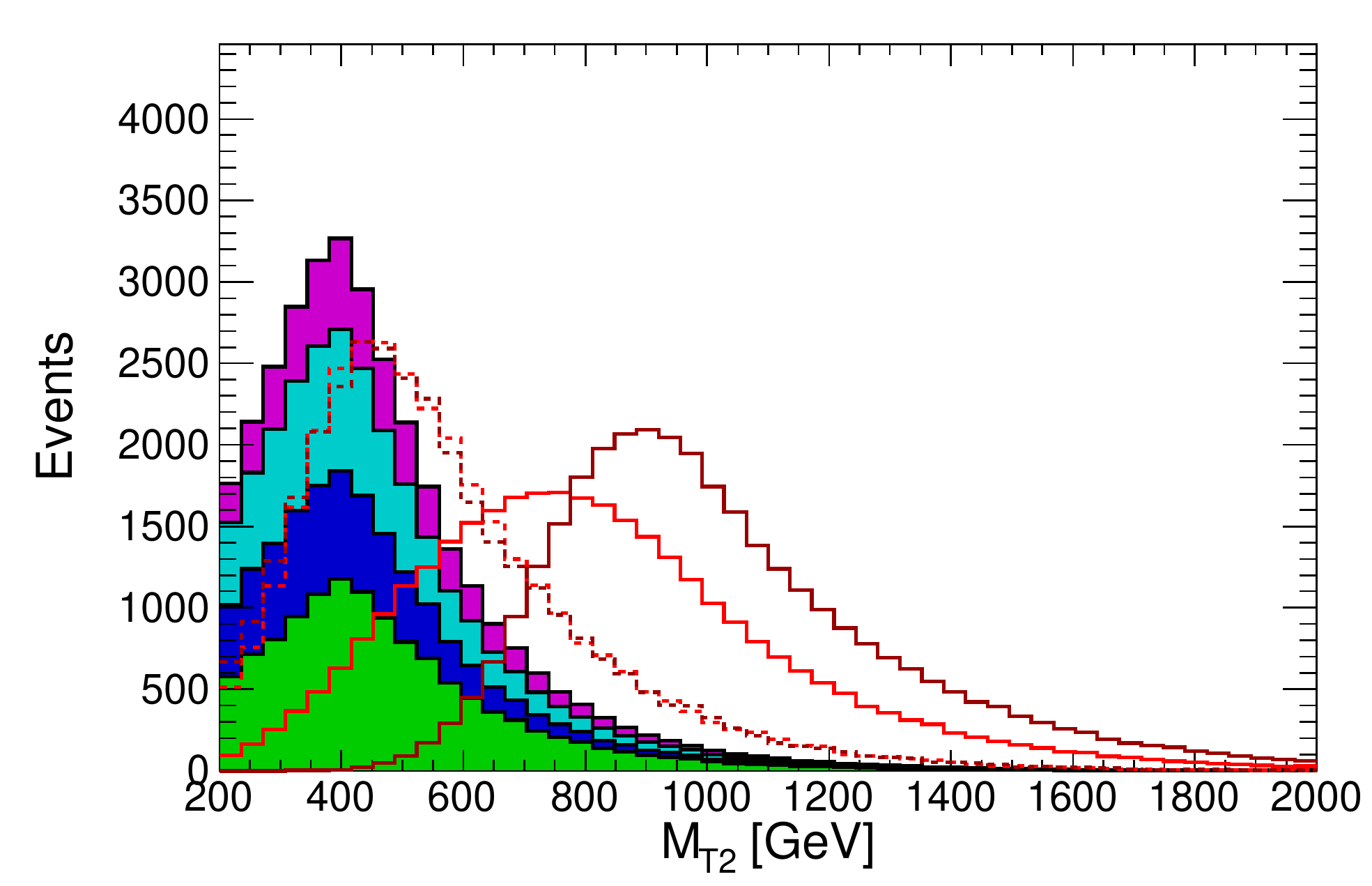}
\includegraphics[width=0.49\textwidth]{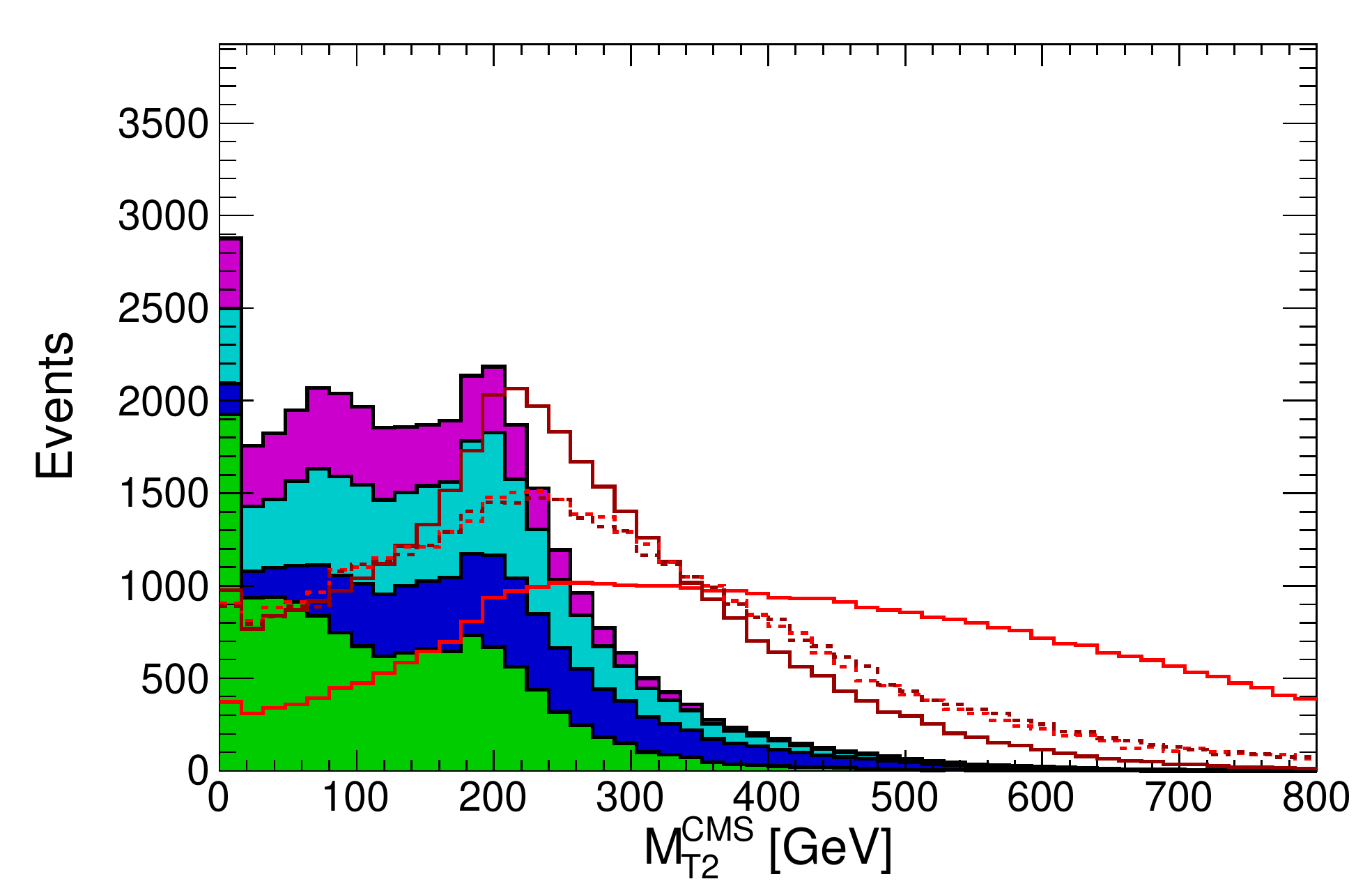}\\
\includegraphics[width=0.49\textwidth]{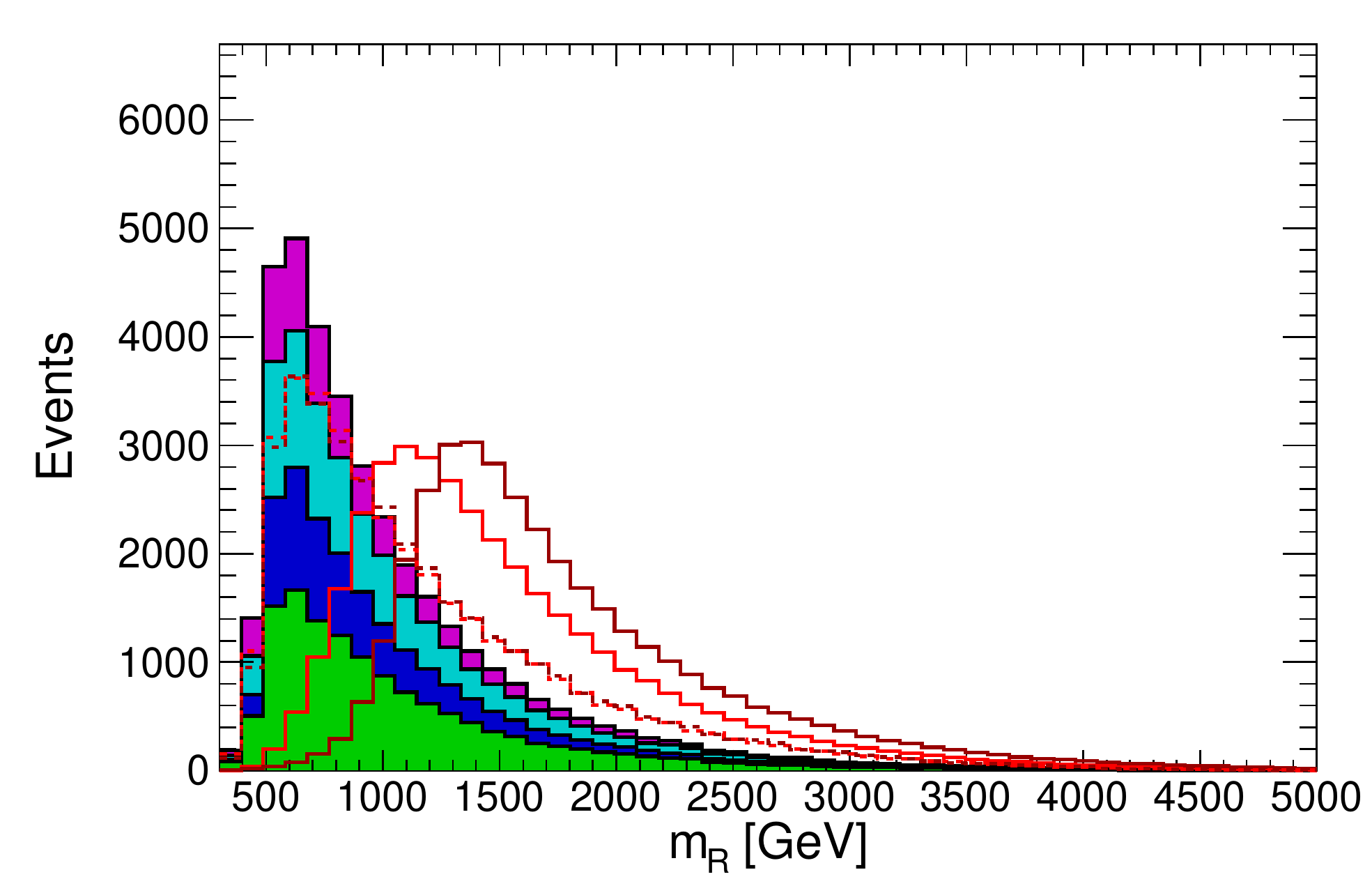}
\includegraphics[width=0.49\textwidth]{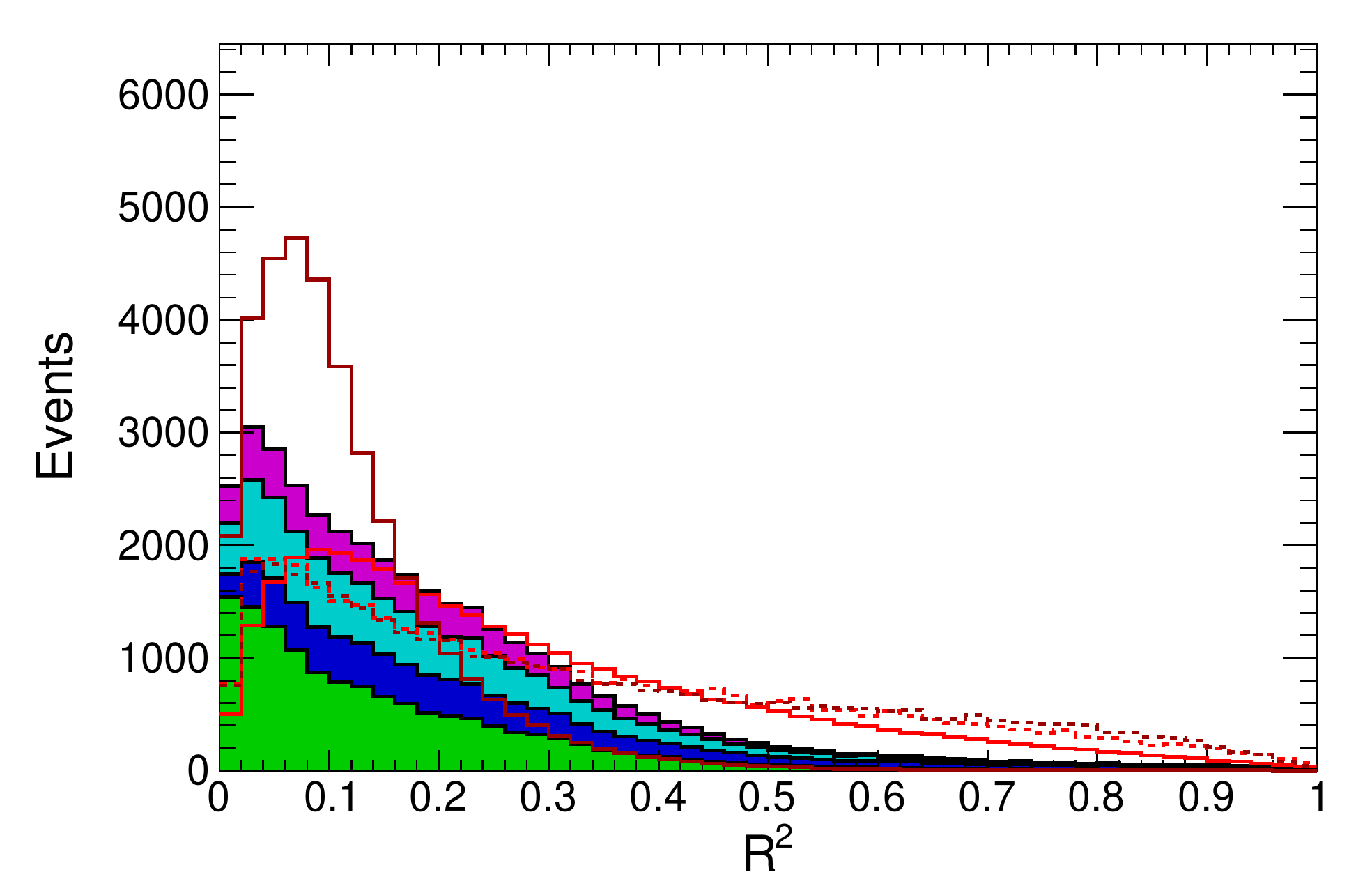}
\caption{\label{fig:1dstacks2} Additional 1D distributions of selected observables.  The legend is given in Fig.~\ref{fig:1dstacks}.  Backgrounds are normalized to 10 fb$^{-1}$.  Signals are normalized to the same yields as the sum of all backgrounds for shape comparison.  The solid (dashed) signal histograms are for uncompressed (compressed) spectra.
for all of the signal models is 1 TeV.}
\end{figure}

The distributions of the variables that are the focus of this study are shown in Figs.~\ref{fig:1dstacks} and \ref{fig:1dstacks2}, after the application of the preselection cuts. 
The statistics are equivalent to  10~fb$^{-1}$ of data at the 13~TeV LHC.
The different background contributions are stacked together: \ZpJ, $W$ + jets, $t\,\overline{t}$ + jets, and QCD.
A variety of signal distributions are also shown in the figures as empty histograms, both with compressed and uncompressed spectra along with a few choices for the number of partons $n$ in the final state.

We validate these plots and also perform additional comparisons that have a tighter phase space requirement against public material from CMS~\cite{Khachatryan:2016kdk}.
We generally find good agreement in the relative normalization of the background contributions and overall agreement within a factor of 2 using the {\tt Delphes} simulation.
The largest discrepancy with ATLAS and CMS public results is the QCD $\MHT$ distribution resulting from fake missing energy which has a noticeably longer tail in our {\tt Delphes} samples.  This can be understood as resulting from jet resolution mismodeling in our simulation.
This implies that the results regarding missing energy variables in QCD could be quantitatively different than what is shown below, although the qualitative conclusions are robust.

\clearpage

Since the relative importance of each of these backgrounds will depend heavily on the selection cuts, we will study the performance of our variables against each background separately.  This helps accommodate the application of our results to searches where cuts on variables other than the ones we consider lead to an alternate background composition. For example, in a search that requires two or more $b$-tagged jets, the dominant background would be $t\,\overline{t}$ while for a compressed spectrum search without $b$-tags, the dominant background would be QCD.   Note that we will not provide results for a ``total" background for this reason. Furthermore, since we have not included $K$-factors the relative cross sections are not robust, and additionally the mismatch found when validating the $\MHT$ QCD tails could be exacerbated by such a naive combination.  

\begin{figure}[h!]
\includegraphics[width=0.4\textwidth]{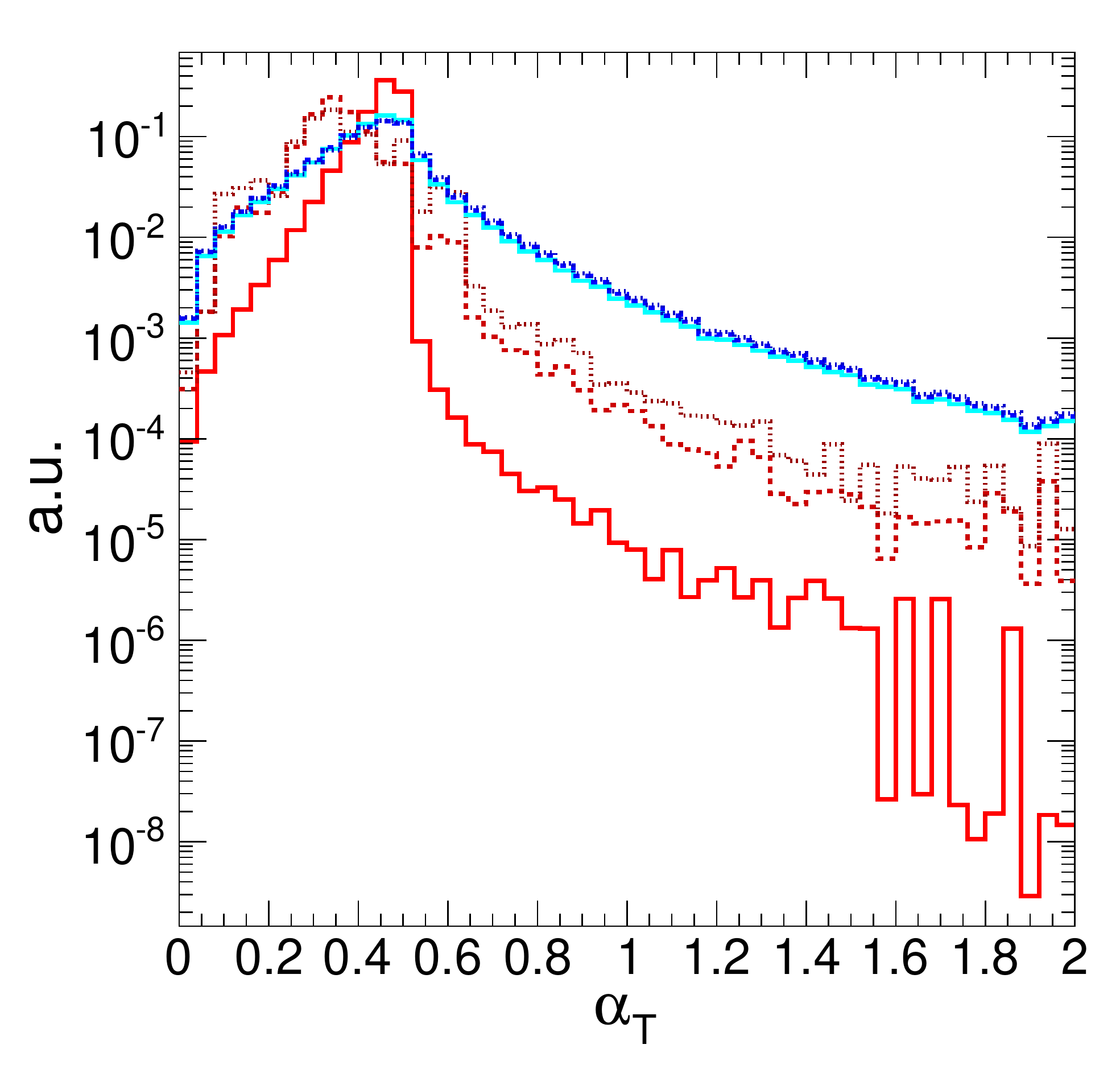}
\caption{\label{fig:alphaT} Distributions of $\alpha_T$ for QCD (red) and gluinos decaying to a single decay parton 
(blue) after various pre-selection requirements. $H_T>$500 GeV (solid lines); $H_T>$500 GeV and $\MHT>200$ GeV (dashed lines); $H_T>$500 GeV and $\MHT>200$, and min($\Delta\phi$)$>$0.4 (dotted lines). The pre-selection cuts shape the background to be similar to the signal.}
\end{figure}

Finally, notice that Fig.~\ref{fig:1dstacks} does not include the $\alpha_T$ distribution.  
We find that this variable is highly correlated with ${\rm min}(\Delta\phi)$ and $\MHT$, and thus loses much of its discriminatory power after applying the pre-selection cuts.  
This effect be inferred from Fig.~\ref{fig:alphaT}, where after applying the $\MHT$ and min($\Delta\phi$) cuts, the QCD distribution in $\alpha_T$ looks much more signal-like.
By considering all three variables, the contribution to the signal region of QCD can be greatly reduced as designed.  Since our interest here is to compare observables in the same region of phase space, we have chosen to use a pre-selection 
which roughly conforms to the ATLAS and CMS $\MHT$/$H_{T}$ triggers.   We leave the study of $\alpha_T$ outside of the preselection region to future work.

\begin{figure}[h!]
\includegraphics[width=0.49\textwidth]{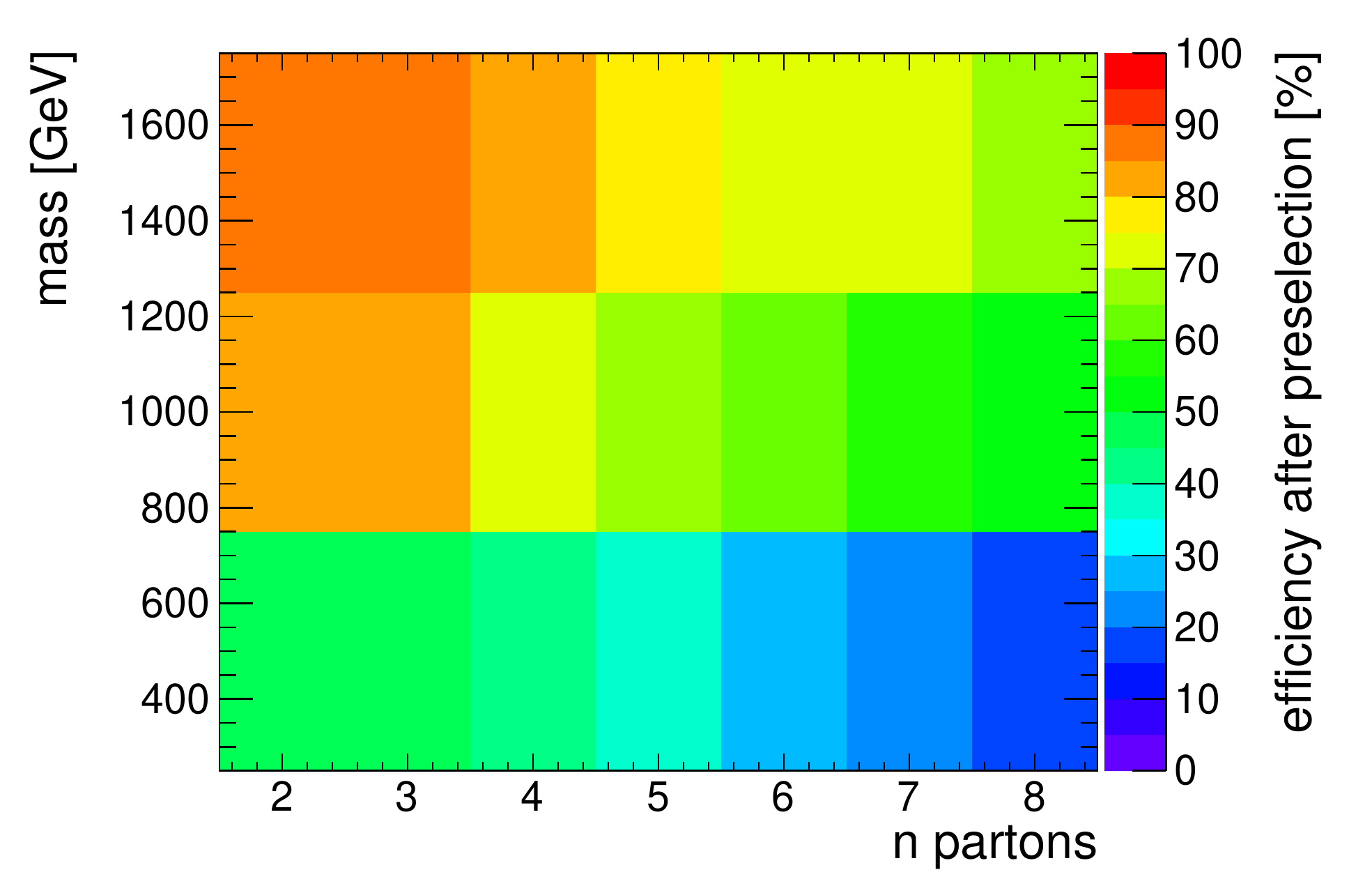}
\includegraphics[width=0.49\textwidth]{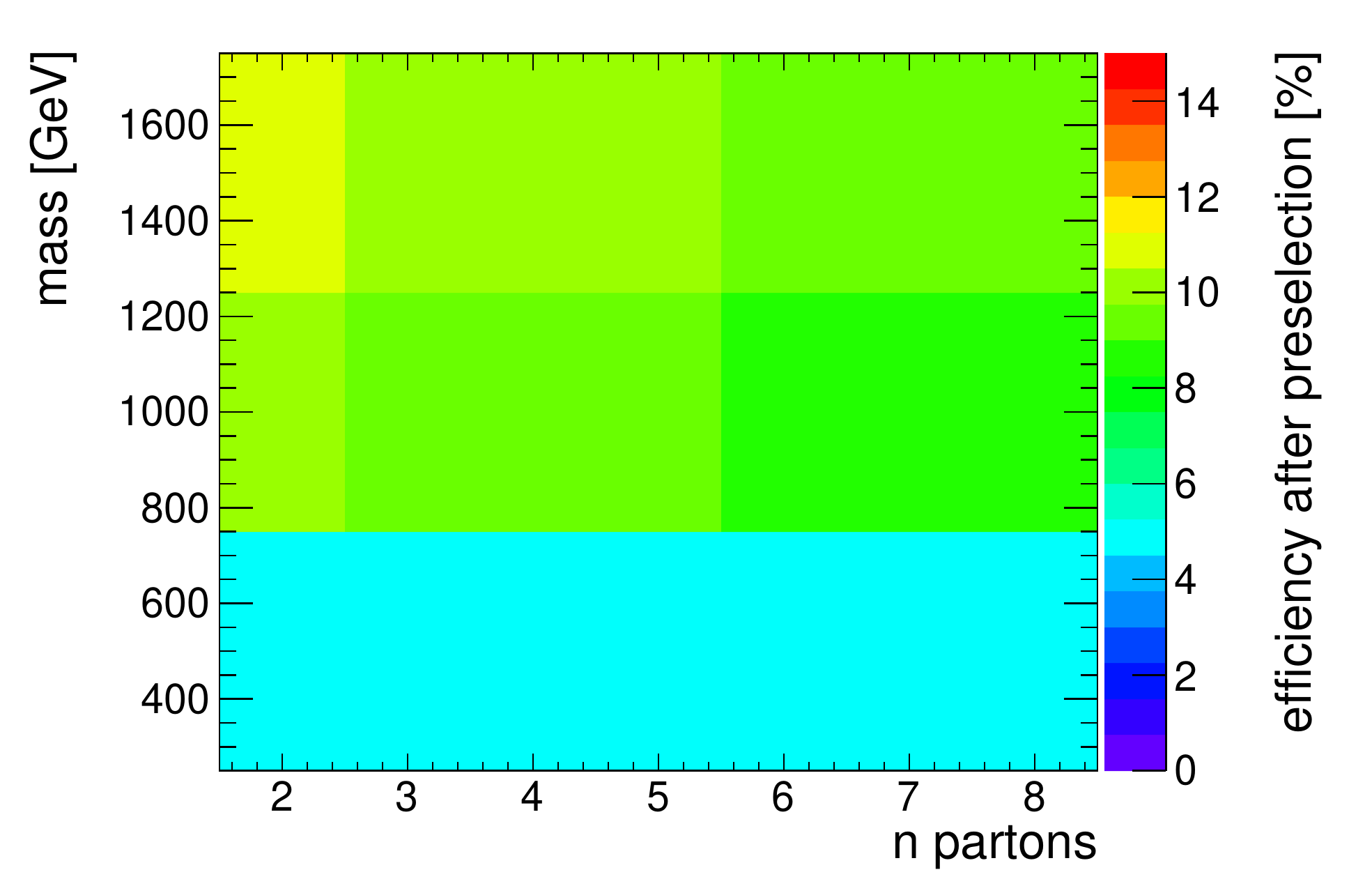}
\caption{\label{fig:abssigeff} Signal efficiencies of $n$-body simplified signal models after preselection cuts.  The \emph{uncompressed} signal points are on the left and the \emph{compressed} signal points are on the right.}
\end{figure}

In the following, we consider the background rejection power for a given set of observables as a function of the number of matrix element partons in the final state and for each of the different background processes.
We fix the signal efficiency {\it with respect to the pre-selection cuts} to $\varepsilon_\text{sig} = 10$\%, and compute the background rejection power $1/\varepsilon_\text{bkg}$, again where the background efficiency is computed with respect to the pre-selection cuts.  
A signal efficiency of 10\% is typical of most searches -- we checked additional signal efficiency points $\varepsilon_{\text{sig}}\lesssim 25\%$ and find that the results do not change qualitatively.
For completeness, Fig.~\ref{fig:abssigeff} shows the absolute selection efficiency after preselection so that it is possible to infer the implications of our results in terms of limits on signal production cross section $\times$ branching ratio.
To get the absolute signal efficiency of the final cuts, multiply the pre-selection efficiency by 10\%.

\subsection{Massless Neutralino Limit}

This section applies our methodology to the study of $n$-body signatures with a massless neutralino.  Backgrounds are considered separately to isolate the essential kinematic features of each signal and background.  We begin with a study of the individual variables of interest, followed by judiciously chosen combinations of two and three observables.

\subsubsection*{\textbf{\textit{One variable at a time}}}
\label{subsubsec:oneVar}

\begin{figure}
        \centering
        \hspace{33pt}\includegraphics[width=0.4\textheight]{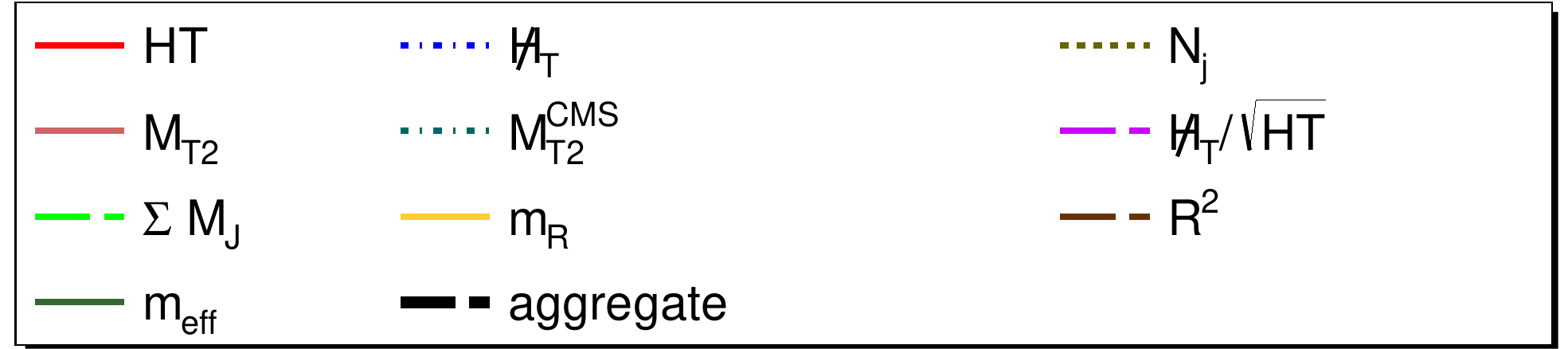}\\[10pt]	
	\includegraphics[height=0.26\textheight]{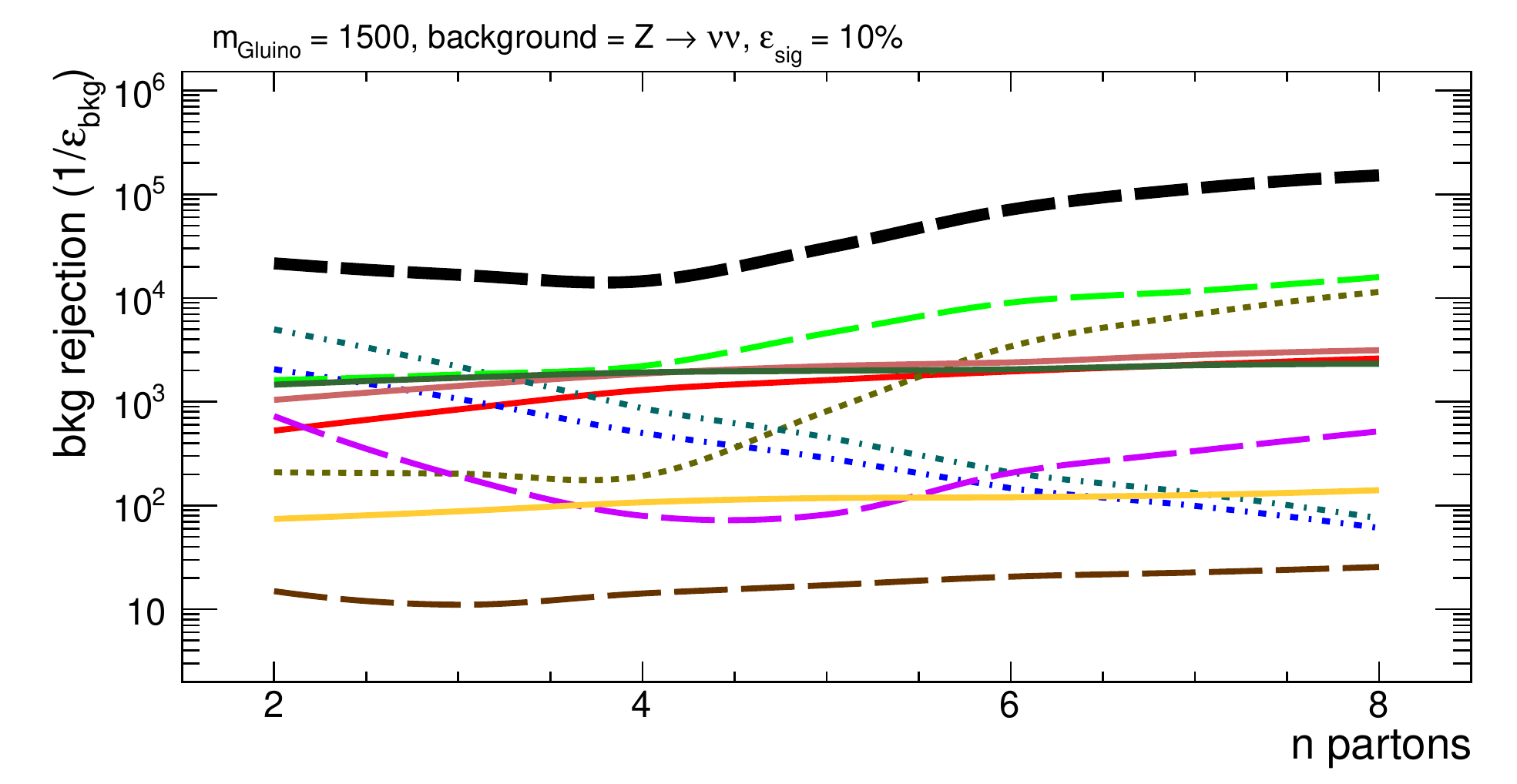}\\
	\includegraphics[height=0.26\textheight]{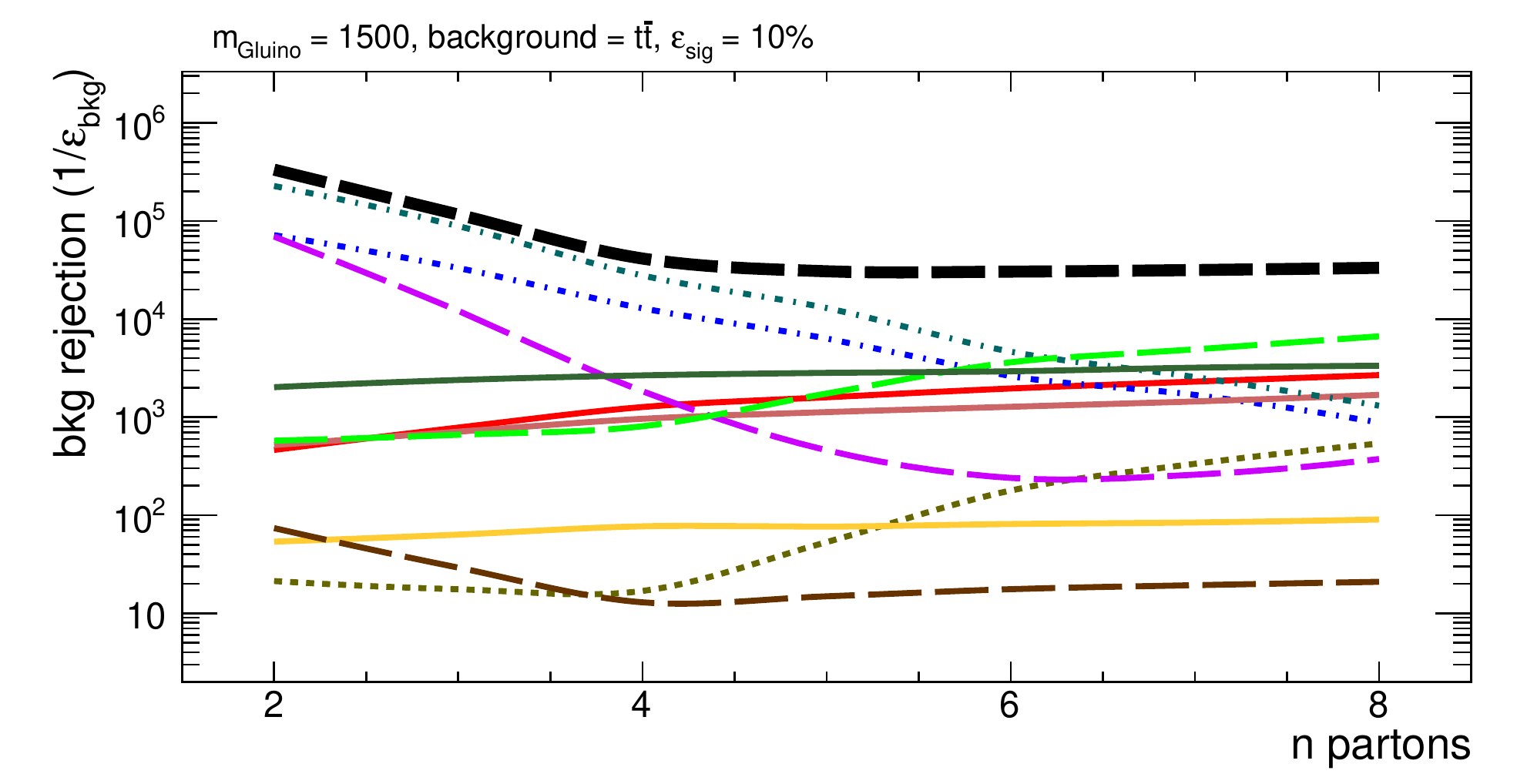}\\
	\includegraphics[height=0.26\textheight]{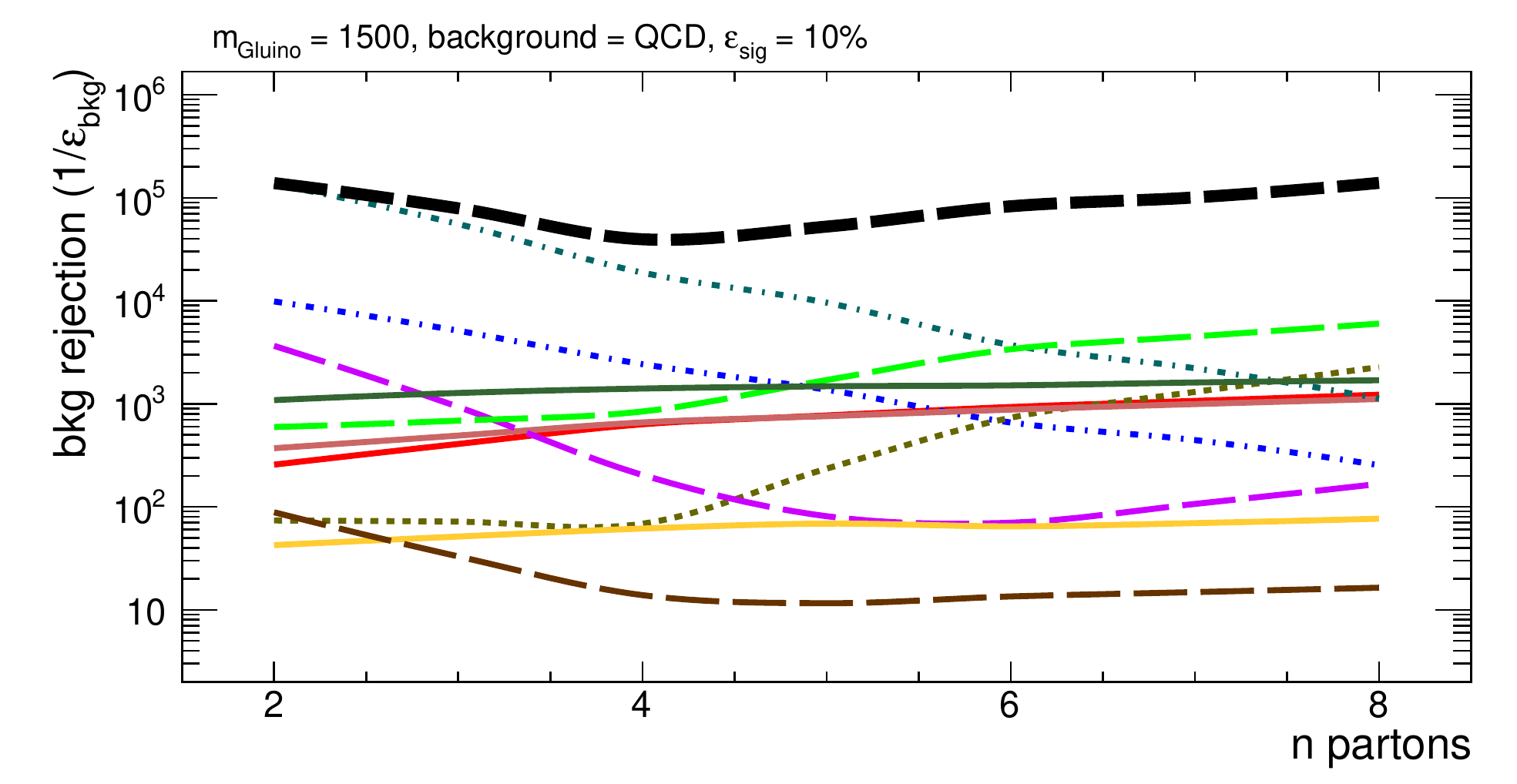}
	\caption{Background rejection efficiency for \emph{uncompressed} signals as a function of the total number of partons, $n$, for $10$\% signal efficiency. The variables shown are given in the legend at the top of the page. From top to bottom we show the rejection efficiency against \ZpJ, $t\,\overline{t}$, and QCD.}
	\label{Fig: 1var_nPartons} 
\end{figure}

We first evaluate the performance of each observable as a function of the total number of decay partons in the final state. The results for a $1.5$ TeV gluino and for the different backgrounds are shown in Fig.~\ref{Fig: 1var_nPartons}. Note that we consider both Razor variables $M_R$ and $R^2$ separately. We also define an aggregate analysis which feeds all the variables given above to the BDT. We regard this as the ``optimal" background rejection rate that is possible, and show it in each plot as a reference.

Based on the behavior of these variables versus number of partons, we can already learn many valuable lessons and define the following variable categories:\footnote{Note that while these categories are somewhat arbitrary, we find them to be useful as a guide for which combinations of variables will lead to the largest gain in sensitivity.}

\begin{itemize}
\item \typea\!: The \textbf{missing energy variables} $\Big\{\vec{\leavevmode\cancel{H}}_T,\MCMS\Big\}$
 are sensitive to the properties of the invisible states, \emph{e.g.}~how many neutralinos in the event, what is their mass, etc.;
\item \typeb\!: The \textbf{energy scale variables} $\Big\{H_T,M_{T2},M_{R},m_\text{eff}\Big\}$ are sensitive to the overall energy scale of the event, \emph{e.g.}~the gluino mass;  
\item \typec\!:  The \textbf{energy structure variable} $\big\{N_j\big\}$: is sensitive to the structure of the visible energy, \emph{e.g.}~how many partons are generated in the decay;
\item \typehybrid\!: The \textbf{hybrid variables} $\Big\{$Razor $R^2$, $\MHT/\sqrt{H_T}$, $M_J \Big\}$ 
exhibit characteristics from multiple types depending on the number of decay partons in the event.\footnote{The hybrid variables can be categorized as Razor $R^2$ $\big[$\typeac\!$\big]$;  $\MHT/\sqrt{H_T}$ $\big[$\typeab\!$\big]$; and $M_J$ $\big[$\typebc\!$\big]$.}
\end{itemize}
The performance of some of the variables obey trends that are independent of the background.
\typea variables perform best at low number of partons since, for low multiplicity final states, each individual final state jet or particle is expected to have a large $p_T$.
As the number of visible objects and visible energy increases, the total energy has to be split between more and more final states. 
\typea variables therefore become progressively supplanted by \typeb and \typec variables.

\begin{figure}
	\centering
      \hspace{15pt} \includegraphics[width=0.55\textheight]{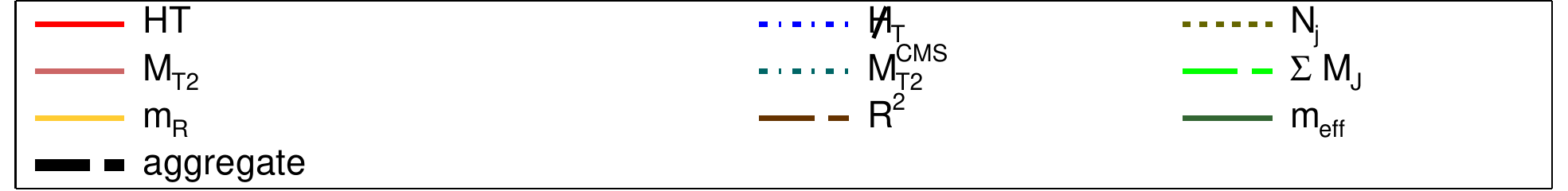}\\[10pt]
	\includegraphics[width=0.34\textheight]{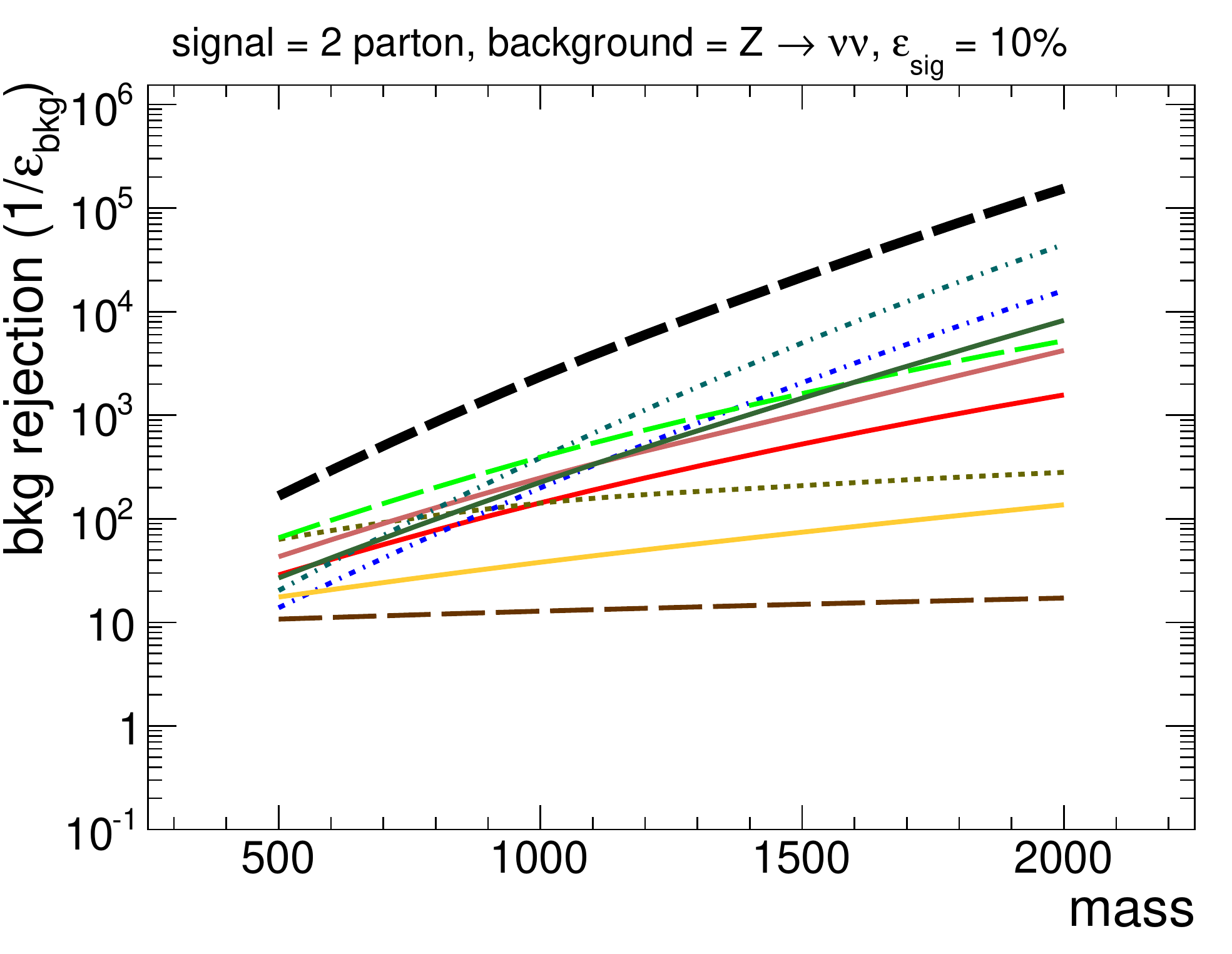}
	\includegraphics[width=0.34\textheight]{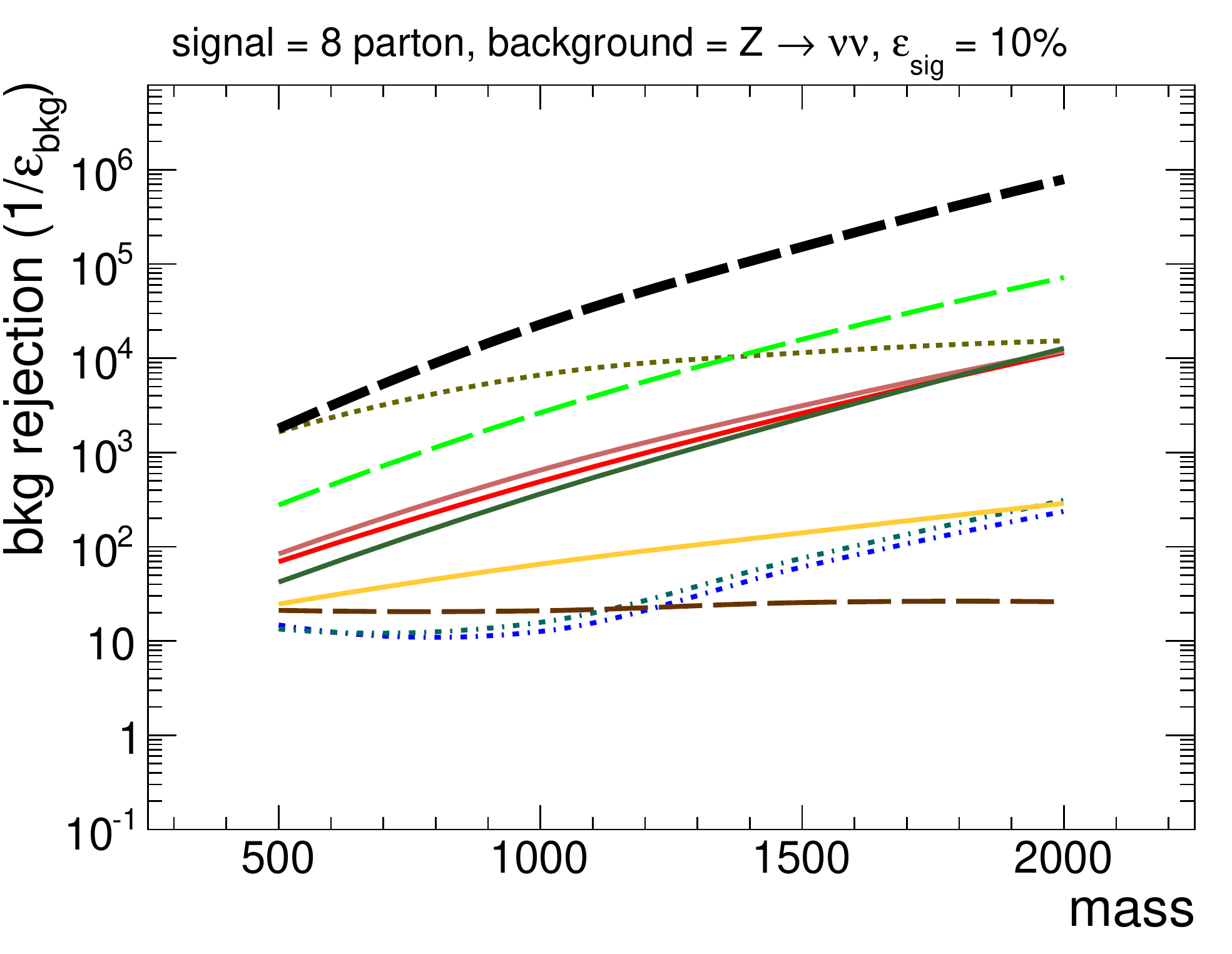}\\
	\includegraphics[width=0.34\textheight]{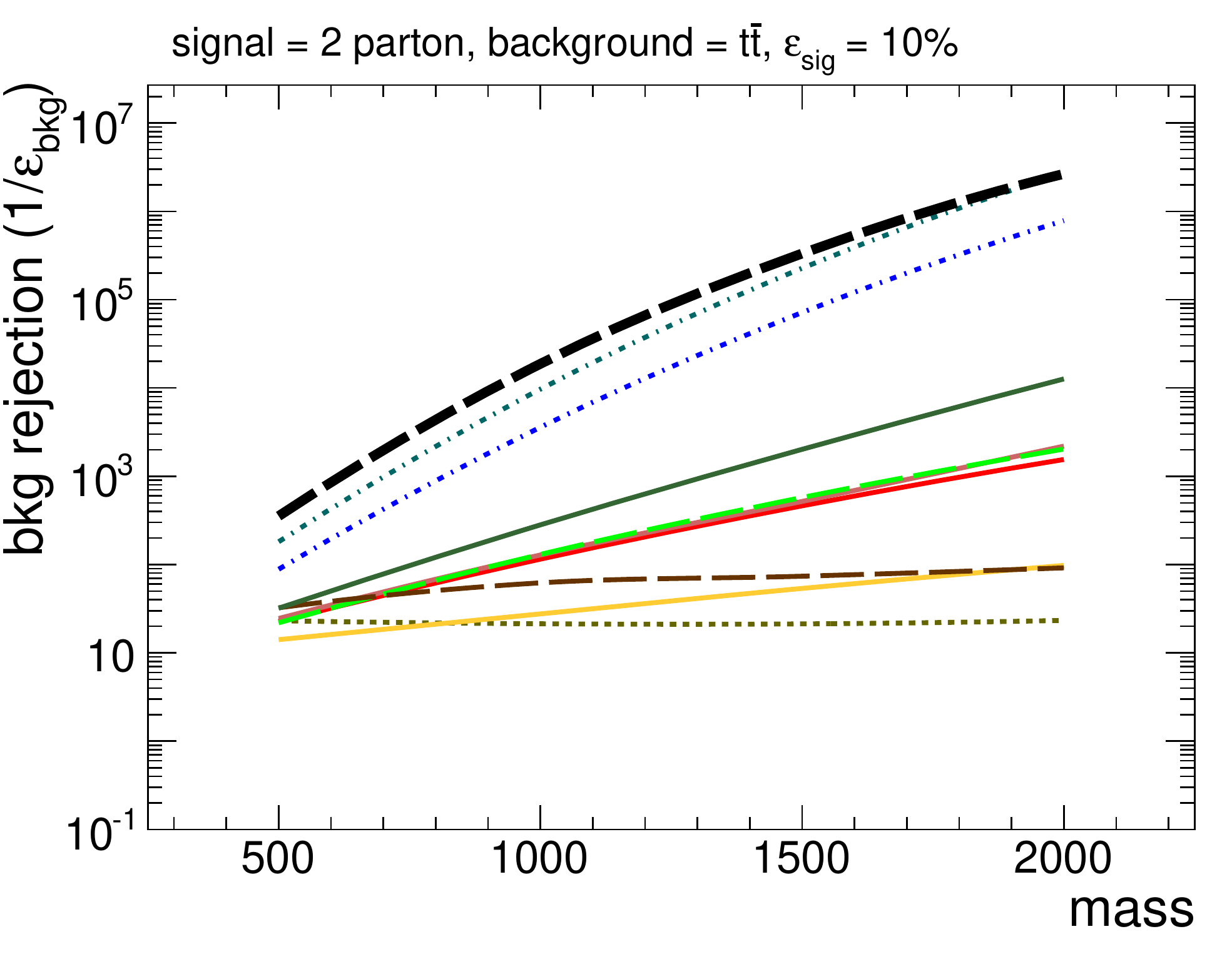}
	\includegraphics[width=0.34\textheight]{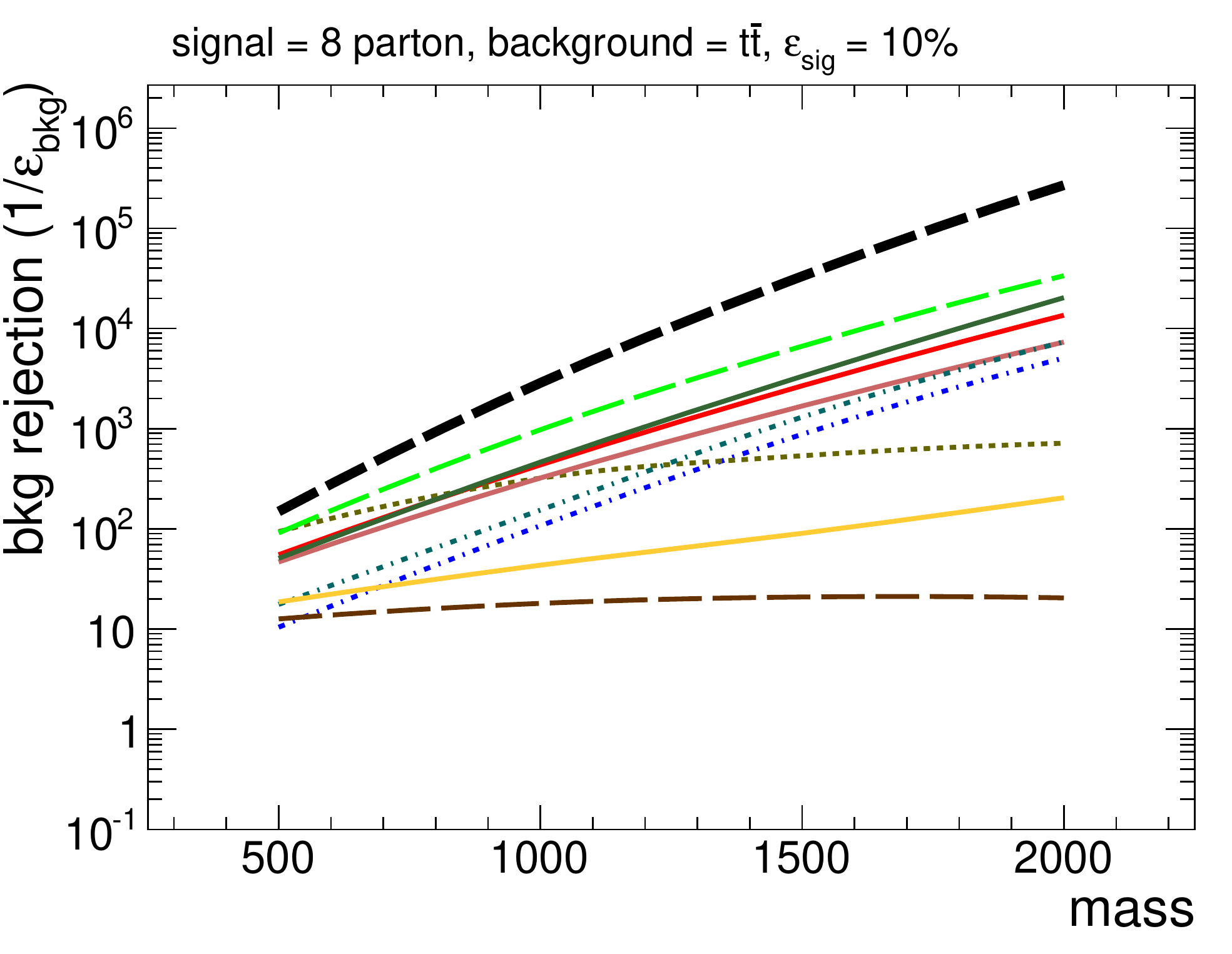}\\
        \includegraphics[width=0.34\textheight]{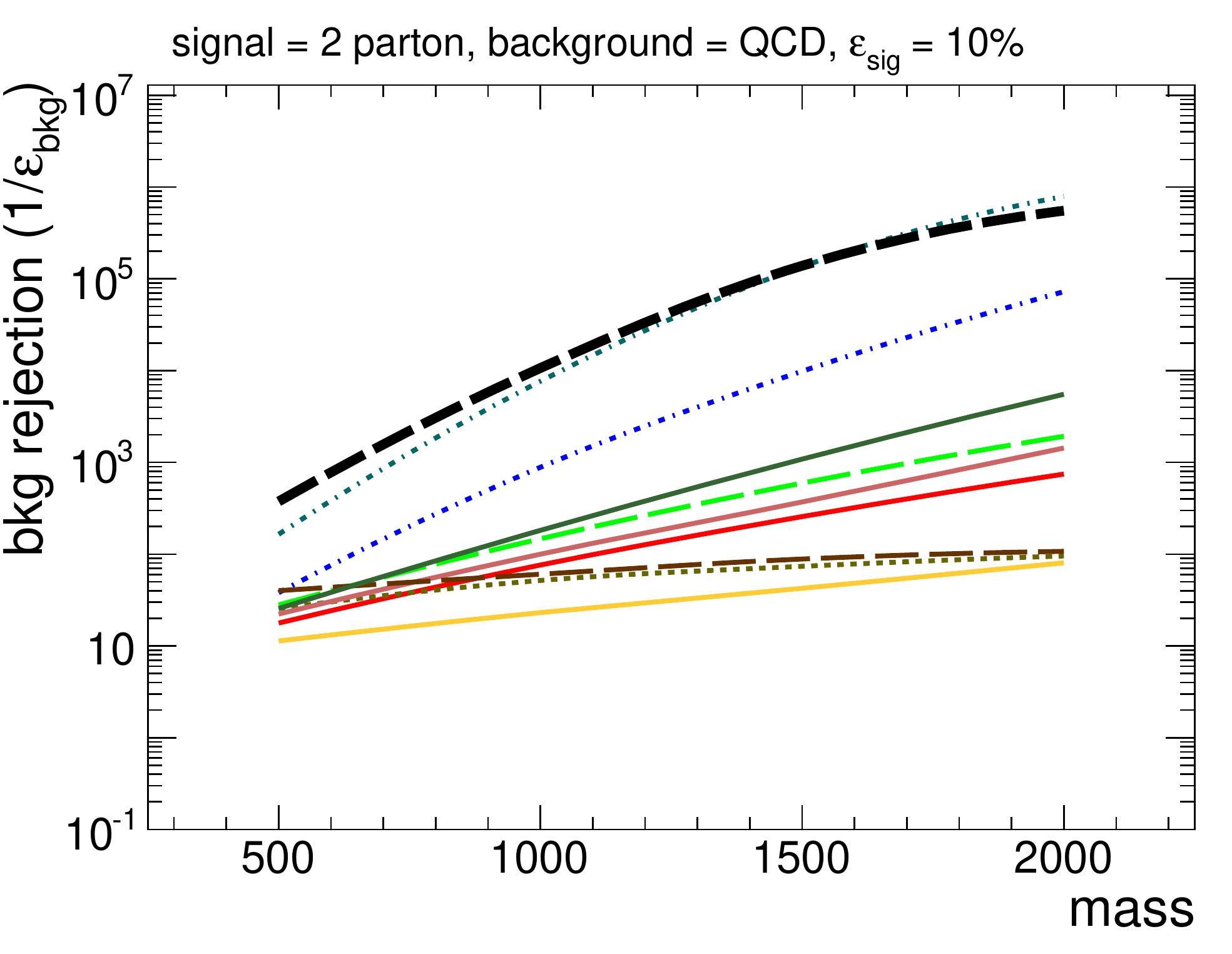}
	\includegraphics[width=0.34\textheight]{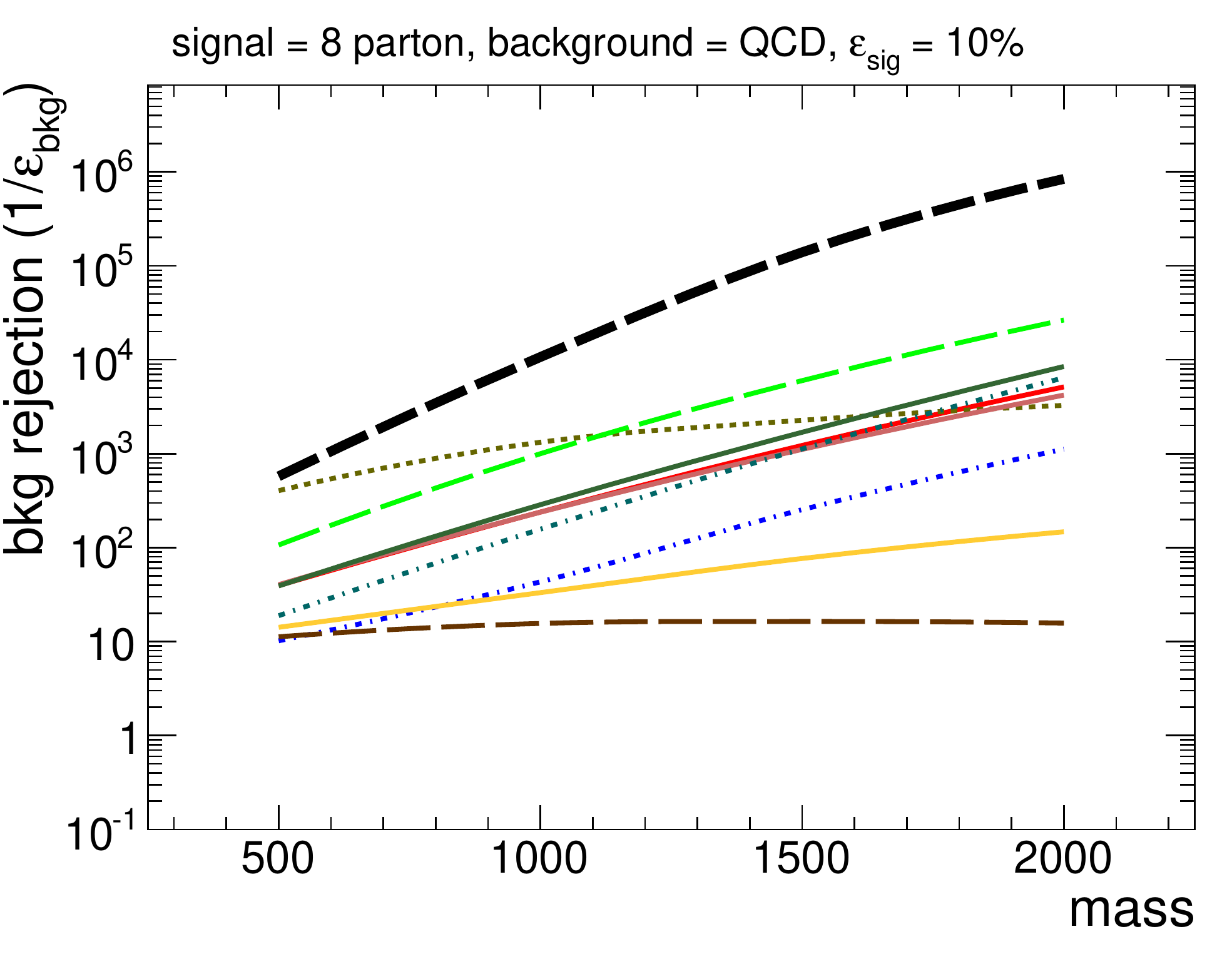}	
	\caption{Background rejection efficiency given a $10$\% signal efficiency for the \emph{uncompressed} signal as a function of the gluino mass for two different choices of the number of partons $n = 2$ (left) and $n=8$ (right). The variables shown are given in the legend at the top of the page.  From top to bottom we show the rejection efficiency against \ZpJ, $t\,\overline{t}$, and QCD.}
	\label{Fig: 1var_mGluino} 
\end{figure}

The relative importance of these different types becomes apparent when we consider performance across different backgrounds.
\typea variables are more important in $t\,\overline{t}$ and QCD events, while in \ZpJ~they are less effective 
due to the large recoil of invisible energy already present in the background.
Therefore, in \ZpJ, \typeb and \typec variables perform better.  
\typec variables tend to be more powerful against \ZpJ~compared to $t\,\overline{t}$.  This can be understood since in $t\,\overline{t}$ events, energy structure is a natural consequence of the multiple scales in the problem.
It is interesting to note that for low $n$ partons, the \typea variables perform very well, and in the particular case of QCD and $t\,\overline{t}$  are near optimal.

We also provide results as a function of the gluino mass as shown in Fig.~\ref{Fig: 1var_mGluino} for the different backgrounds under consideration.  We find that for a low number of partons, the performance of \typea variables improves quickly with mass, though this trend is mitigated as $n$ partons increases.  \typec variables do not have a strong dependence on the gluino mass as they are more sensitive to the structure of the energy.   Focusing on the individual variables we can infer the following lessons:
\begin{itemize} 
\item For uncompressed spectra, $\MCMS$ tends to perform better than $\MHT$.  This is unsurprising since the variable is optimized for a massless neutralino.
\item $H_T$ and $M_{T2}$ perform very similarly while $M_R$ tends to perform worse in terms of background rejection; this is expected as $M_R$ is highly complementary to $R^2$ and will be examined further below.  Meanwhile, $m_\text{eff}$ tends to do the same or slightly better than $H_T$ at low $n$ partons where performance gains are largest for $\MHT$.  However, $m_\text{eff}$ is not superior to $\MHT$ itself.  
\item $R^2$ tends to perform like a \typea variable at low $n$ partons and like a \typeb variable a high $n$ partons although the performance overall is worse; again this is expected since the real power of Razor comes from the exploiting $R^2$ and $M_R$ together.
\item $\MHT$/$\sqrt{H_T}$ tends to perform like a \typea variable at low $n$ partons and like a \typeb variable at high $n$ partons; though it never performs better than both types.
\item $M_J$ tends to perform like a \typeb and \typec variable becoming more \typec-like at high $n$ partons.  It typically does better than both $H_T$ and $N_j$ except at the highest $n$ partons for lower gluino masses.
\end{itemize}

Studying each variable separately shows that unsurprisingly no single variable maximizes the performance throughout all of the phase space considered here. Although the performance of observables like $H_T$ and $\MCMS$ has a weak dependence on $n$, variables such as $\MHT$ or $N_j$ exhibit much better discriminating power, but only for some categories of signals. 

Inclusive searches aimed at a large variety of signatures therefore consider a minimal set of discriminating variables that cover complementary regions of the parameter space. Building such a set requires understanding the correlations between the different variables, which cannot be captured by the previous study. In the following sections, we study the discriminating power of algorithms that take into account more than one variable.

\subsubsection*{\textbf{\textit{Correlations between variables}}}
\label{subsubsec: correlations}

At this point, we have explored the performance of the variables individually, and used the results to classify them into three basic categories (plus hybrid).  Yet, no single variable was a clear winner for the full phase space explored for all values of $n$.  Therefore, it is interesting to explore how complicated of an approach is required to asymptote to the ``ideal" aggregate result for all signals and backgrounds.   This section is devoted to exploring combinations of variables that will lead to an improved discrimination power.  

\begin{figure}[h!]
	\centering
\hspace{20pt}\includegraphics[width=0.27\textheight]{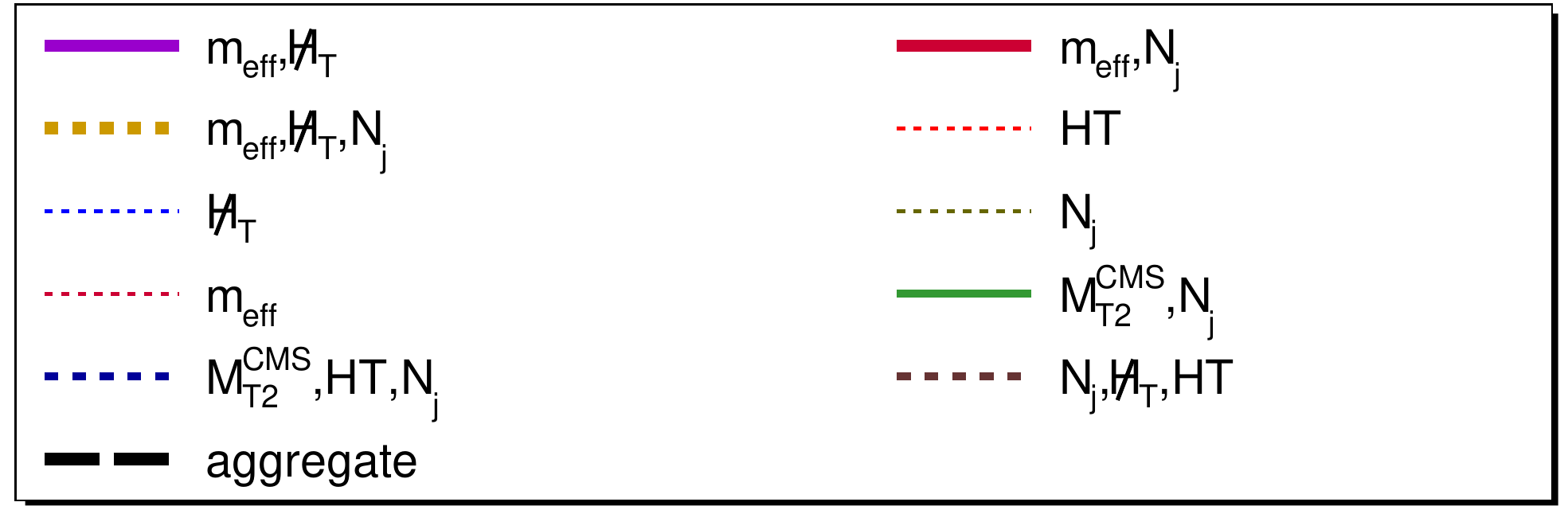} \hspace{30pt}
	\includegraphics[width=0.27\textheight]{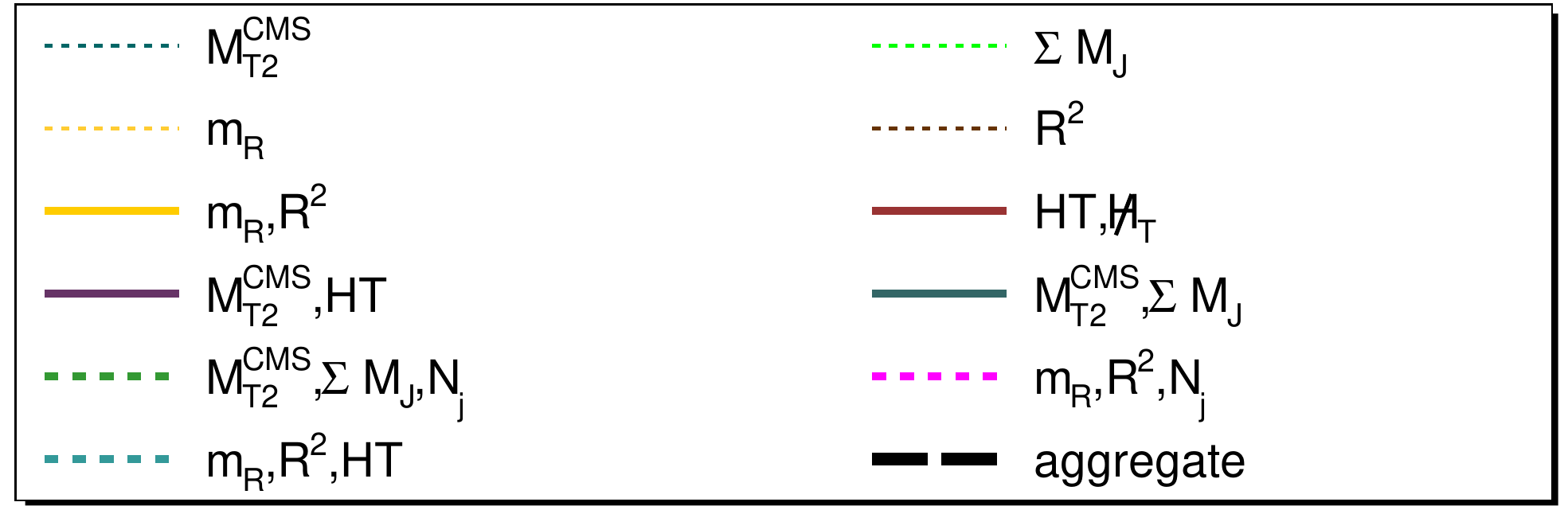}\\[10pt]
	\includegraphics[width=0.32\textheight]{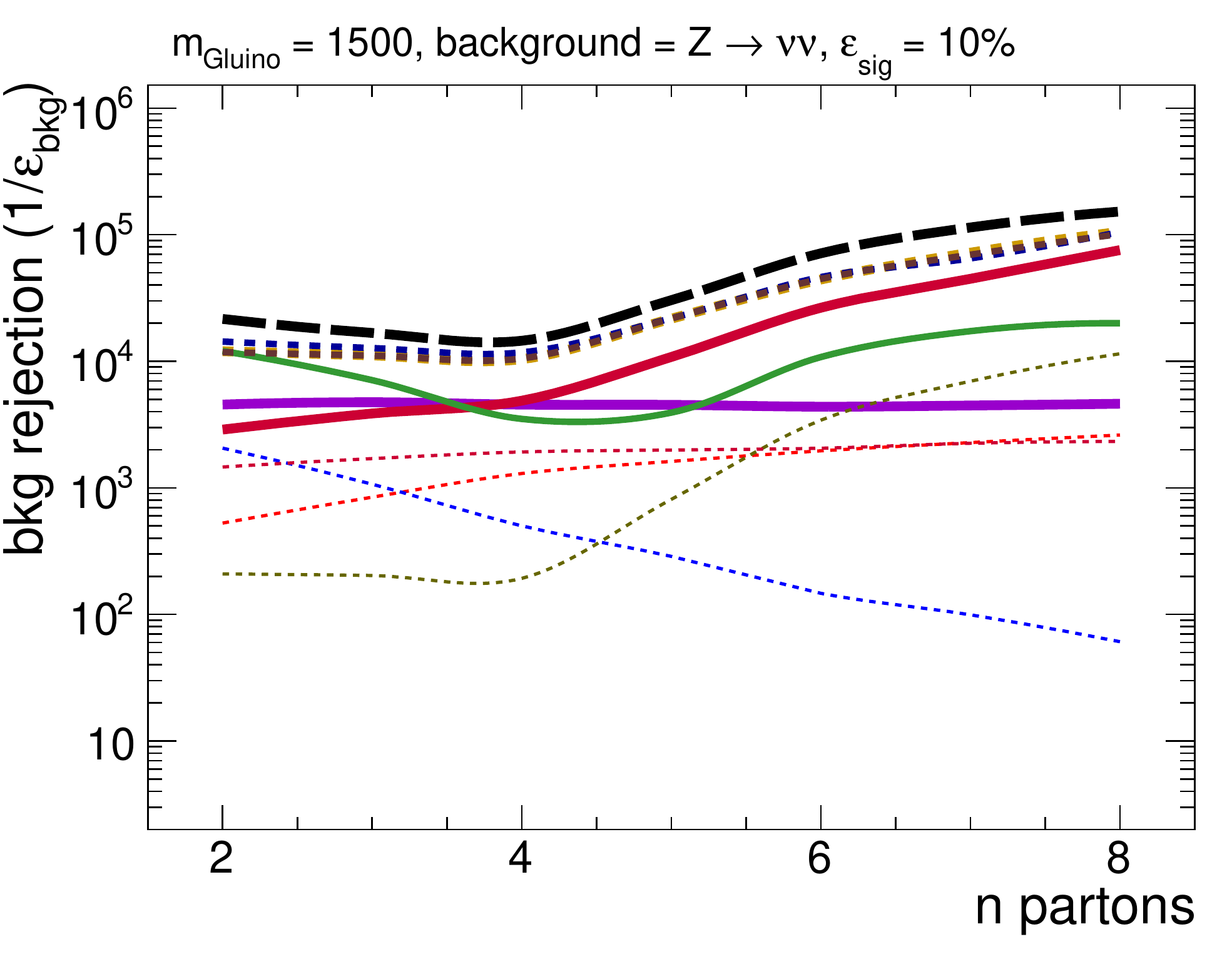}
	\includegraphics[width=0.32\textheight]{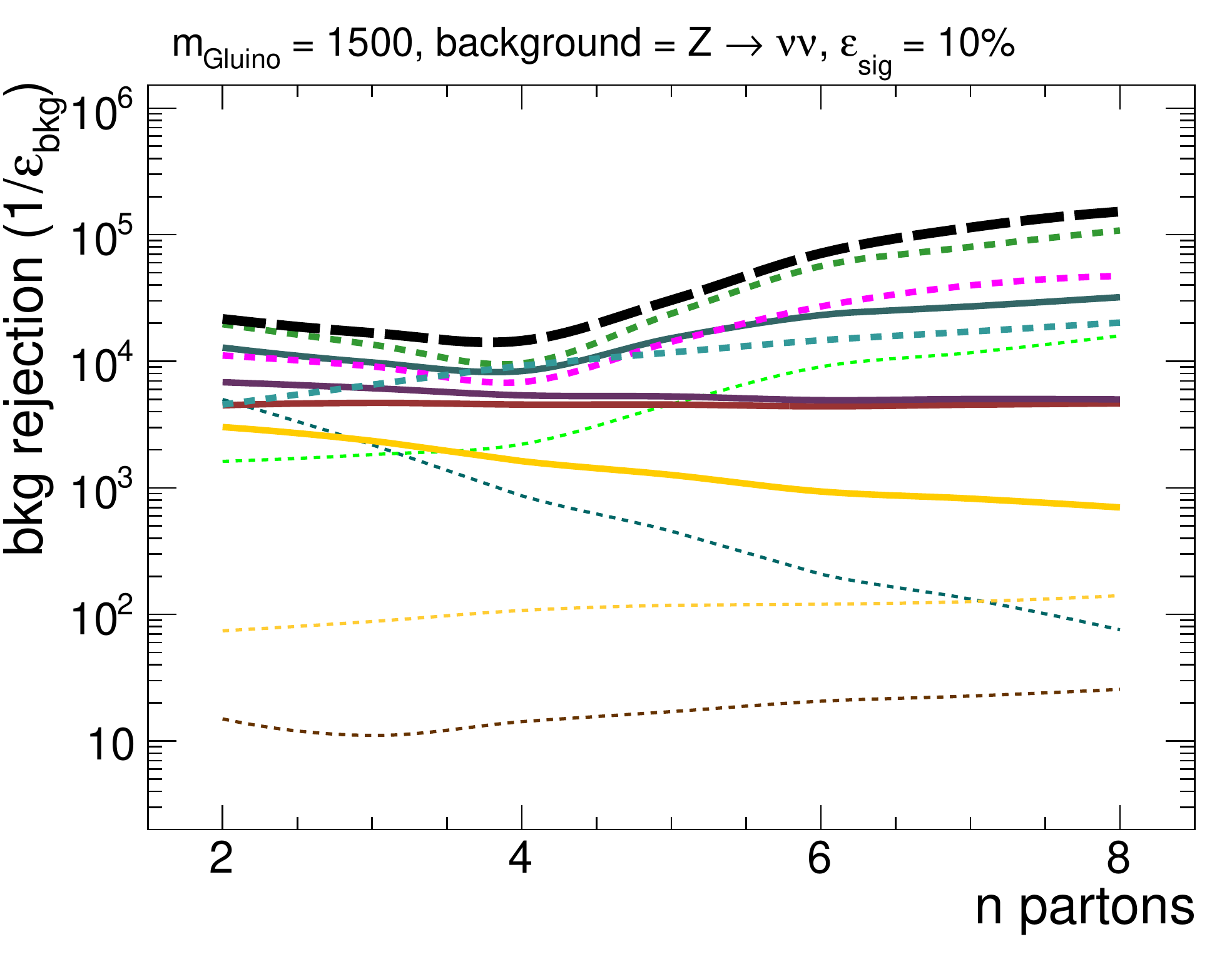}\\	
	\includegraphics[width=0.32\textheight]{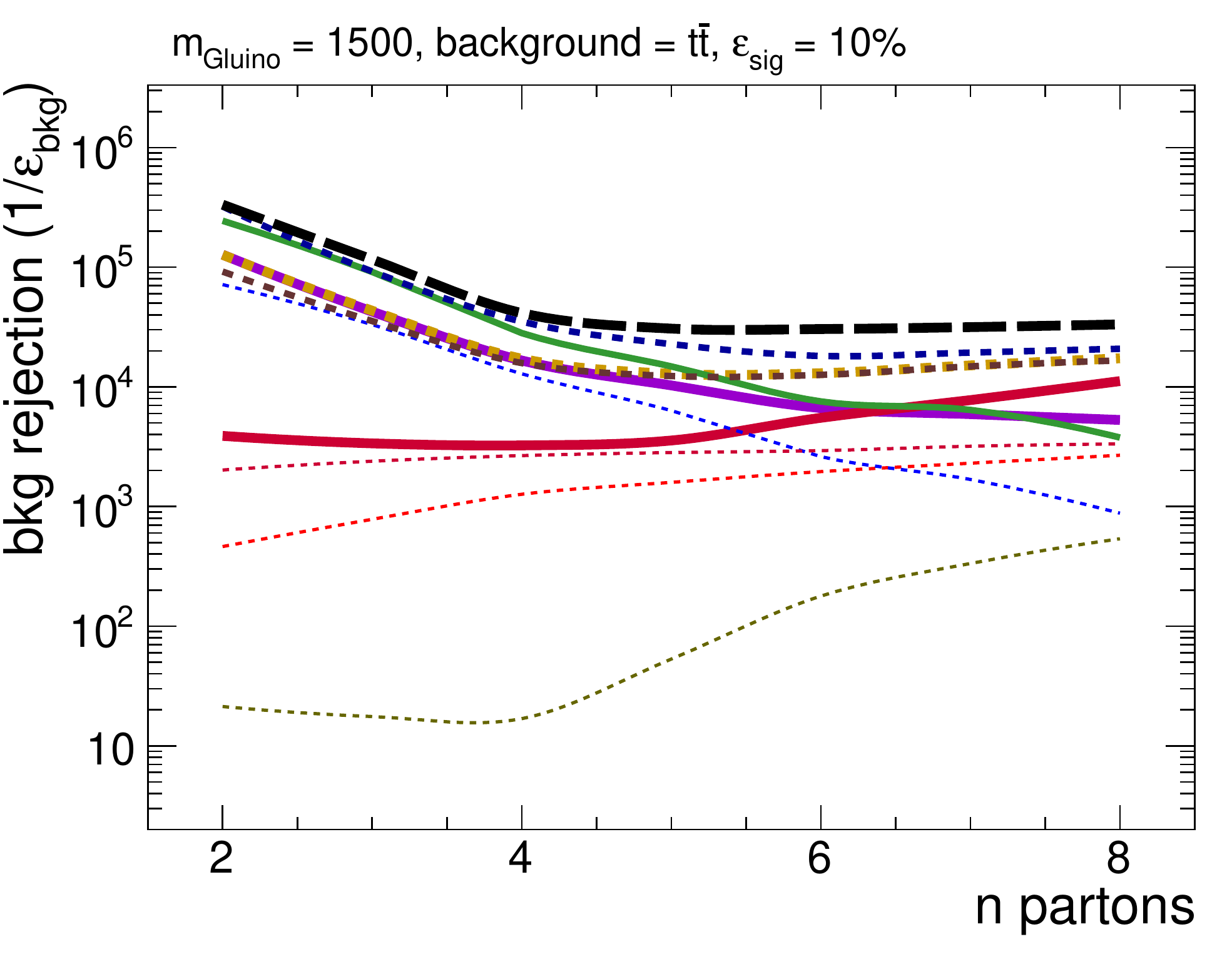}
	\includegraphics[width=0.32\textheight]{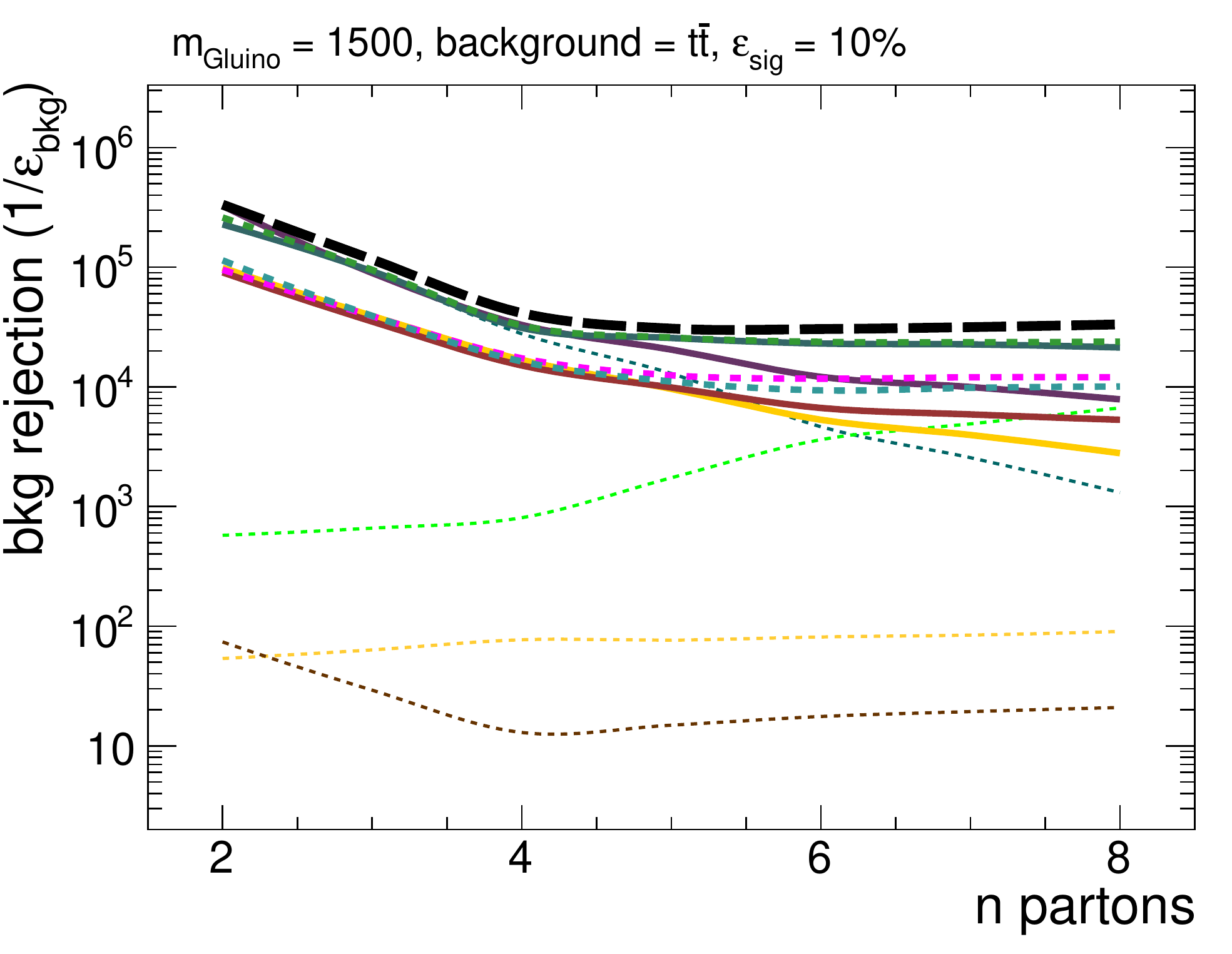}\\
	\includegraphics[width=0.32\textheight]{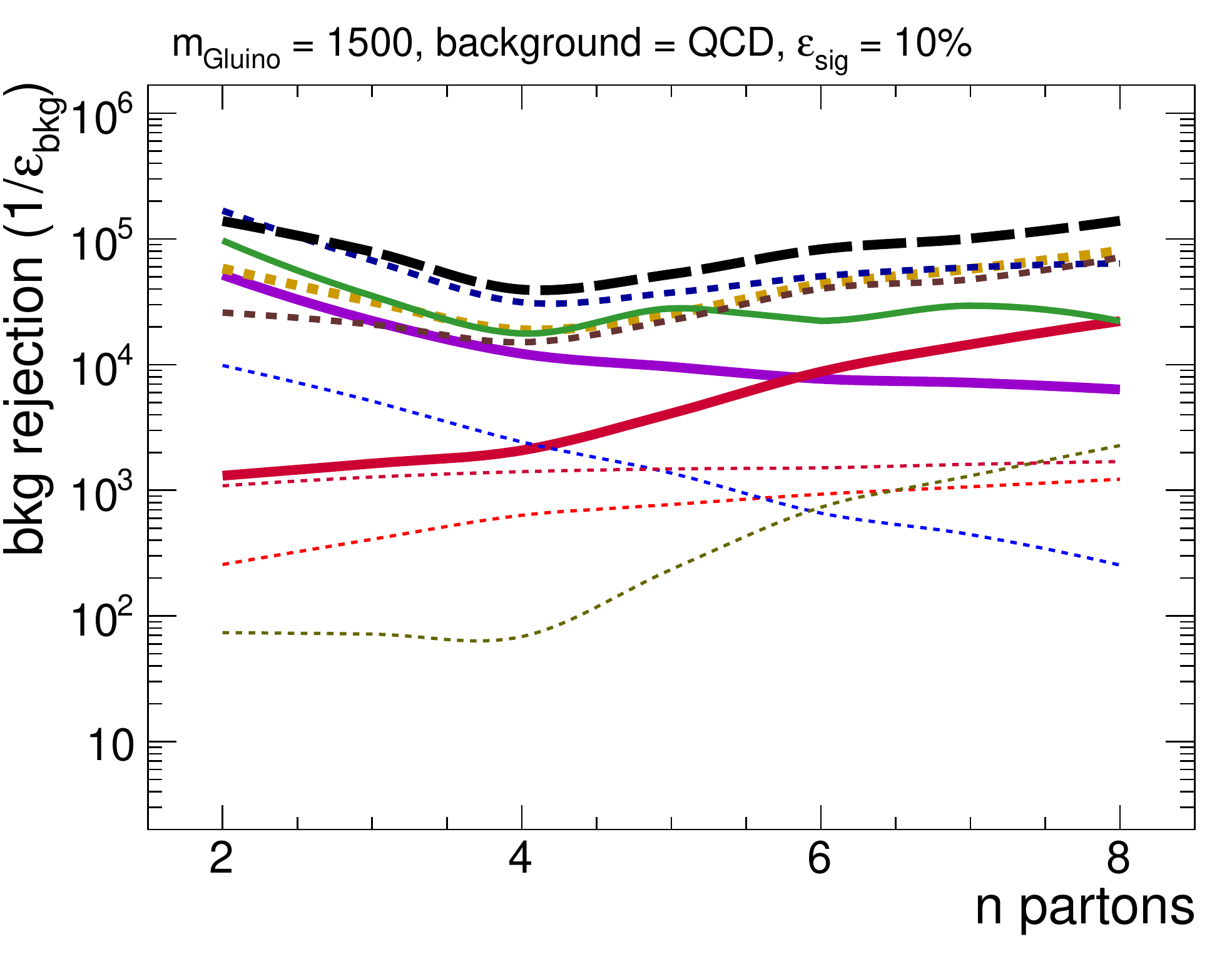}
	\includegraphics[width=0.32\textheight]{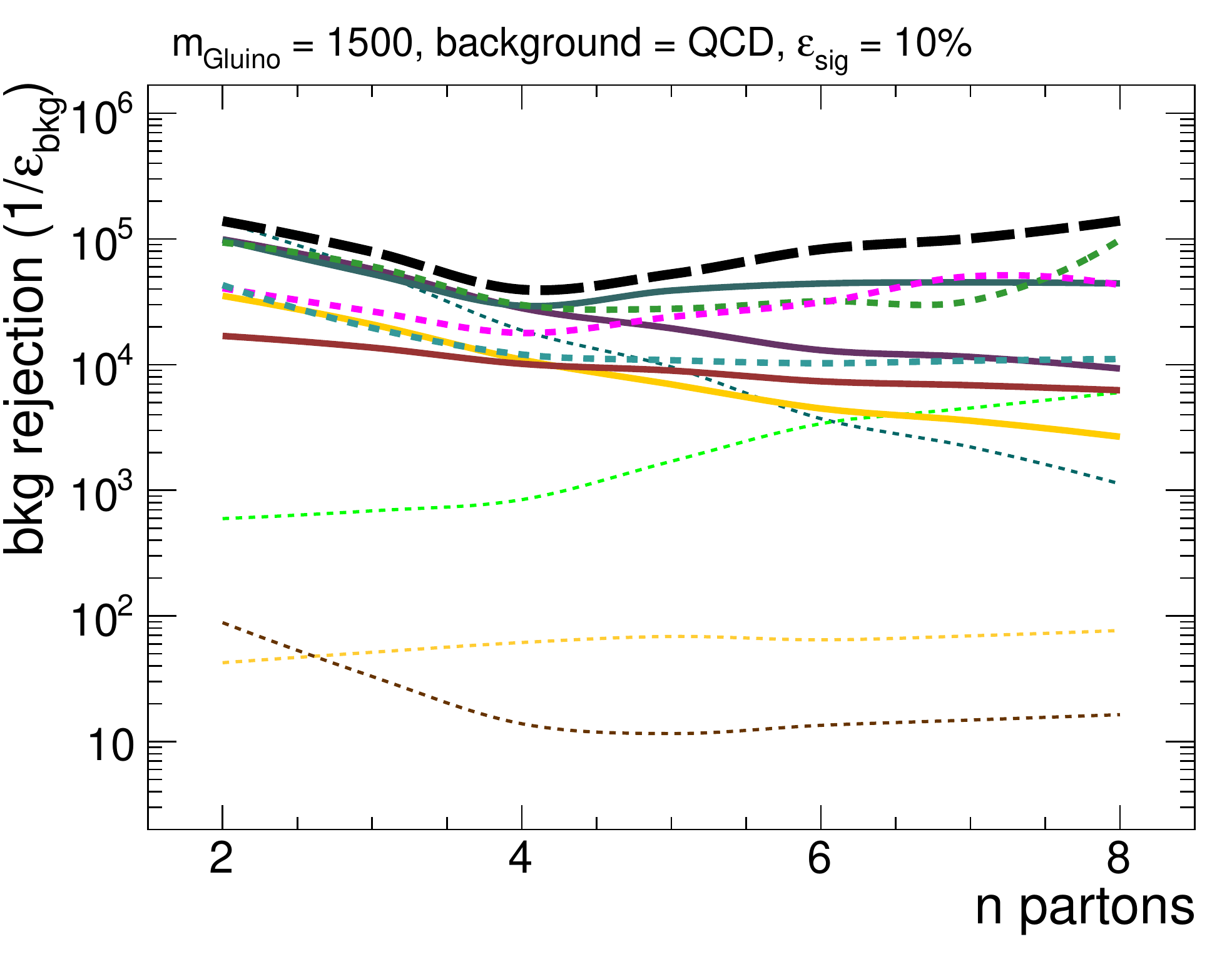}
	\caption{Background rejection efficiency for the \emph{uncompressed} signals as a function of the total number of partons $n$, for $10$\% signal efficiency. The variable combinations shown are given in the legend at the top of the page.  (One-, Two-, Three-) variable curves are given as (dotted, solid, dashed).  The left column shows the results for $H_T$ through $(\MCMS, N_j)$, and the right shows $(\MCMS, M_J, N_j)$ through $(m_R, R^2, N_j)$ to keep the figures from being too cluttered.  From top to bottom we show the rejection efficiency against \ZpJ, $t\,\overline{t}$, and QCD.}
	\label{Fig:ManyVar} 
\end{figure}

In order to organizing the huge number of possibilities, we start by taking one variable from each category to generate two- or three-variable combinations.
By analyzing the pairwise discriminating power, we can understand which variables are least correlated thereby leading to the best complementarity when designing a search.  The results of these explorations are given in Fig.~\ref{Fig:ManyVar}, where we show $1/\epsilon_\text{bkg}$ as a function of $n$ partons for a set of multivariable combinations.  
In order to reduce the many possible combinations, we show the selection of combinations with the most striking features.   As was done above, performance is separated by \ZpJ~(top), $t\,\overline{t}$ (middle), and QCD (bottom) backgrounds.  
The left and right columns of the figure show different variable combinations in an attempt to minimize clutter.
The single variables are given as dotted lines, the two-variable combinations as solid lines and the three-variable combinations as dashed.  
The aggregate, denoting a BDT combination of all variables, is shown in black long-dashed.\footnote{We note that for some comparisons, particularly against QCD, there are slight inconsistencies, \emph{e.g.} adding a variable slightly decreases discrimination power.  This is the result of non-zero statistical errors (which can be exacerbated by the presence of rare high weight background events) along with systematic errors associated with slight over/under training of the BDT.  We do not attempt to quantify these effects, but caution the reader to keep these issues in mind so as to not over-interpret these results.  The qualitative behavior shown in these plots is robust.} 

\clearpage

From the figures, it is clear that no two-variable combination is optimal across all regions of phase space and all backgrounds.  However, the doublet $\big(\MCMS, M_J\big)$ does out perform all other two-variable combinations essentially across the full range -- this can be understood by realizing that $M_J$ is a hybrid of \typeb and \typec variables. This combination is only deficient at the very highest $n$ partons and perhaps in the \ZpJ~case where the interplay between $H_T$ and $N_j$ is not fully captured in $M_J$.  For the other two-variable combinations we find the following general tends: 
\begin{itemize}
\item \typeab\!: deficient at high $n$ partons where $N_j$ is more important;
\item \typeac\!: deficient at medium to high $n$ partons where visible energy becomes more important;
\item \typebc\!: deficient at low $n$ partons where missing energy variables are most dominant.
\end{itemize}

Moving on to the three-variable combinations, we see that adding $N_j$ to $\big(\MCMS, M_J\big)$  provides nearly optimal performance for the full range of $n$-partons shown here. There are additional three-variable combinations which are near optimal over the full uncompressed phase space in $n$ partons and for different backgrounds.  The near optimal three-variable combinations all involve one of each type of variable.  
Because $\MCMS$ is the best performing of the \typea variables for the uncompressed spectra (see Fig.~\ref{Fig: 1var_nPartons}), we find that triplets which include it do the best: $\big(\MCMS, H_T~{\rm or}~M_J, N_j\big)$.  

Another conclusion one can draw from Fig.~\ref{Fig:ManyVar} is a compelling confirmation that $M_J$ outperforms $H_T$, especially for high multiplicity final states.  Ignoring correlations with other observables, these two variables are approximately proportional to each other~\cite{Hook:2012fd}:
\begin{align}
H_T \simeq M_J \,\frac{\sqrt{1+\big(\kappa\,R\big)^2}}{\kappa\,R},
\end{align}
where $\kappa \simeq \sqrt{\alpha_s}$ when the jet mass is the result of the QCD parton shower, as compared to $\kappa \sim 1$ for events characterized by high multiplicity signals which manifest accidental substructure.  For events in a fixed $H_T$ bin, $\big(M_J\big)_\text{signal} \gg \big(M_J\big)_\text{background}$, thereby providing enhanced signal discrimination.  The results of Figs.~\ref{Fig: 1var_nPartons} and \ref{Fig:ManyVar} clearly demonstrate this effect showing that $M_J$ does as well or better than $H_T$ both alone and in combinations for the three dominant backgrounds considered here.  The case to use $M_J$ is especially strong since it already been demonstrated to be a useful variable in the complex environment of the LHC~\cite{Aad:2015lea,CMS-PAS-SUS-15-007}, and is also amenable to analytic study, \emph{e.g.}~\cite{Chien:2012ur, Dasgupta:2012hg, Jouttenus:2013hs, Dasgupta:2013ihk}.  Ideally, $M_J$ would replace $H_T$ for the wide class of jets + $\MHT$ searches whose phase space is covered by $n$-body Simplified Models, and can be done so while maintaining the core strategy implemented by many existing approaches for these beyond the Standard Model searches.  However, it is worth acknowledging that once one goes to multi-dimensional variables the improvement in replacing $H_T$ by $M_J$ is not as dramatic and would require a careful job of finding a new class of control regions, along with a reassessment of systematic errors.

Next, we turn our focus to the $m_\text{eff}$ results.  Recall in Fig.~\ref{Fig: 1var_mGluino}, we found that $m_\text{eff}$ turned out to be slightly better performing than $H_T$, particularly at low $n$ partons.  
However, when we add additional variables to these \typeb variables in order to find the optimal triplet, the final performance is extremely similar.
For example, the triplets $\big(m_\text{eff}, \MHT, N_j\big)$ and $\big(H_T, \MHT, N_j\big)$ have essentially the same discriminating power.

\begin{figure}[h!]
	\centering
\includegraphics[width=0.49\linewidth]{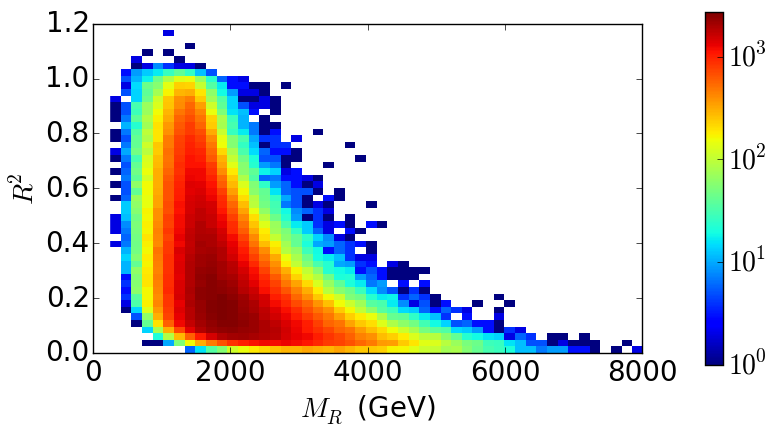}
\includegraphics[width=0.49\linewidth]{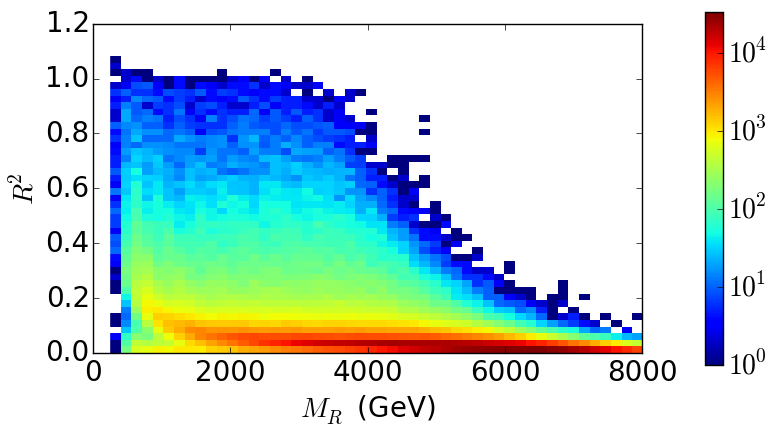}
	\caption{Density profile distributions of the Razor variables $R^2$ and $M_R$ for a 1 TeV gluino signal with $n=4$ (left) and the QCD background (right).  The color scale is in arbitrary units.}
\label{Fig:razor}
\end{figure}

To end this section, we will comment on the Razor variables.  The combination $\big(M_R,R^2\big)$ does significantly better than each variable individually.  
This implies that they are highly complementary, as expected by the design of the variable and now seen explicitly.
This is also illustrated in Fig.~\ref{Fig:razor} where the 2-D profile distributions of $R^2$ and $M_R$ are shown for the QCD background (right panel) and 
an uncompressed 1~TeV gluino signal with $n=4$ (left panel).
The optimal cut is non-rectangular in the Razor 2D plane such that a 1D projection of either $R^2$ or $M_R$ is not expected to be a particularly good discriminator.  However, as can be seen from Fig.~\ref{Fig:ManyVar}, the two-variable combination of $\big(M_R,R^2\big)$ behaves like a \typea missing energy-like variable which decreases in performance as a function of $n$ parton, although Razor does display less degradation than a \typea variable.  
This motivates combinations with either $N_j$ or $H_T$ to see if it would then perform as well as other three-variable choices.  The better combination appears to be $\big(M_R,R^2,N_{\rm jets}\big)$, which is fairly close to optimal although it does not perform as well as $\big(\MCMS, H_T~{\rm or}~M_J, N_j\big)$.

\newpage

\subsection{Degenerate Gluino-Neutralino Limit}
\label{sec:Degenerate}
\begin{figure}
	\centering
        \hspace{33pt} \includegraphics[width=0.4\textheight]{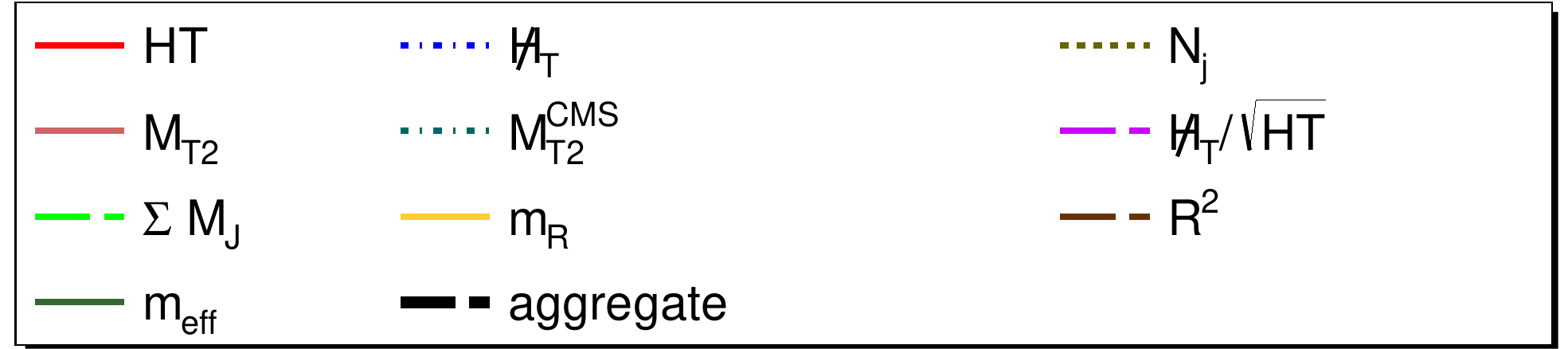}\\[10pt]	
	\includegraphics[height=0.26\textheight]{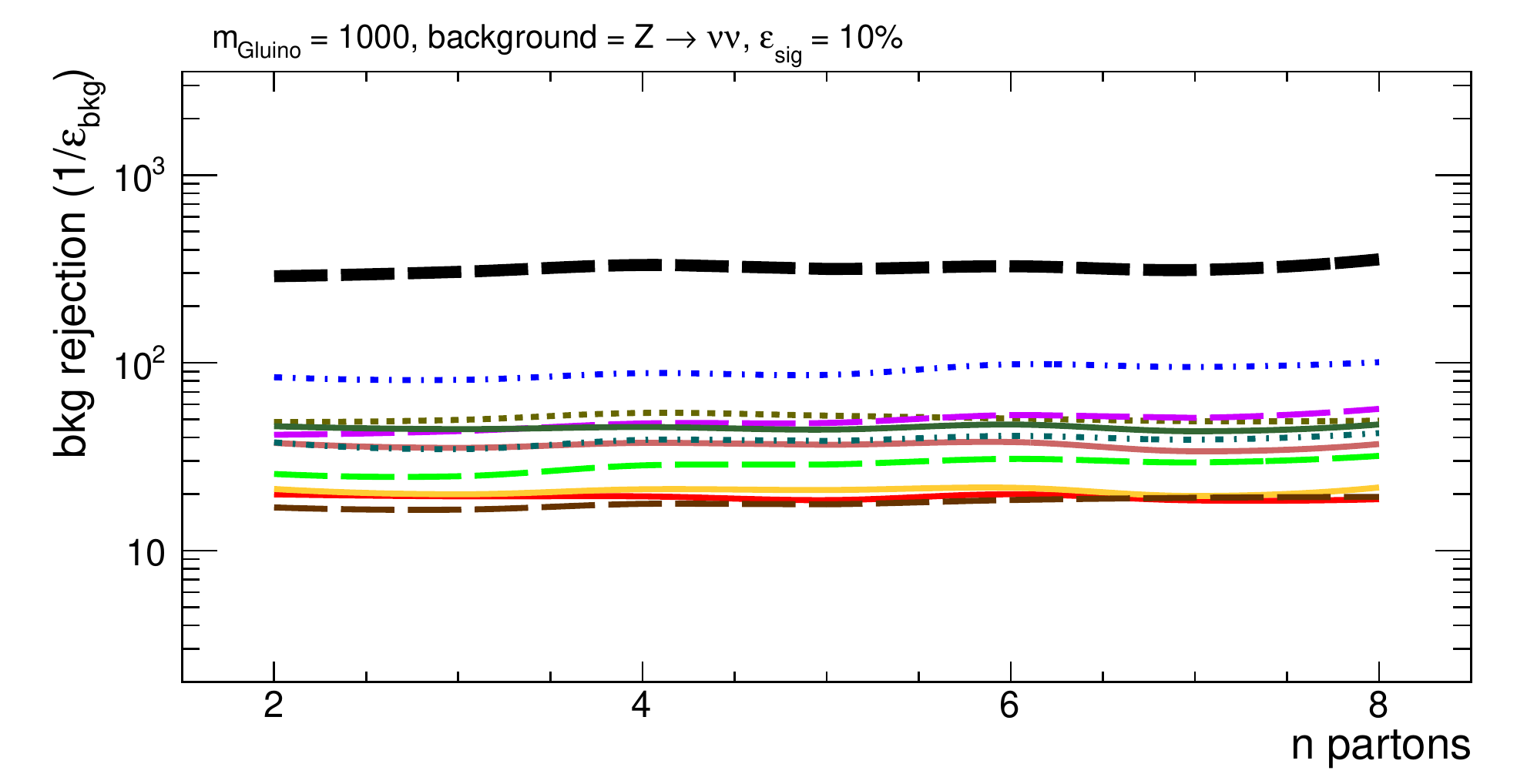}\\
	\includegraphics[height=0.26\textheight]{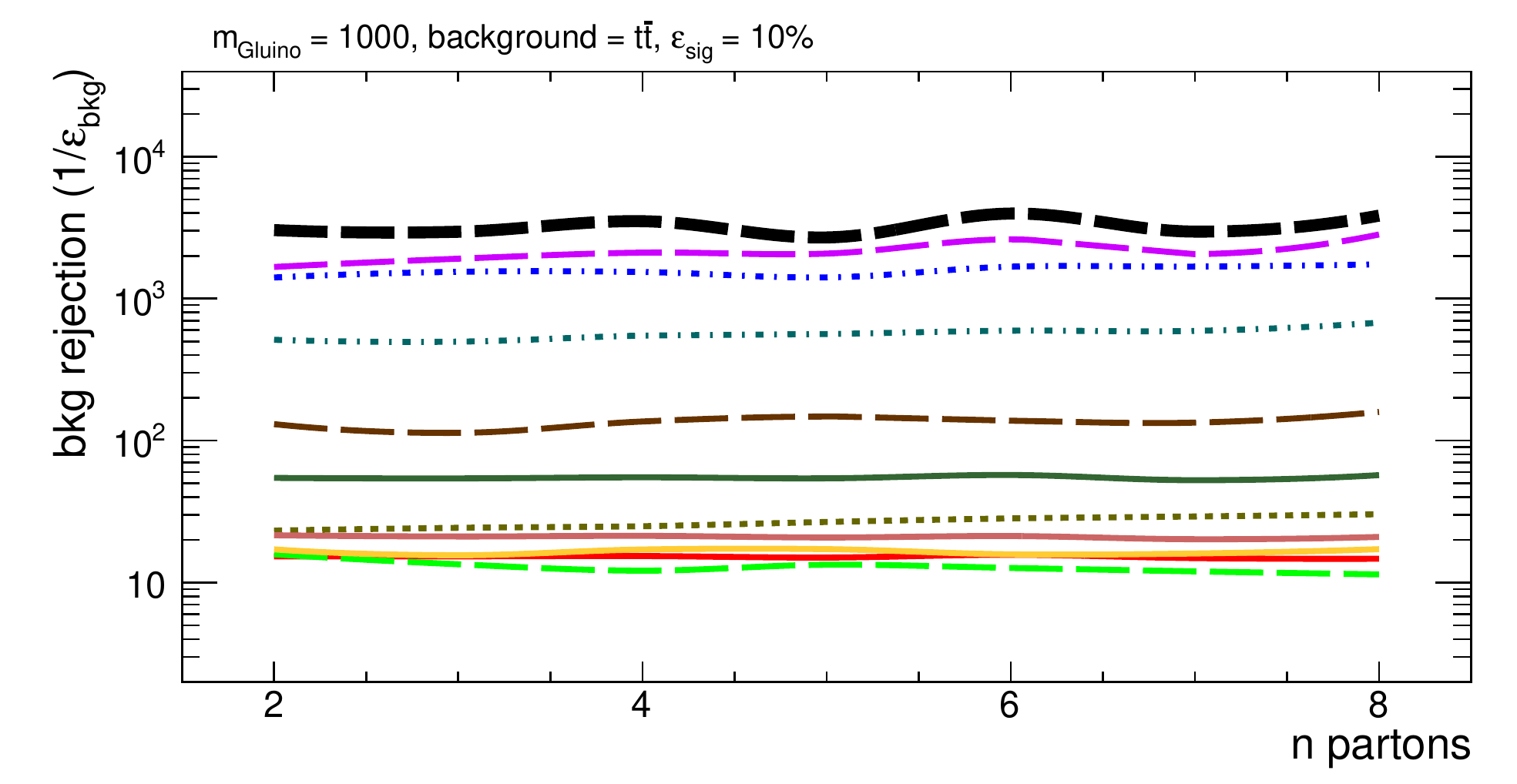}\\
		\includegraphics[height=0.26\textheight]{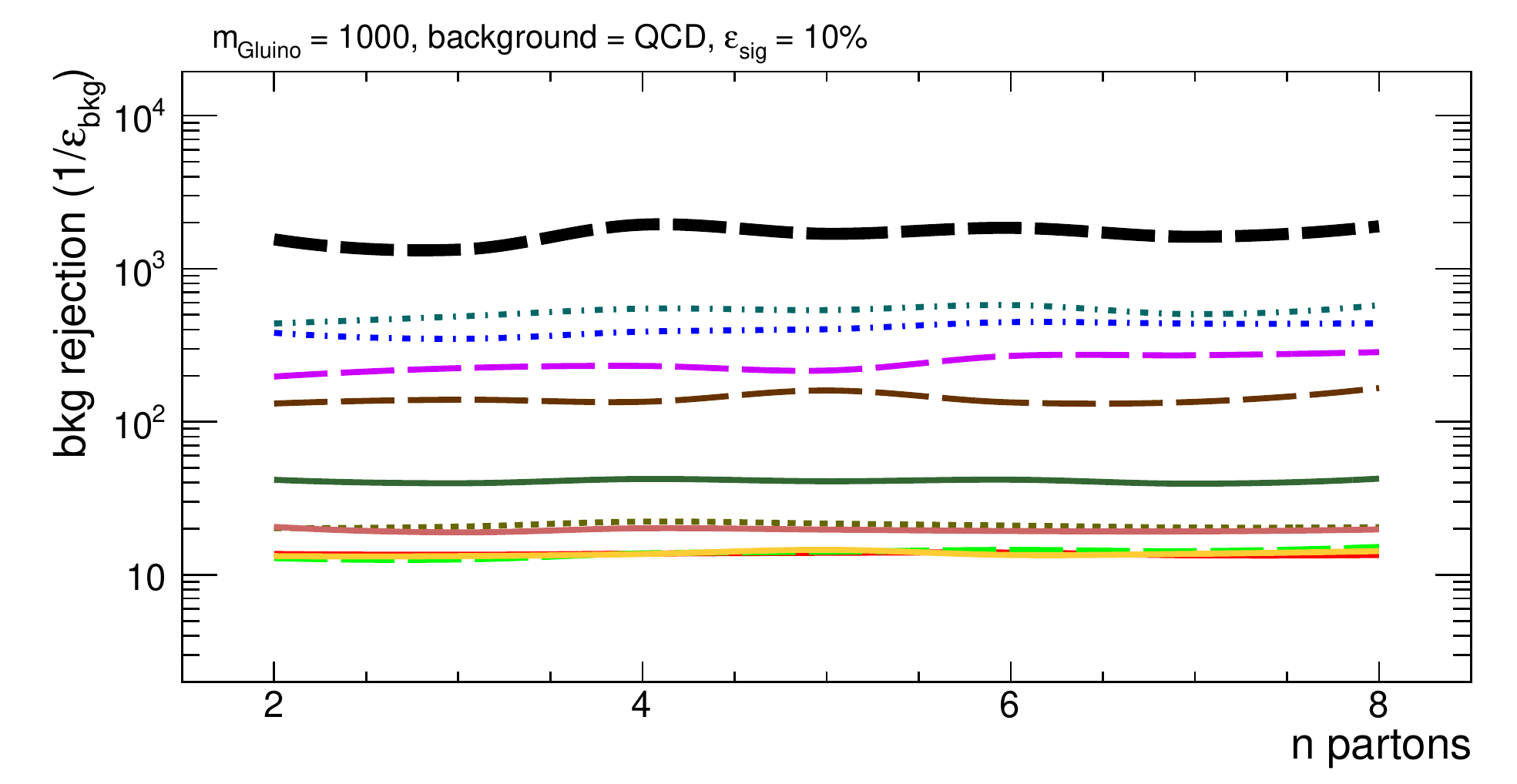}
	\caption{Background rejection efficiency for \emph{compressed} signals as a function of the total number of partons, $n$, for $10$\% signal efficiency.  The variables shown are given in the legend at the top of the page. From top to bottom we show the rejection efficiency against \ZpJ, $t\,\overline{t}$, and QCD.}
\label{Fig:1varComp}
\end{figure}

We now consider compressed topologies with a $5$\% splitting between the gluino and neutralino masses. With such a small splitting, if the gluinos are not boosted against additional objects, the final state jets and $\MHT$ are expected to be soft and difficult to distinguish from the SM background. We therefore include topologies where the pair-produced gluinos are boosted against at least one ISR jet by producing matched signal samples. We consider the background rejection rate as a function of the number of final state partons $n$, for a signal efficiency of $10$\%, a $1$ TeV gluino and a $950$ GeV LSP. Figure~\ref{Fig:1varComp} shows the individual performance of each of the variables listed in Sec.~\ref{sec:Observables}. Unlike in the uncompressed scenario, the discriminating power of the different observables is practically independent of the number of visible gluino decay products. This result is explained by the large neutralino mass, which causes most of the gluino momentum to manifest as missing energy. Conversely, the light jets coming from the gluino decay get most of their momentum from the gluino-neutralino mass splitting and will therefore be too soft to pass the jet selection criteria.  In this compressed scenario, all the signals can then be reduced to a single topology with a small number of hard partons accompanied by $\MHT$.

Looking more closely at the individual variables in Fig.~\ref{Fig:1varComp}, there are some striking differences with respect to the uncompressed spectra.  
The missing energy-like \typea variables all behave similarly when discriminating against QCD, but $\MHT$ is more powerful than $\MCMS$ for both \ZpJ~and $t\,\overline{t}$ backgrounds.  
However, it is true that the most important single variables are still of \typea or a related hybrid.  
For example, $\MHT/\sqrt{H_T}$ is quite powerful against $t\,\overline{t}$ and $R^2$ distinguishes itself from the other \typeb and \typec variables.  
It is also important to note that the \typea variables are not very close to optimal in the case of the QCD and \ZpJ~backgrounds.
This means that additional information from the visible energy and its structure can be utilized to additionally distinguish signal from background.

We have also checked the gluino mass dependence of the single variables and the performance is relatively independent of these variations, in contrast with the uncompressed spectra case.   Specifically, the discrimination power changes by a factor of $\sim 2$ when changing the mass from 500-1500~GeV while in the uncompressed spectra case, the discrimination power changes by more than 2 orders of magnitude (see Fig.~\ref{Fig: 1var_mGluino} above). 
This is somewhat expected as the visible energy resulting from the decay of the compressed system does not drastically change as a function of mass.

\begin{figure}
	\centering
        \hspace{20pt}\includegraphics[width=0.27\textheight]{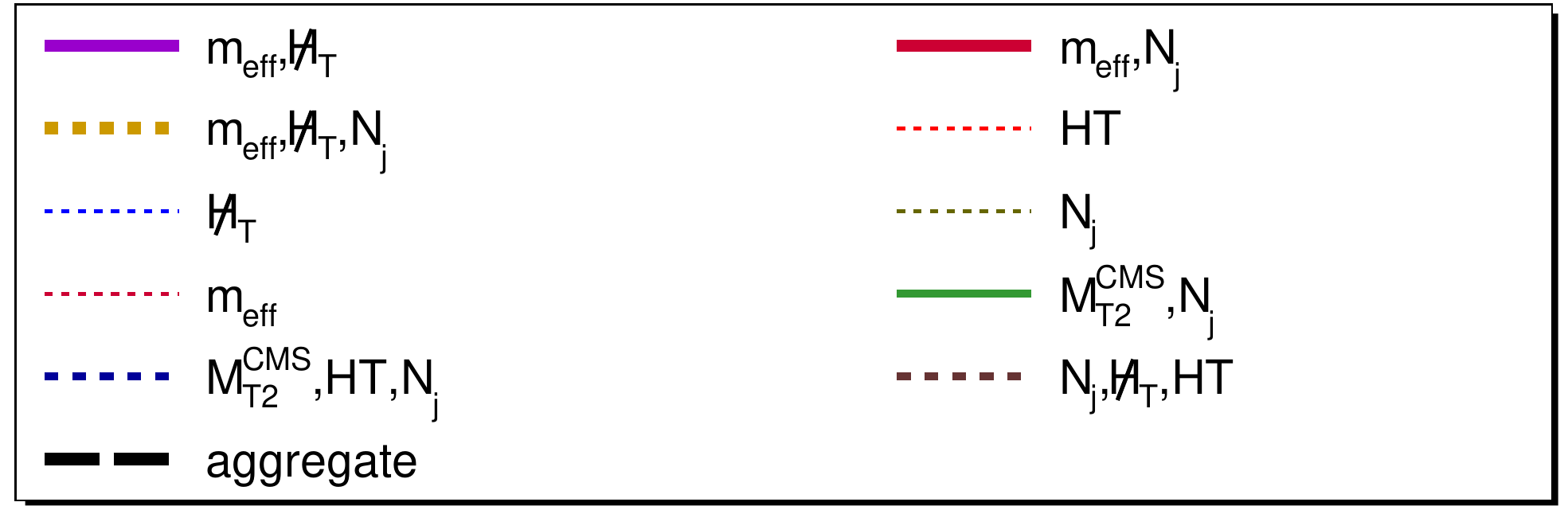}\hspace{30pt}
        \includegraphics[width=0.27\textheight]{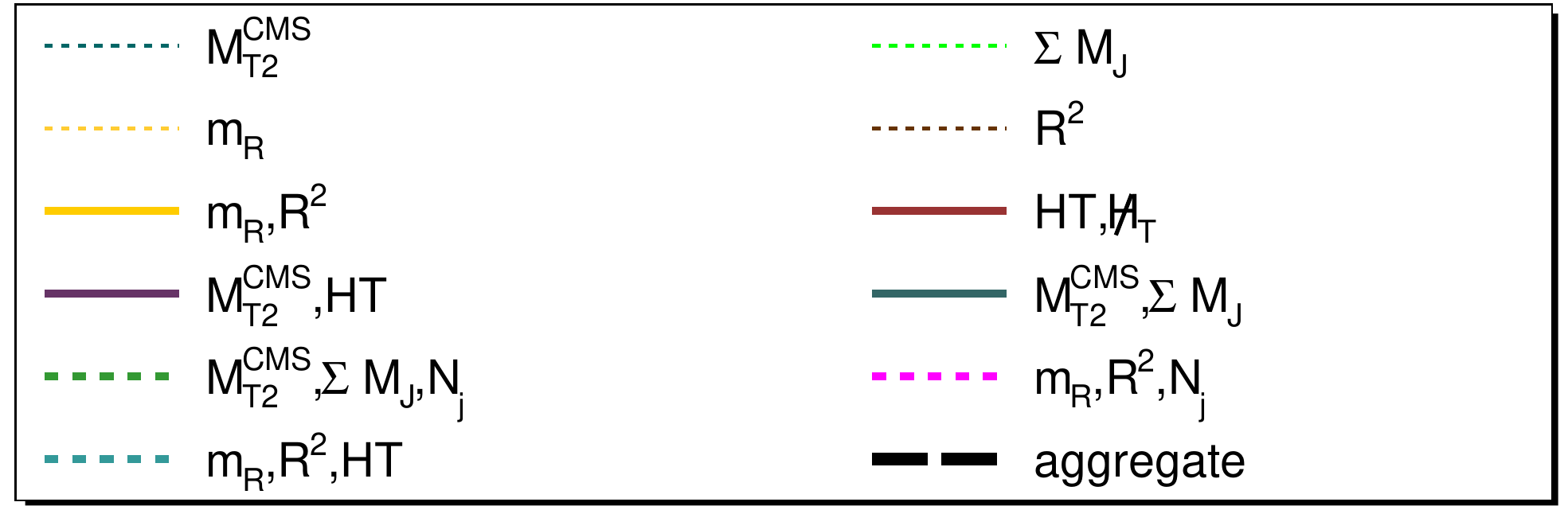}\\[10pt]
	\includegraphics[width=0.32\textheight]{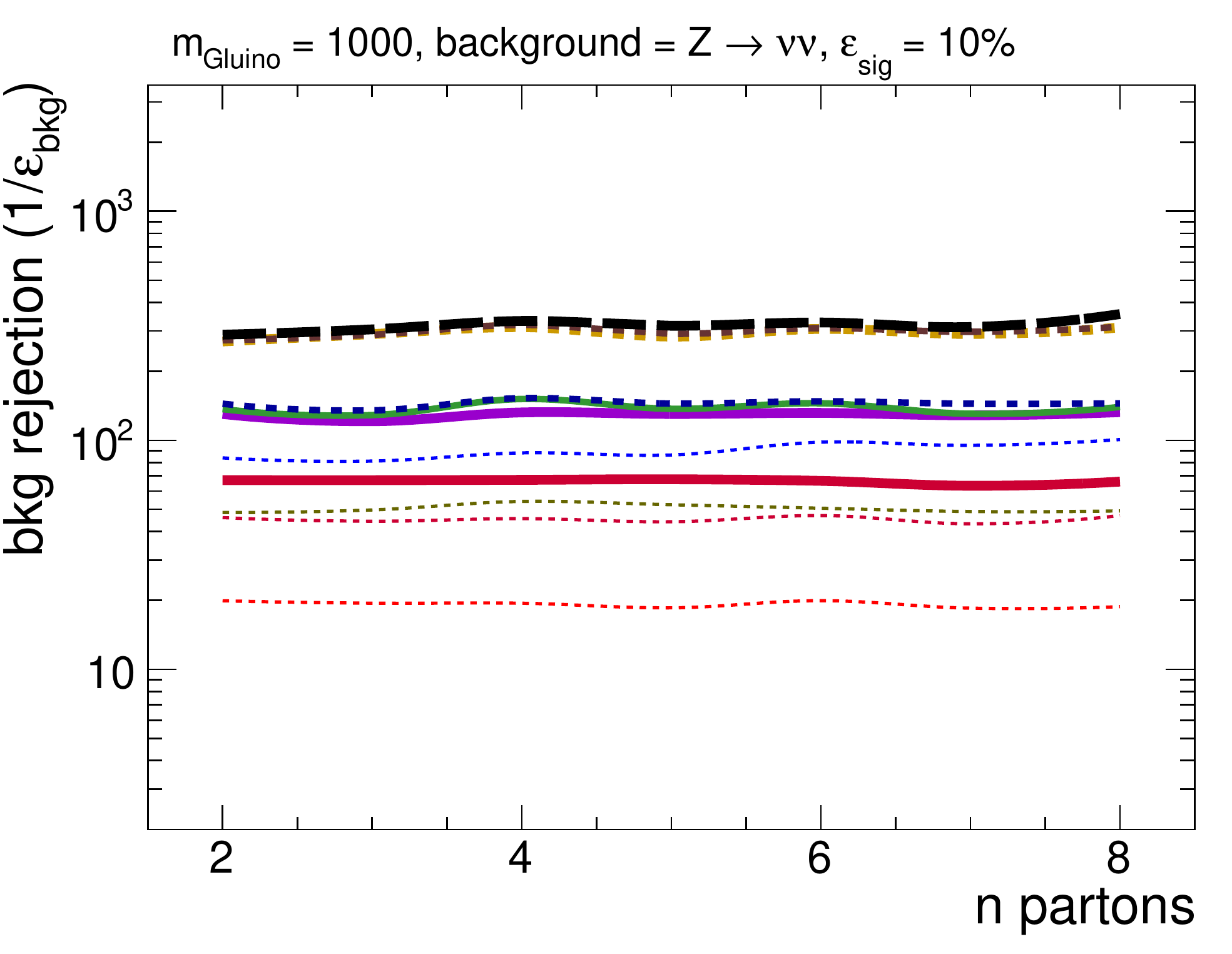}
	\includegraphics[width=0.32\textheight]{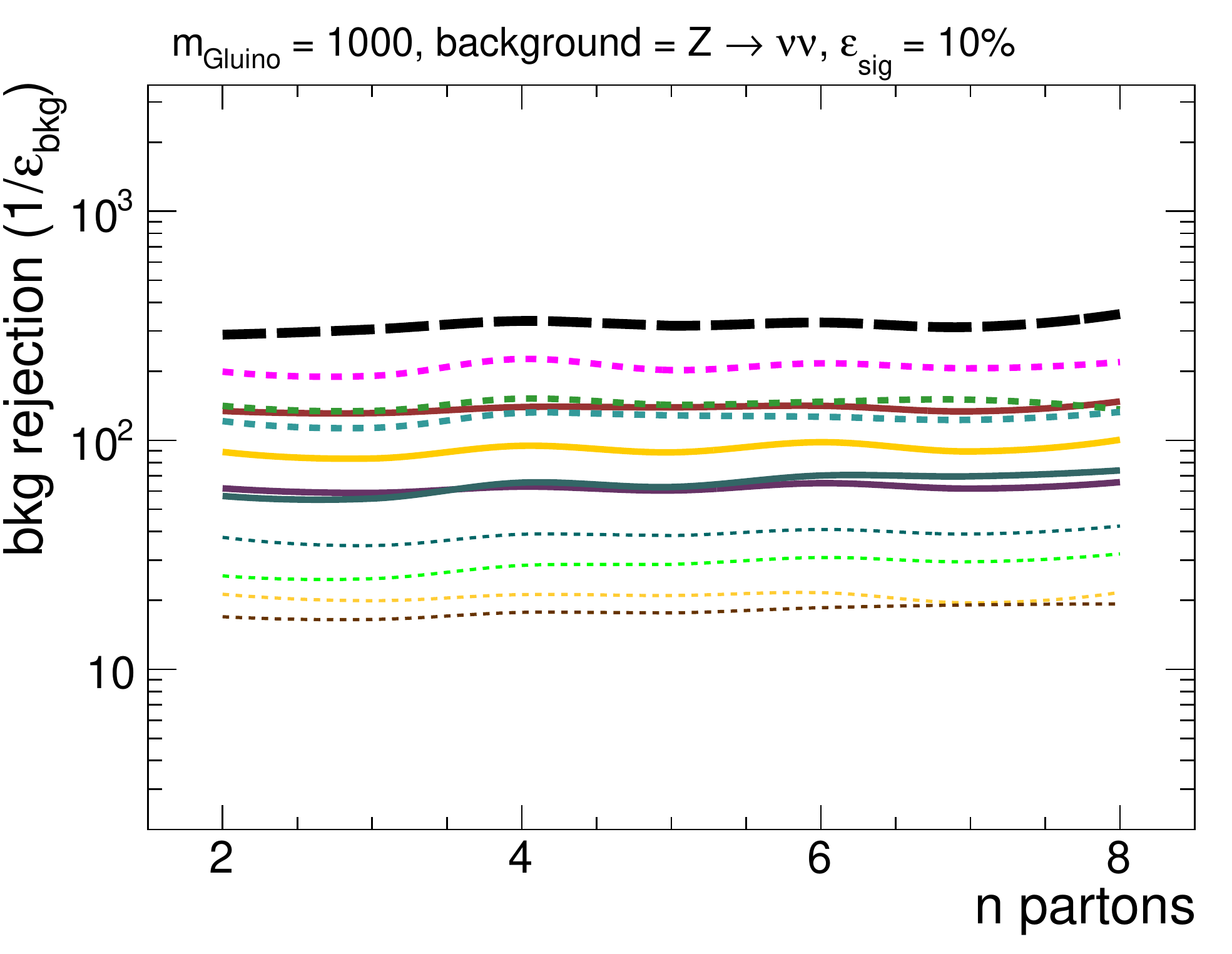}\\
	\includegraphics[width=0.32\textheight]{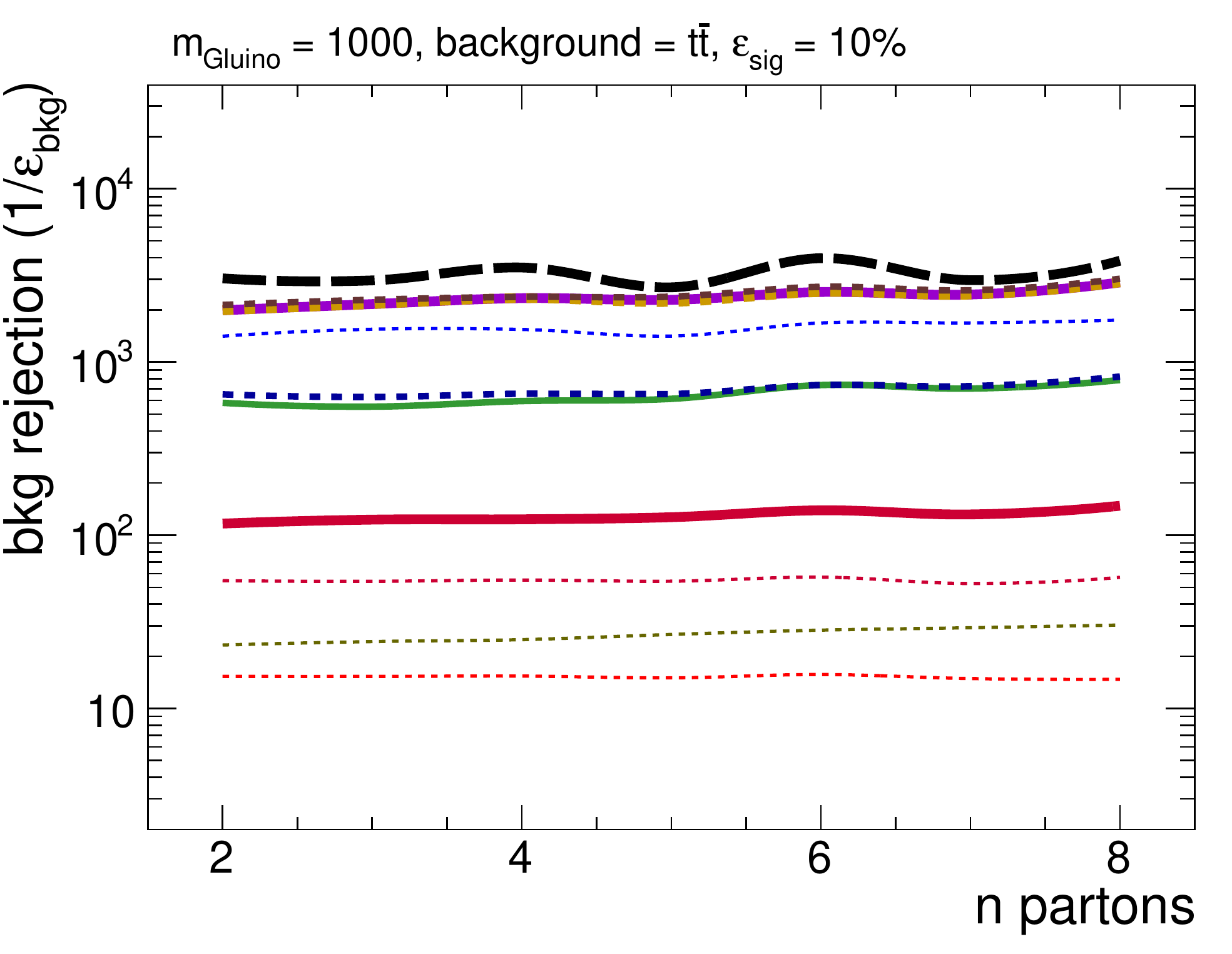}
	\includegraphics[width=0.32\textheight]{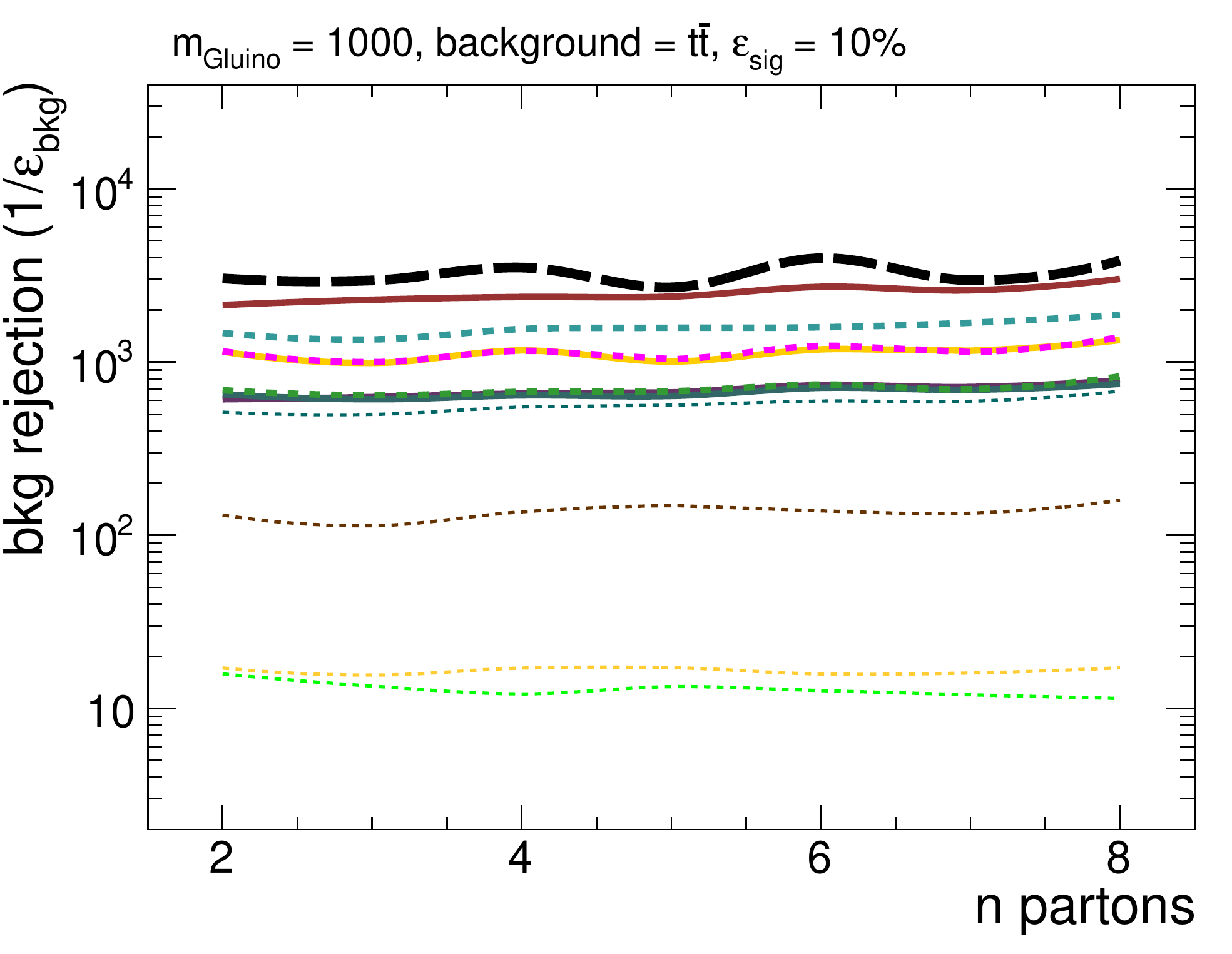}\\
	\includegraphics[width=0.32\textheight]{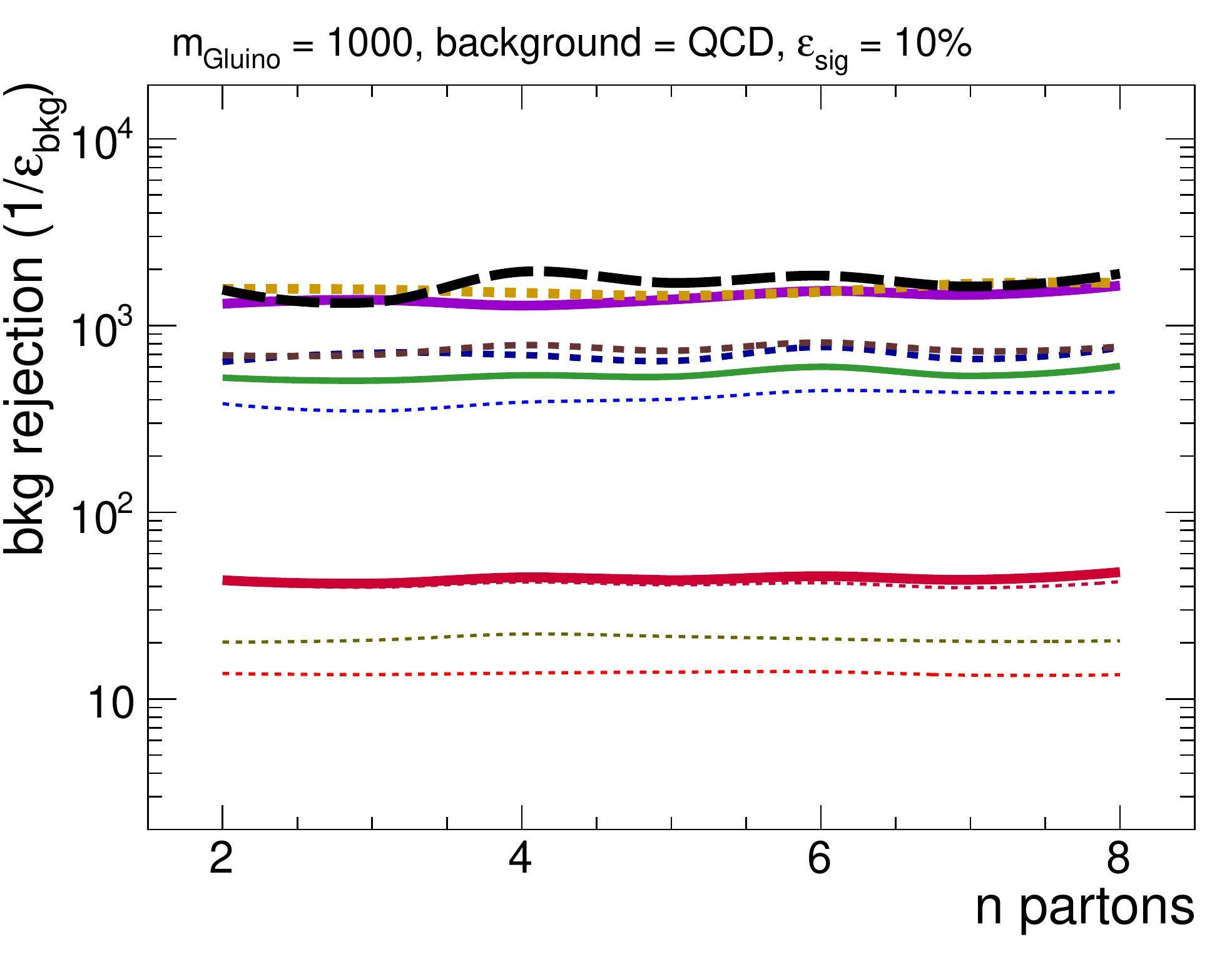}
	\includegraphics[width=0.32\textheight]{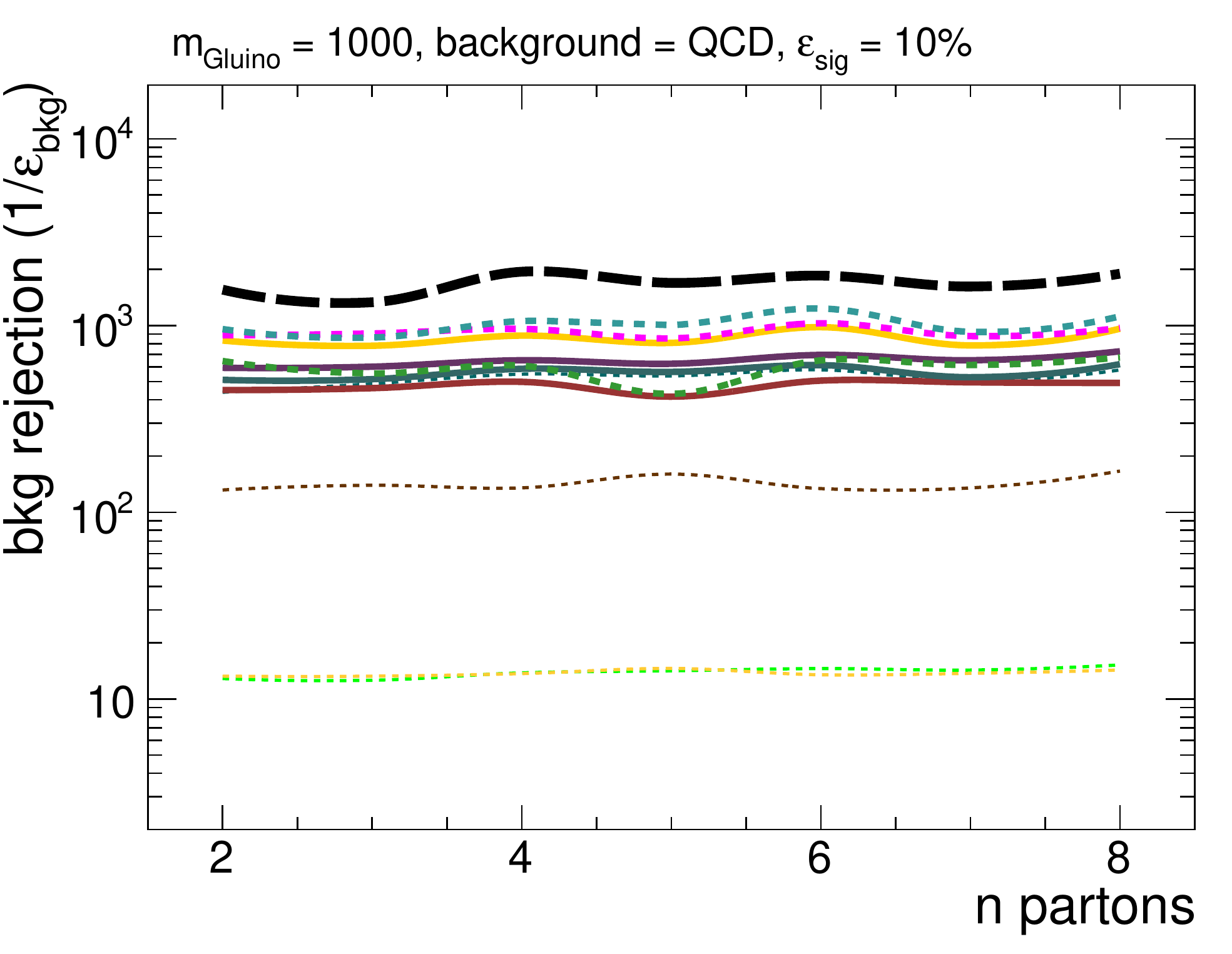}	
	\caption{Background rejection rate for the \emph{compressed} signals as a function of the total number of partons $n$, for $10$\% signal efficiency. The variable combinations shown are given in the legend at the top of the page.  (One-, Two-, Three-) variable curves are given as (dotted, solid, dashed).  The left column shows the results for $H_T$ through $(\MCMS, N_j)$, and the right shows $(\MCMS, M_J, N_j)$ through $(m_R, R^2, N_j)$ to keep the figures from being too cluttered.  From top to bottom we show the rejection efficiency against \ZpJ, $t\,\overline{t}$, and QCD.}
\label{Fig:nvarComp}
\end{figure}

As we begin to look at multi-variable combinations as shown in Fig.~\ref{Fig:nvarComp}, we come to the same general conclusions as in the uncompressed spectra case.  No two-variable combination achieves near optimal performance across the range $n$ partons and various backgrounds.  However, it is possible to essentially maximize sensitivity using three-variable combinations.  Among these, $\big( H_T, \MHT, N_j \big)$ and  $\big( m_\text{eff}, \MHT, N_j \big)$ stand out, and in particular do better than combinations involving $\MCMS$. 
It is interesting that in the case of QCD, $m_\text{eff}$ yields additional discrimination power over $H_T$ within the top-performing triplets.
Combinations including Razor  also achieve a good performance.  However, the best performing triplet involving Razor depends slightly on the background under consideration: $\big( M_R, R^2, H_T\big)$ or $\big( M_R, R^2, N_j\big)$.

The main lessons for the compressed study is similar to those learned in the uncompressed case.  Optimal performance for all three backgrounds considered is achieved by combining variables from each of the three classes.   For some backgrounds, two-variable combinations are nearly optimal but if one is interested in near-ideal discrimination across all backgrounds three variables are required.

\clearpage

\section{Conclusions}
\label{sec:Conc}
The $n$-body extension of Simplified Models provides a class of signal injections with which one can model a wide range of possible final-state phase-space within a unified phenomenological framework.  This has many applications in collider searches for beyond the Standard Model physics, and is particularly well suited for seeking out final-state topologies which require additional optimization beyond the searches that are currently being performed.  The focus in this work was to utilize this tool in order to assess the discriminating power for many of the ever-growing number of variables used for searches in the classic jets + missing energy final state.  This was an ideal forum to explore the utility of the $n$-body extended approach since the only observables in these searches stem from a single class of object: jets of visible hadronic energy.  

A large number of variables were considered: $H_T$, $\MHT$,$\MHT / \sqrt{H_T}$, $N_j$, $M_J$, $m_\text{eff}$, Razor, $M_{T2}$, and $\MCMS$.  As was expected and shown in Fig.~\ref{Fig: 1var_nPartons}, no variable can do the job alone.  A winning strategy derives from placing cuts on maximally uncorrelated observables in order to generate signal regions where background events are very rare.  Boosted Decision Trees were used in order to access the strength of correlations between observables.  Once a choice of variables was made, the BDT was trained to distinguish signal from background using only these inputs.  Then the resulting machine takes events and generates an output which yields the optimal background rejection efficiency as a function of a target signal efficiency.  The result is a quantitative assessment of performance.

Analyzing single variables alone led us to a classification scheme based on their trends as a function of the number of final state partons, assuming an uncompressed spectrum:  ``missing energy"-type, ``energy scale"-type, and ``energy-structure"-type.  This scheme grouped the variables based on the region of phase space where each provides the best discrimination against the backgrounds.  Not all of the observed behaviors were intuitive.   For example, $M_{T2}^\text{CMS}$ behaves like missing energy, although it is slightly better optimized for uncompressed signals as compared to compressed ones.  For uncompressed signals, $M_J$ and $m_\text{eff}$ tend to be slightly better performing than $H_T$.  However, the general lesson is that an ideal search strategy requires at least one variable from each class.  Differences in both the power and behavior of combinations as a function of $n$ partons or $m_{\tilde{g}}$  are reduced when analyzing triplets that include variables of each type.  As expected, the combination of classic variable types $\big( \MHT,H_T,N_j \big)$  performs very well in most cases.  However in some instances it is not fully optimal, and other triplets should be considered when performing searches in the future, \emph{e.g.} $\big(\MCMS, H_T~{\rm or}~M_J, N_j\big)$.

While this study reveals many properties of the search variables and their correlations for a large range of jets + $\MHT$ signals, we have not attempted to realistically include sources of error.  In particular, the use of BDTs obscures the exact nature of the ``signal region," thereby making it difficult to assess the quality of agreement between the Monte Carlo predictions and the measured backgrounds in a control region.  This is a standard issue with using machine learning tools, and we are not advocating to replace the traditional cut-and-count approach.  Instead, our point of view is that one can use this technology to quantitatively evaluate the performance of variables with a particular emphasis on their correlations. 

There are a variety of future directions which will be interesting to explore. The $n$-body framework could be extended to other searches for supersymmetry, such as those for $R$-parity violation, as well as more directed searches involving heavy flavor tags and also electroweak production. This framework is also clearly useful for non-SUSY new physics searches as well.  Along with extending the framework in theory space, it would be interesting to realistically quantify the effects of systematic and other errors on our conclusions.

With the LHC on the cusp of delivering up to 100~fb$^{-1}$ of data in the next few years, understanding the optimal ways to search for new physics has never been more important. We have provided a new framework for organizing and studying the collider phenomenology of a variety of beyond the Standard Model scenarios, which can be utilized to more deeply understand the breadth of results from the LHC, whatever they may be.  And once a hint of new physics begins to emerge, $n$-body extended Simplified Models will be very useful as a signal injection.  This will allow us to quantitatively unravel the properties of whatever decays are generating the anomalous signal, ultimately yielding new insights into the nature of our Universe.

\vspace{-15pt}
\section*{Acknowledgements}
\vspace{-10pt}
We thank Jeff Richman, Scott Thomas, Keith Ulmer, Hannsjoerg Weber, and Si Xie for useful conversations and comments to this manuscript.  
TC is supported by an LHC Theory Initiative Postdoctoral Fellowship, under the National Science Foundation grant PHY-0969510, and also thanks the KITP at UCSB, where this work was competed and the related support from the National Science Foundation under grant NSF PHY11-25915.  MD is supported by the Australian Research Council, and whose research was undertaken partly at the  Munich Institute for Astro- and Particle Physics (MIAPP), part of the DFG cluster of excellence ``Origin and Structure of the Universe".  SEH is supported by the Cluster of Excellence Precision Physics, Fundamental Interactions and
Structure of Matter (PRISMA-EXC 1098) and the Mainz Institute for Theoretical Physics.  JH, NT, and AW are supported by Fermi Research Alliance, LLC under Contract No. De-AC02-07CH11359 with the United States Department of Energy.  JH and AW are specifically supported by an Early Career Award (FNAL 14-05) from the Department of Energy, Office of Science, Office of High Energy Physics.

\appendix
\section*{Appendix}
\section{Review of Boosted Decision Trees}
\label{app:bdts}
Decision trees are a method of separating a parameter space into signal and background (or noise-like) regions.  They have been used in particle physics for over a decade (early examples include particle identification at MiniBOONE~\cite{Roe:2004na,Yang:2005nz} and the search for single top-quark production at the Tevatron~\cite{Abazov:2006gd}). Decision trees operate through a recursive partitioning of parameter space into signal and background-like regions which are determined through the use of training datasets. In that sense they represent the optimal cut-and-count discrimination between signal and background that can be performed. 

Informally, a single decision tree can be imagined as a cut-flow through a series of nodes. Each node corresponds to a cut in a particular variable, with events being partitioned into different bins as they progress further down the tree. The end (or terminal) nodes of the tree correspond to signal or background-like regions, depending on whether they contain a majority of signal or background events from the training data-set used. However, single trees can be unstable,  in the sense that the cuts chosen at each node are sensitive to the details of the training dataset. A more powerful a approach derives from the use of a multiplicity of trees -- effectively a vote by committee. Such a collection is a called a boosted decision tree. This also has the advantage that events which were misclassified by the original single tree can now be up weighted, leading to greater attention from succeeding trees.  Essentially, it this series of trees collectively act to minimize a predetermined loss function. An example is a least-squares loss function for fitting an unknown multidimensional function, although often other functions can be chosen which lead to greater stability against outlying points. We now describe this more formally, before providing an illustrative example using the razor variable. A good introduction to machine learning techniques is~\cite{Hastie} (which we have adapted the following from), while the original ideas of boosted decision trees can be found in~\cite{Friedman1,Friedman2}.

\begin{center}
\textbf{Formalism}
\end{center}
\vspace{-10pt}
A tree is defined as a series of nodes, each of which corresponds to a cut on an observable calculated from the input data -- in our case these correspond to the event observables such as $M_J$, $H_T$, $M_{T2}$ and so forth. More formally, a tree partitions the parameter space into a set of disjoint rectangular regions $R_j$, which are represented by the final (terminal) nodes of the tree. Each region is associated with a constant $\gamma_j$, which indicates whether that node or region of parameter space is considered signal-like or background-like. For classification into two classes these are usually taken to be $\{-1,1\}$. Then any event which falls into the region $R_j$ is assigned value $\gamma_j$. If we define the indicator function $I\big(x\in R_j\big)$ to be 1 if $x \in R_j$ and 0 otherwise, we can represent the a decision tree $T$ by
\begin{equation}
T\big(x;\Theta\big) = \sum_{j=1}^{J} \gamma_j \,I\big(x\in R_j\big)\,,
\end{equation}
where the parameters of the tree are $\Theta=\big(R_j,\gamma_j\big)$, and the number of regions $J$ is a meta-parameter which is usually $4\leq J\leq 8$. 
The numerical optimization problem which must then be solved is to find the regions $R_J$ and constants $\gamma_J$. These parameters are set by requiring that they minimize a loss function $L$ over a large set of training data whose properties are already known (that is to say, whether a given event is signal or background), so that the chosen parameters are
\begin{equation}
\hat{\Theta} = \arg\min_{\Theta}\sum_{j=1}^{J} \sum_{x_i\in R_j} L\big(y_i,\gamma_j\big) \, .
\end{equation}
This is a difficult problem in numerical optimization, and so approximate solutions  are usually used to find the regions $R_j$ and $\gamma_j$, which we will describe below. 

The output of a single decision tree can be quite sensitive to minor changes in the training sample. Furthermore, since the decisions at each node are only locally optimal, there is no guarantee that the globally optimal decision tree is obtained in this way. It is also possible to overfit the training data using complex trees. To avoid these issues we use boosted decision trees in our study. 

Boosting starts with a group of individual 'weak learners' such as single trees whose output may be only slighter better than random guessing. Then by weighting their outputs, a much better 'strong learner'  can be constructed, whose output is very well correlated with the true classification of any unknown event. In our work we use the gradient boosting algorithm as implemented in the TMVA class within $\texttt{ROOT}$~\cite{Brun:1997pa}.

A boosted tree model can thus be represented as a sum of trees,
\begin{equation}
f_M(x) = \sum_{m=1}^{M} T_m\big(x;\Theta_m\big) \, .
\end{equation}
We do not attempt to solve for all trees simultaneously, but rather do so in forward stage-wise manner: \textit{i.e.}, we solve for one tree at a time, where each tree is fit to the residual of the training data and the sum of all previous trees.  In other words, the parameters $R_{jm}$ and $\gamma_{jm}$ of the $m^\text{th}$ tree are determined by minimizing 
\begin{equation}
  \hat{\Theta}_m = \arg\min_{\Theta_m} \sum_{i=1}^{N} L\Big(y_i, f_{m-1} + T\big(x_i;\Theta_m\big)\Big)
  \label{eq:tree_m}
\end{equation}
where the sum is over the elements in the training dataset and $(m-1)^\text{th}$ boosted model. For example, if we wished to fit a sum of trees to a function using a squared-error loss function, the $m^\text{th}$ tree would be the tree that best predicts the residuals $y_i-f_{m-1}(x_i)$, and the constant $\gamma_{jm}$ would be given by the mean of the residuals in each region $R_{jm}$. Such trees can be constructed relatively quickly. For more general loss differentiable loss functions simple fast algorithms do not exist for solving~\eref{eq:tree_m}.

The forward stage-wise boosting strategy outlined above is very computationally greedy: it seeks to maximally minimize~\eref{eq:tree_m} at each step of the process. To do this in practice, we calculate coefficients of the negative of the gradient of the loss function $L$ at for each stage $m$:
\begin{equation}
 g_{im} = \left[ \frac{\partial L(y_i,f(x_i))}{\partial f(x_i)} \right]_{f(x_i)=f_{m-1}(x_i)}  
\end{equation}
The approximate solution to~\eref{eq:tree_m} is then given by fitting a tree to the negative gradients using a squared-error loss function
\begin{equation}
  \tilde{\Theta}_m = \arg \min_{\Theta} \sum_{i=1}^{N} \big( -g_{im} - T(x_i;\Theta) \big)^2 \, .
\end{equation}
As noted above, these trees can be constructed quickly. On the other hand, the regions $\tilde{R}_{jm}$ which result from the above process are not necessarily the same as those $R_{jm}$ which solve the exact problem in~\eref{eq:tree_m}. Of course, the forward stage-wise boosting procedure we employ is itself also approximation to the exact result (if it could be constructed).

For our classification problem of discriminating between signal from the $n$-body extended Simplified Models and SM backgrounds we  use the binomial log-likelihood loss
\begin{equation}
L(y,f(x))= \log \Big(1+ e^{-2\,y\,f(x)} \Big) \, ,
  \end{equation}
which is also implemented in the TMVA package in \texttt{ROOT}. This is known to be more robust than the common exponential loss function $L(y,f(x))=e^{-F(x)\,Y}$, since misclassified points and outliers effectively are assigned a linear penalty, as opposed to an exponential one.

\begin{center}
\textbf{Example: Razor Variables}
\end{center}

We now show a simple application of the BDTs to differentiate gluino events from QCD backgrounds using the razor variables $M_R$ and $R^2$ only. As shown in Fig.~\ref{Fig:razor} and mentioned in Sec.~\ref{subsubsec: correlations}, combining these two variables together provides a much higher discriminating power than using either of them individually. Additionally, simple rectangular cuts in the two-dimensional parameter space would overlook some subtle features of the signal and background distributions such as the increase in the typical value of $R^2$ at low $M_R$.

For this simple study, we use the \texttt{scikit-learn} module from \texttt{Python 2.7.9}. We consider a training sample composed of an event mix of $10^5$ unweighted QCD and $\widetilde{g}\rightarrow j + \widetilde{\chi}$ events. We use gradient boosting to generate different numbers of decision trees with maximal depth $J=4$. In order to limit overfitting, we multiply $T(x_i;\Theta_m)$ in Eq.~(\ref{eq:tree_m}) by a ``learning rate'' coefficient $\alpha = 0.1$. Introducing a small learning rate is a standard regularization procedure when using gradient boosting. Here, we use the Huber loss function, that is a combination of a squared-error and a least absolute deviation loss function.  When applied to a given $(M_R, R^2)$ doublet, the final classifier will output a number $0\le r\le 1$ that is close to $0$ if the event is background-like and close to $1$ if the event is signal-like. 

\begin{figure}[h!]
\centering
\includegraphics[width=0.48\linewidth]{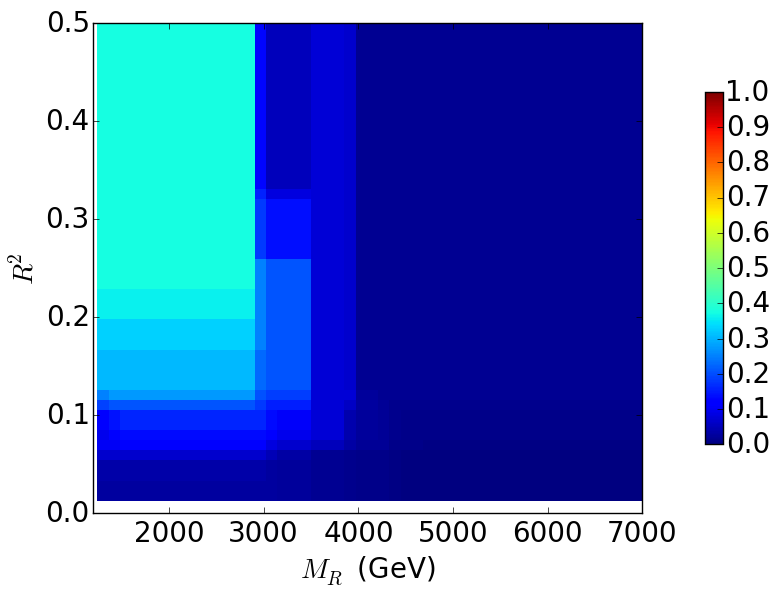}
\includegraphics[width=0.48\linewidth]{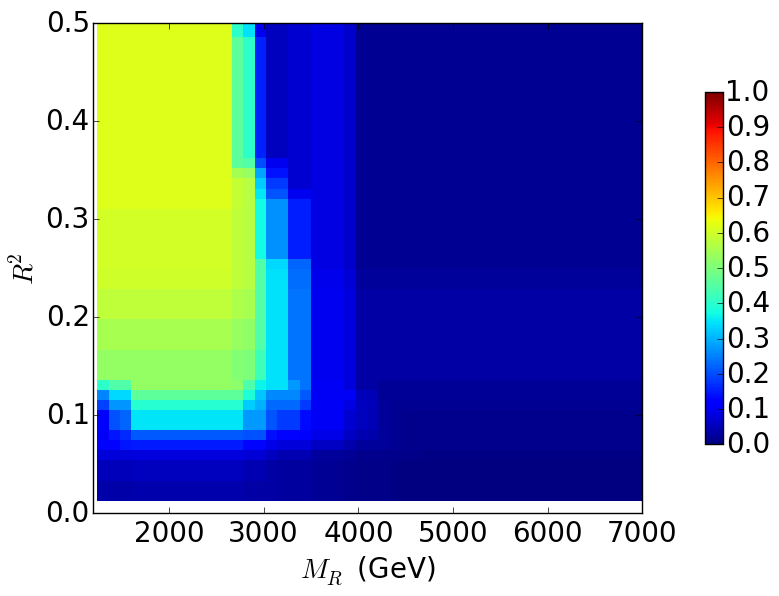}
\includegraphics[width=0.48\linewidth]{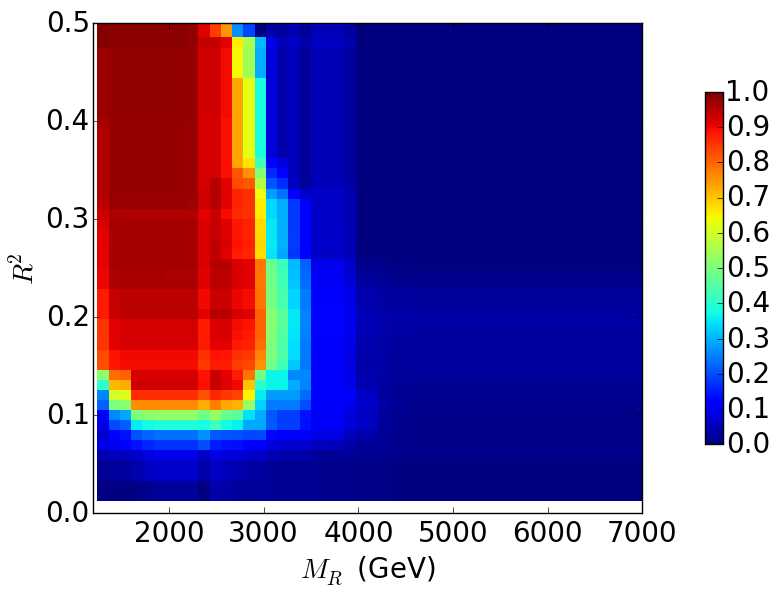}
\includegraphics[width=0.48\linewidth]{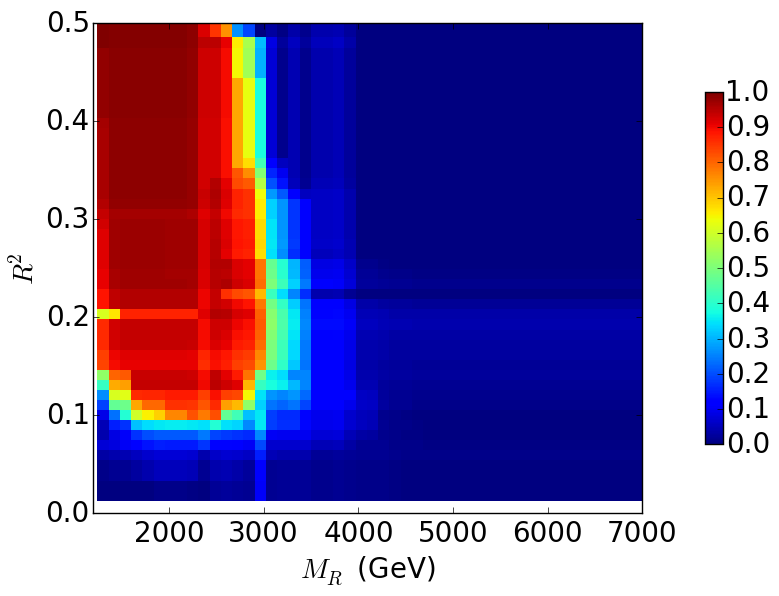}
\caption{ Color plots of the $r$ parameter in the $(M_R, R^2)$ space for classifiers involving $5$ (top left), $10$ (top right), $50$ (bottom left) and $100$ (bottom right) decision trees. The parameter $r$ should be close to $1$ for signal and to $0$ for background.}
\label{Fig: BDT} 
\end{figure}

Figure~\ref{Fig: BDT} shows color plots of $r$ in the $(M_R, R^2)$ space for classifiers involving $n = 5, 10, 50$ and $100$ trees. In regions with a large number of training events, comparing these $r$ plots to the density profiles in Fig.~\ref{Fig:razor} demonstrates that the BDT successfully captures the characteristic features of the signal and the background. In particular, when the number of trees is large, one can clearly see the red region departing from the $y$-axis for $R^2 \lesssim 0.15$. A more quantitative estimate of the performance of the BDTs can be obtained by applying the $n = 100$ classifier to a separate test sample of $4\times 10^5$ evenly mixed signal and background events. Tagging the events with $r < 0.5$ as background and the events with $r > 0.5$ as signal, our classifier correctly identifies $95\%$ of the background events and $80\%$ of the signal events. A better accuracy could be obtained by increasing the size of the training sample.  Also note that the partial dependence plot associated to the $n = 100$ classifier exhibits strange features that do not appear in the profiles shown on Fig.~\ref{Fig:razor}. With such a large numbers of trees, the classifier starts overfitting the training samples and loses part of its predictive power.   This is a concrete demonstration of the issues involved in optimizing the BDT parameters discussed previously.

\pagebreak
\section{Consistency of \emph{n}-body Decay Operators}
\label{App: UV}

\begin{figure}[h!]
\centering
\includegraphics[scale=0.55]{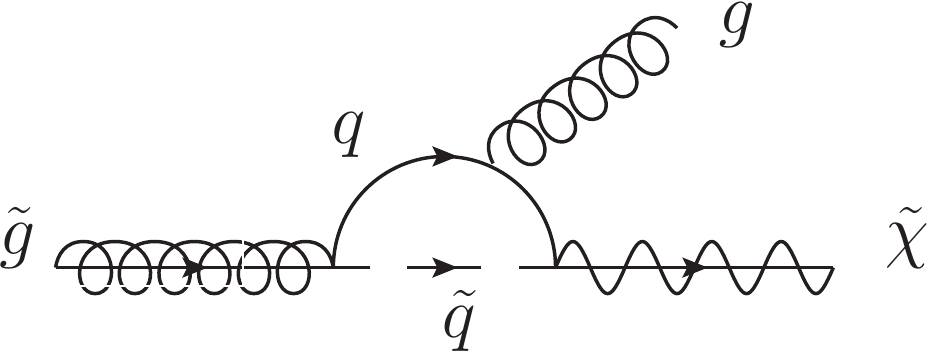}$\quad\quad\quad$
\includegraphics[scale=0.6]{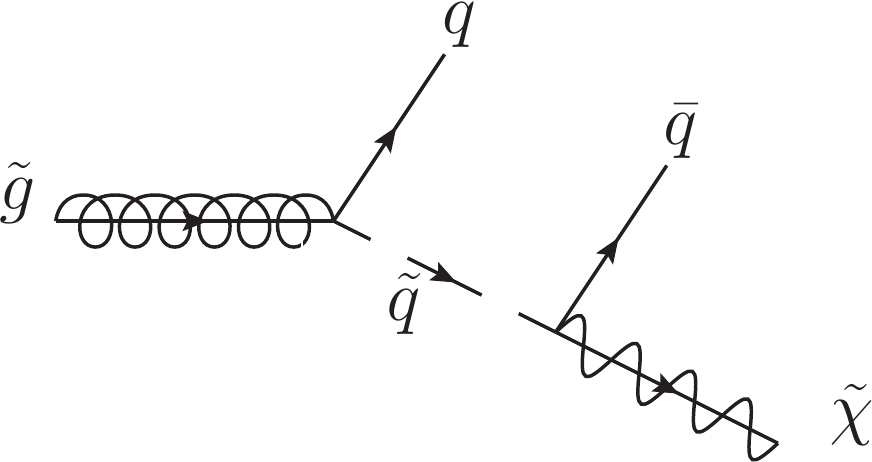}
\caption{Illustrative Feynman diagrams corresponding to the possible UV completions for the two-body (left) and three-body (right) decays of the gluino ($\mathcal{L}_1$ and $\mathcal{L}_2$ operators).}
\label{Fig: l1} 
\end{figure}

It is reasonable to wonder if the $n$-body operators in Eqs.~(\ref{Eq:l1})-(\ref{Eq:l5}) can be realized in any complete new physics scenario.  There are two issues that will be discussed: are there any regions of parameter space where a given $\mathcal{L}_i$ would model the dominant decay mode of the gluino, and what would be the corresponding lifetime of the gluino.  We begin with the $\mathcal{L}_1$ and $\mathcal{L}_2$ operators which can yield the dominant decay modes in models that include one extra heavy scalar particle, as shown in Fig.~\ref{Fig: l1}. A well-known example of such model is splitSUSY~\cite{Wells:2004di,ArkaniHamed:2004fb,Giudice:2004tc}, where the squarks are decoupled.  The relative importance of $\mathcal{L}_1 \sim 1/\Lambda$ compared to $\mathcal{L}_2\sim 1/\Lambda^2$ depends on the mass scale of the heavy particle.  For splitSUSY with a $1$~TeV gluino, $\mathcal{L}_1$ starts dominating over $\mathcal{L}_2$ only for squark masses heavier than $10^9$~GeV~\cite{Gambino:2005eh}.  Note that in this region of parameter space, the gluino tends to be sufficiently long lived to warrant a different class of search strategy.\footnote{For completenesses, we note that $\mathcal{L}_1$ has been included as one of the Simplified Models studied in the prompt gluino searches of~\cite{Aad:2015iea}.}  This example already demonstrates the point of this appendix -- the $\mathcal{L}_i$ operator should not be interpreted as a physical gluino decay mode, but rather as a topology representative of $(i+1)$-body decays in general. For example, our results for this operator can be used as a qualitative proxy with which to analyze signals of the form $\widetilde{q} \rightarrow q\, \Nlino$ as shown in Appendix~\ref{sec:OnShellIntermediateStates}.  

\begin{figure}[h!]
\centering
\includegraphics[scale=0.57]{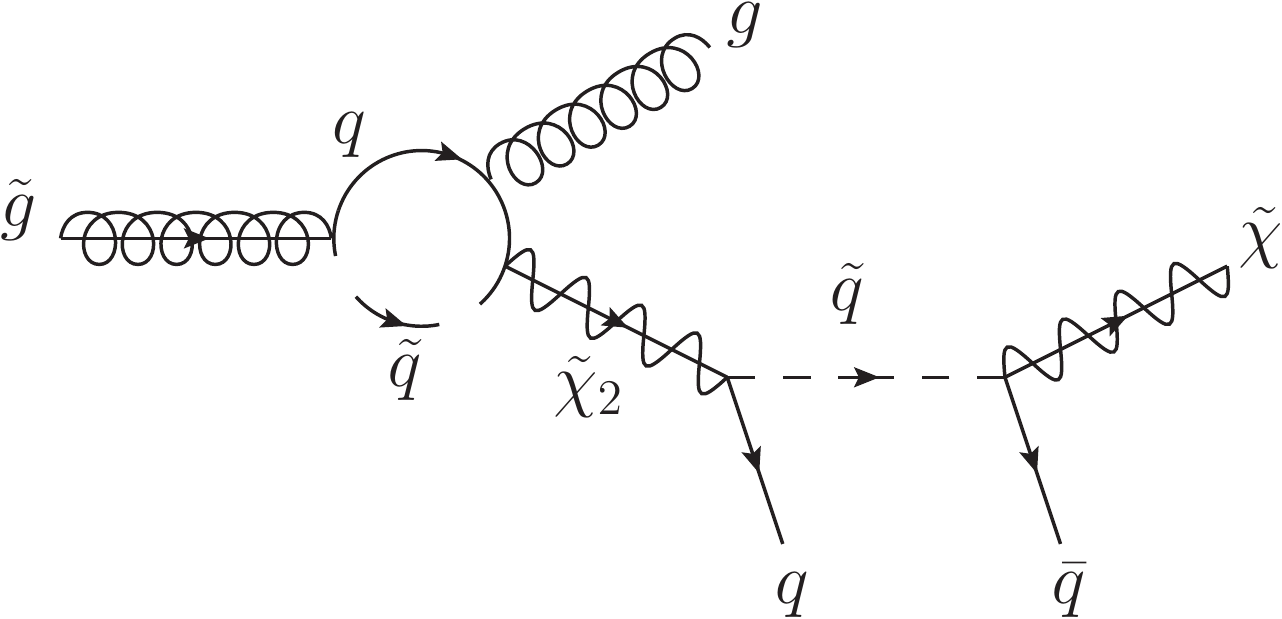} \\[20pt]
\includegraphics[scale=0.6]{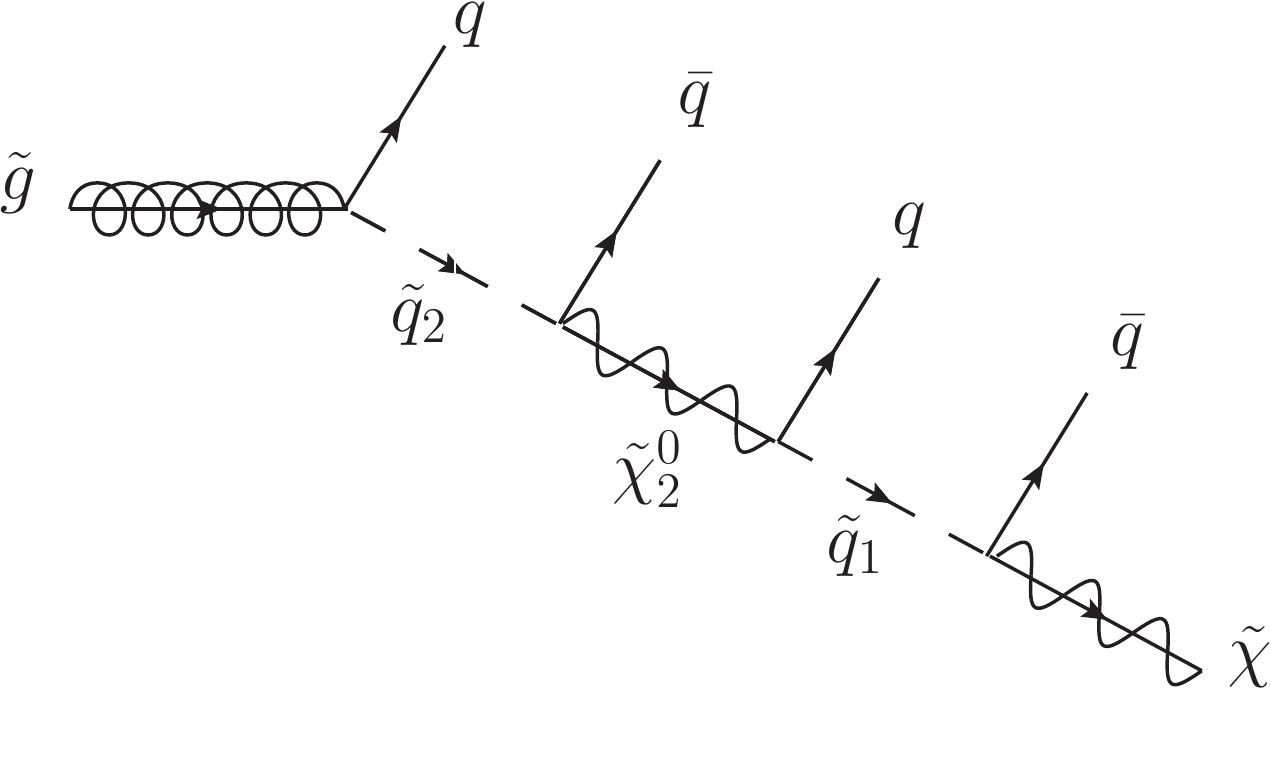}\\
\includegraphics[scale=0.6]{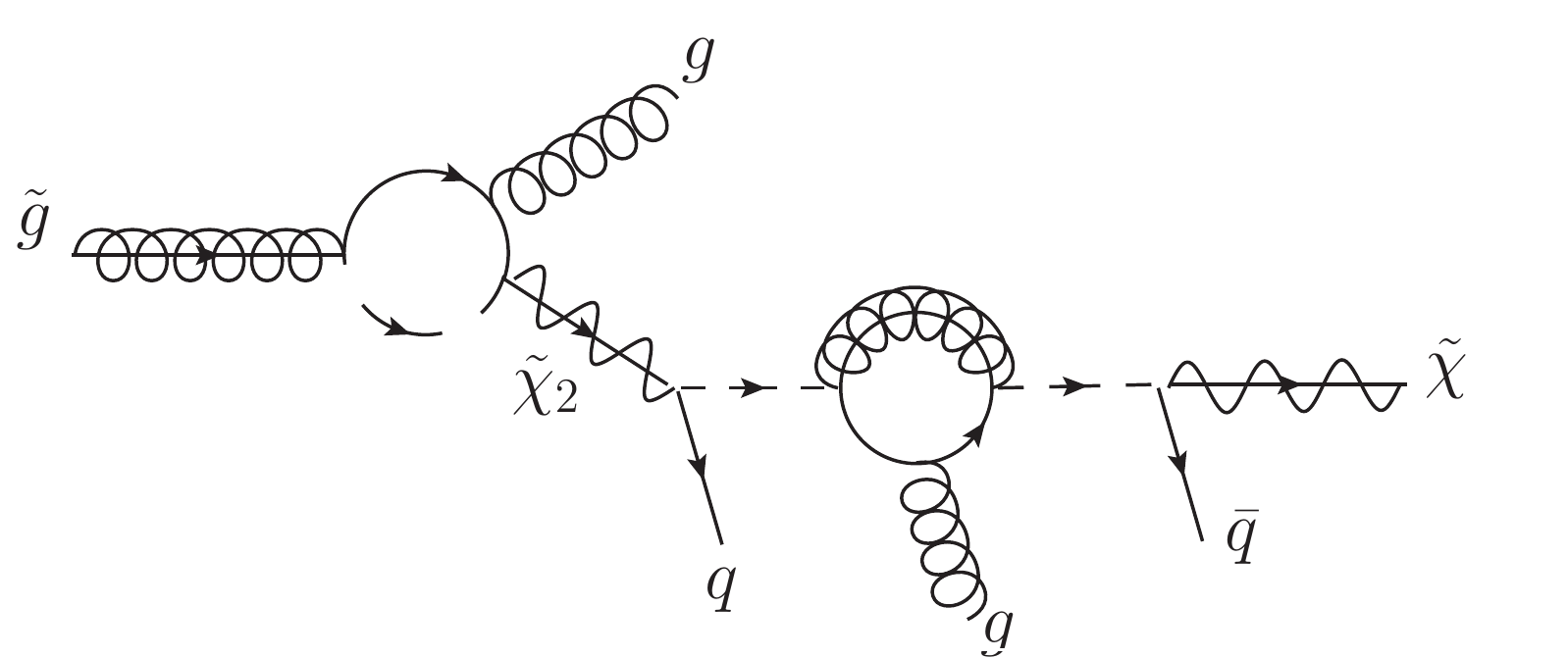}
\caption{Top: Illustrative Feynman diagrams corresponding to a UV completion for the four-body decay of the gluino at loop level (the $\mathcal{L}_3$ operator). Gluino decay modes involving gluons in the final state can be generated at tree-level, however the corresponding operators imply the existence of lower-order effective operators (here $\mathcal{L}_2$). 
Center and Bottom: Illustrative Feynman diagrams corresponding to the UV completions for the five-body decay of the gluino. The process for a gluino decay to four quarks and a neutralino ($\mathcal{L}_4$ operator, center) can be used to study gluino cascade decays through new particles. The operator involving two gluons at loop-level (bottom) will be highly suppressed. Note that, although we use the SUSY notation for the names of the new particles, the relevant couplings are not necessarily described in standard SUSY models.}
\label{Fig: l3l4} 
\end{figure}

Similar reasoning can be applied to the $\mathcal{L}_3$, $\mathcal{L}_4$ and $\mathcal{L}_5$ operators.  However, it is important to note that concrete UV completions might involve taking couplings that do not obey the SUSY relations, even though we continue to call them gluons and squarks for convenience.  These operators have a relatively high mass dimension, and most of the time will either give subdominant contributions to the gluino decays or will be associated with long-lived gluinos. Operators involving gluons, in particular, will either be loop-suppressed, as shown in Fig.~\ref{Fig: l3l4}, or will imply the existence of lower order operators without the presence of the gluon. For example, if the gluon in the $\mathcal{L}_3$ operator is generated at tree-level, the gluino decays will be dominated by $\mathcal{L}_2$. Although operators other than $\mathcal{L}_2$ will not describe dominant SUSY processes, they can be used to study either exotic processes with similar topologies or long decay chains involving intermediate on-shell states. As shown in Fig.~\ref{Fig: l3l4}, for example, the $\mathcal{L}_4$ operator can have UV completions analogous to gluino cascade decays through one or more heavy squarks and/or electroweakinos.  The operators studied here therefore do not apply exclusively to direct gluino decays but also to the general classes of models targeted by multijet plus $\slashed{E}_T$ searches such as~\cite{Aad:2016jxj,ATLAS-CONF-2015-062,Aad:2014kra,Aad:2014bva,Aad:2014wea,Khachatryan:2016kdk,Khachatryan:2015wza,Khachatryan:2015vra,Chatrchyan:2014lfa}. Our study can also be applied to cascade decays involving top quarks or gauge bosons, such as the ones investigated in~\cite{Aad:2015iea,CMS:2014dpa}. A more in-depth discussion of the effects of intermediate on-shell states is shown in Appendix~\ref{sec:OnShellIntermediateStates}.

\pagebreak
\begin{center}
\textbf{Lifetime Estimates}
\end{center}

High dimensional operators such as $\mathcal{L}_4$ or $\mathcal{L}_5$ or operators involving gluons in the final state -- and therefore loop interactions -- will be associated with highly suppressed gluino decay rates. If no other decay channel is open, the gluinos could therefore be long-lived, and the search strategies studied here will be less relevant. Note however that most of our results for these operators  will remain applicable to signals involving long decay chains through on-shell particles.
We provide a naive estimate of the lifetime associated with each $\mathcal{L}_i$.  The effective operator corresponding to $\tilde{g}\rightarrow k\, (q + \overline{q}) + \tilde{\chi}$ (where $k$ counts the number of quark-anti-quark pairs), takes the form
\begin{align}
	\mathcal{L} = \frac{y^{2\,k}}{\Lambda^{3\,k-1}} (\overline{q}\, q)^{k}\, \widetilde{g}\,\Nlino\,,
\end{align}
with a corresponding decay width
\begin{align}
	\Gamma \sim m_{\tilde{g}} \,\frac{y^{4 \,k}}{2 (2\,\pi)^{4\, k - 1}}\left(\frac{m_{\tilde{g}}}{\Lambda}\right)^{2(3\,k - 1)}.
\end{align}
Again, we emphasize that the $y$ couplings might not be related to Standard Model couplings by SUSY.  For operators that involve an additional gluon in the decay, 
\begin{align}
	\mathcal{L} = \frac{y^{2\,k} g_s}{16\,\pi^2\,\Lambda^{3\,k+1}} (\overline{q}\, q)^{k} G^{\mu\nu}\, \widetilde{g}\,\sigma_{\mu\nu}\,\Nlino,
\end{align}
the width is of the form
\begin{align}
	\Gamma \sim \frac{ m_{\tilde{g}}}{(16\,\pi^2)^2} \frac{y^{4 (k + 1)}\,g_s^2}{2 (2\,\pi)^{4\, k + 1}}\left(\frac{m_{\tilde{g}}}{\Lambda}\right)^{2(3\,k + 1)}\,,
\end{align}
where we have assumed that the gluino interacts with the gluon at one-loop; as discussed above the tree-level operators with gluons will never dominate over the lower point decay without this extra state.

Note that the lifetime of a particle in its rest frame is given by
\begin{align}
	\tau  = \frac{1~\mathrm{GeV}}{\Gamma}\,\big(6.5\times 10^{-25}\big)\,\mathrm{s}.
\end{align}
The lifetimes derived from Eqs.~(\ref{Eq:l1})-(\ref{Eq:l5}) for a gluino at rest are then given by 
\begin{align}
\tau_{\mathcal{L}_1} &\sim y^{-4}\left(\frac{1\text{ TeV}}{m_{\tilde{g}}}\right) \left(\frac{\Lambda}{10~\mathrm{TeV}}\right) \times 10^{-21}\,\, \text{s}\,;\\
\tau_{\mathcal{L}_2} &\sim  y^{-4}\left(\frac{1\text{ TeV}}{m_{\tilde{g}}}\right)\left(\frac{\Lambda}{10~\mathrm{TeV}}\right)^2 \times 10^{-23}\,\, \text{s}\,;\\
\tau_{\mathcal{L}_3} &\sim  y^{-8}\left(\frac{1\text{ TeV}}{m_{\tilde{g}}}\right)\left(\frac{\Lambda}{10~\mathrm{TeV}}\right)^4 \times 10^{-9}\,\, \text{s}\,;\\
\tau_{\mathcal{L}_4} &\sim  y^{-8}\left(\frac{1\text{ TeV}}{m_{\tilde{g}}}\right)\left(\frac{\Lambda}{10~\mathrm{TeV}}\right)^5 \times 10^{-13}\,\, \text{s}\,;\\
\tau_{\mathcal{L}_5} &\sim  y^{-12}\left(\frac{1\text{ TeV}}{m_{\tilde{g}}}\right)\left(\frac{\Lambda}{10~\mathrm{TeV}}\right)^7 \times 10\,\text{s}\,.
\end{align}
  Even with order one couplings and for heavy gluinos, $\mathcal{L}_5$ will lead to macroscopic decay lengths due to the large number of final states along with the loop suppression associated with the gluon emission. Again, due to the gluon loop factor, $\mathcal{L}_3$ will be more suppressed than $\mathcal{L}_4$ and will be associated with gluino decay lengths longer than a millimeter for couplings less than one or gluinos lighter than a PeV. For gluino masses of a TeV or larger, $\mathcal{L}_4$ will lead to tracks shorter than a millimeter for couplings of order one, but the lifetime of the gluino will strongly increase for smaller couplings. Finally, the operators $\mathcal{L}_1$ and $\mathcal{L}_2$ will lead to a short-lived gluino even for weak couplings.

\section{Mapping Onto \emph{n}-body Decays}
\label{sec:OnShellIntermediateStates}

In this appendix we quantify the effects of intermediate on-shell states on the kinematic distributions of the variables studied throughout this paper. We first consider a two-body decay of the form
\begin{align}
	A \rightarrow j + X
\end{align} 
where $j$ is either a quark or a gluon and assumed to be massless. We denote the four-momentum of $A$ as $(E, p)$. There are two configurations in which the momentum of $X$ in the lab frame can be simply estimated:
\begin{itemize}
\item \textbf{The Wide (W) scenario}, with $M_A \gg M_X$: in this case, the energy of $A$ is evenly split between the jet and $X$.
	\item \textbf{The Compressed (C) scenario}, with $M_A - M_X \ll M_A$: in this case, the jet $p_T$ will be negligible and the momentum of $X$ will be similar to the momentum of $A$.  
\end{itemize}
Consider the decay $\widetilde{g} \rightarrow q\,\overline{q}\,\Nlino$, through the diagram shown on the right-hand side of Fig.~\ref{Fig: l1}. If the squark is on-shell, this decay can be decomposed into two two-body decays, with $(A, X)$ being $(\tilde{g},\tilde{q})$ in the first step and $(\tilde{q},\tilde{\chi})$ in the second. Studying each of these successive two-body decays within both the \textbf{C} and \textbf{W} scenarios described above, we can estimate the values of $\MHT$ and $H_T$ for the $\mathbf{CC}, \mathbf{CW},\mathbf{WC}$ and $\mathbf{WW}$ mass hierarchies. The results are shown in Table~\ref{Tab: onshell}, where we compare the estimates derived in these four scenarios to the typical $\MHT$ and $H_T$ values obtained when the squark is off-shell.

\begin{figure}[h!]
	\centering
	\includegraphics[width=0.3\linewidth]{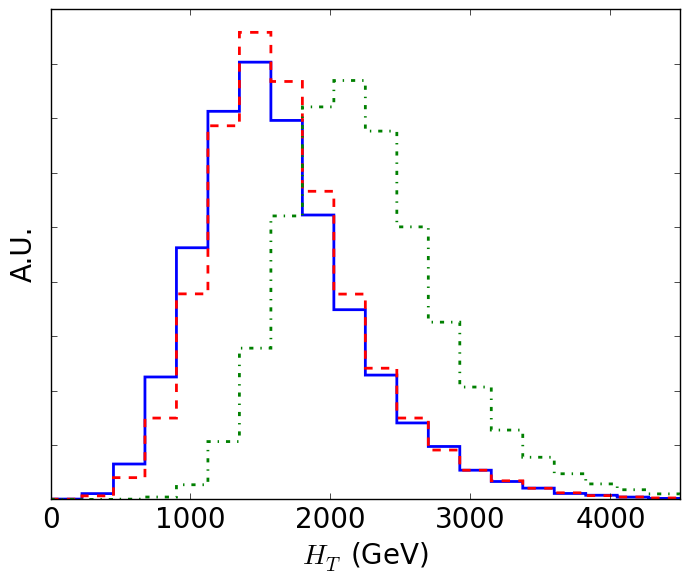}
	\includegraphics[width=0.3\linewidth]{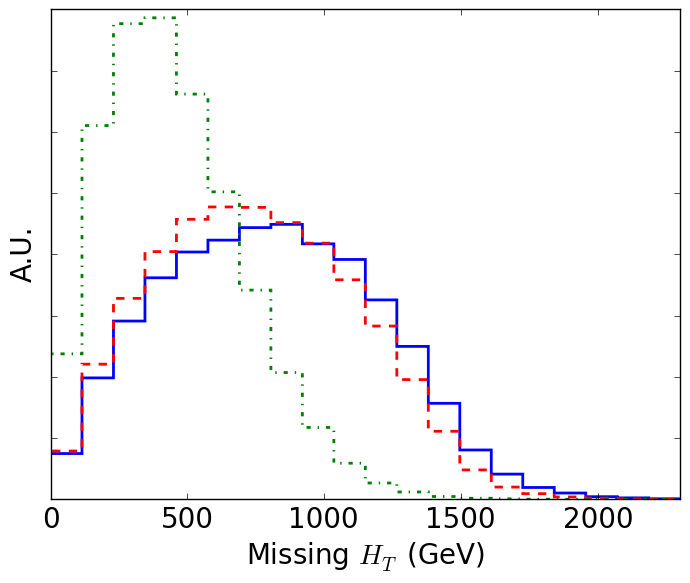}
	\includegraphics[width=0.305\linewidth]{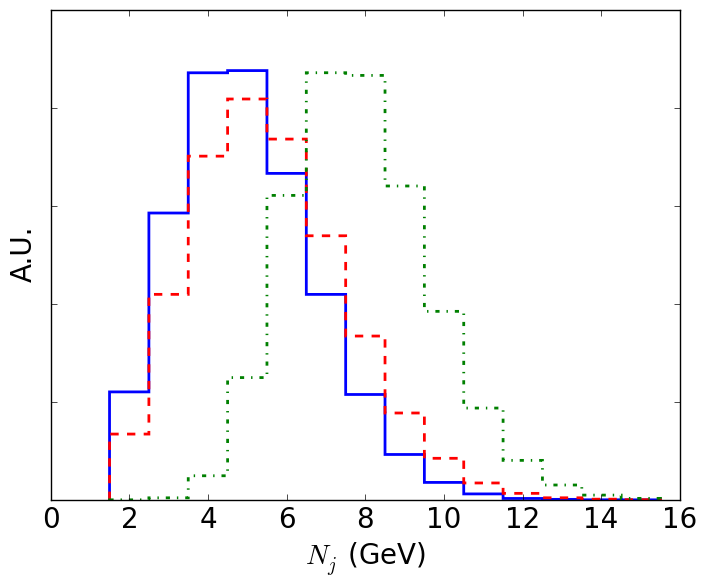}
	\caption{\label{Fig: stop} $H_T$ (top left), $\MHT$ ( top right) and $N_j$ (bottom) distributions for $\widetilde{g}\rightarrow j+\Nlino$ (dashed red),  $\widetilde{g}\rightarrow j+\Nlino$ (dot-dashed green) and $\widetilde{t}\rightarrow t+\Nlino$ (solid blue). We have only included hadronic top decays. Here, the stop mass is $1.5$~TeV and the neutralino mass is $1$~GeV.}
\end{figure}

As shown in this table, the $n$-body formalism that we have adopted throughout this paper accurately describes the kinematics when one set of particles in the cascade is compressed.  For contrast, when the neutralino is much lighter than the gluino, some discrepancies can appear from the presence of on-shell intermediate particles when comparing these decays to the $n = 4$ case. However, the typical values of $\MHT$ and $H_T$ for the $\mathbf{WC}$ and $\mathbf{CW}$ scenarios -- for which the highest disagreement is observed -- are similar to the values associated with the gluino effective two-body decay operator $\mathcal{L}_1$ when the neutralino is light. For these \textbf{WC} and \textbf{CW} mass hierarchies, the process dominating the kinematics is indeed the gluino two-body decay to a quark and a squark. Figure~\ref{Fig: stop} illustrates this effect using the decay of a heavy stop to a top quark and a neutralino as an example. Although the top quark decays to three jets, the $\MHT$ and $H_T$ distributions for the stop decay are very different from the ones corresponding to a gluino four-body decay, which is the naive guess for the relevant $n$-body decay by simply counting the number of decay partons. These distributions are however similar to the ones obtained for a gluino two-body decay, as can be inferred from the figure.

\begin{table}[t!]
	\def\arraystretch{1.5}
	\begin{tabular}{cccc}
		\toprule
		Scenario & $\quad$Mass Hierarchy$\quad$ & $\MHT$ & $H_T$\\
		\midrule
		$\mathbf{CC}$ & $m_{\tilde{g}}\sim m_{\tilde{q}}\sim m_{\tilde{\chi}}$ & $p_{\tilde{g}}$ & $0$\\
		$\mathbf{CW}$ & $m_{\tilde{g}}\sim m_{\tilde{q}}\gg m_{\tilde{\chi}}$ & $E_{\tilde{g}}/2$ & $E_{\tilde{g}}/2$\\
		$\mathbf{WC}$ & $m_{\tilde{g}}\gg m_{\tilde{q}}\sim m_{\tilde{\chi}}$ & $E_{\tilde{g}}/2$ & $E_{\tilde{g}}/2$\\
		$\mathbf{WW}$ & $m_{\tilde{g}}\gg m_{\tilde{q}}\gg m_{\tilde{\chi}}$ & $E_{\tilde{g}}/4$ & $3E_{\tilde{g}}/4$\\
		off-shell $\mathbf{C}$ & $m_{\tilde{g}}\sim m_{\tilde{\chi}}$ & $p_{\tilde{g}}$ & $0$\\
		off-shell $\mathbf{W}$ & $m_{\tilde{g}}\gg m_{\tilde{\chi}}$ & $E_{\tilde{g}}/4$ & $3E_{\tilde{g}}/4$\\
		\bottomrule
	\end{tabular}
	\caption{Estimates for $\MHT$ and $H_T$ for $\widetilde{g}\rightarrow q\,\overline{q}\,\Nlino$ for different possible squark and neutralino mass configurations.}
	\label{Tab: onshell}
\end{table}

\begin{figure}[h!]
	\centering
	\includegraphics[width=0.3\linewidth]{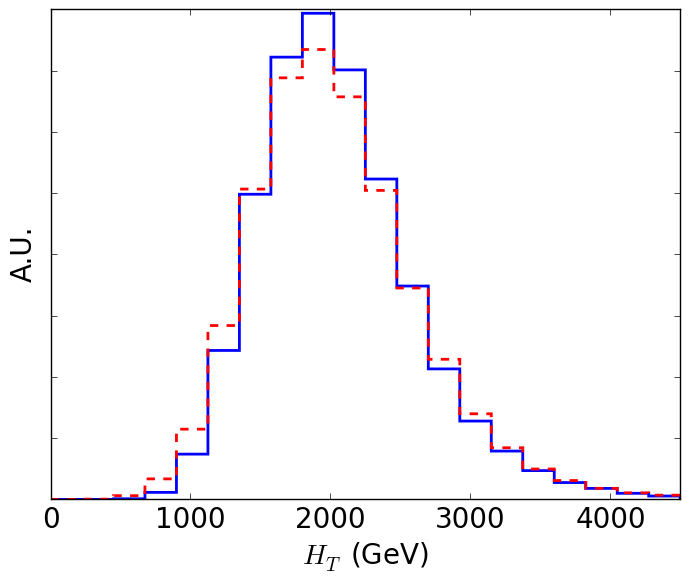}
	\includegraphics[width=0.3\linewidth]{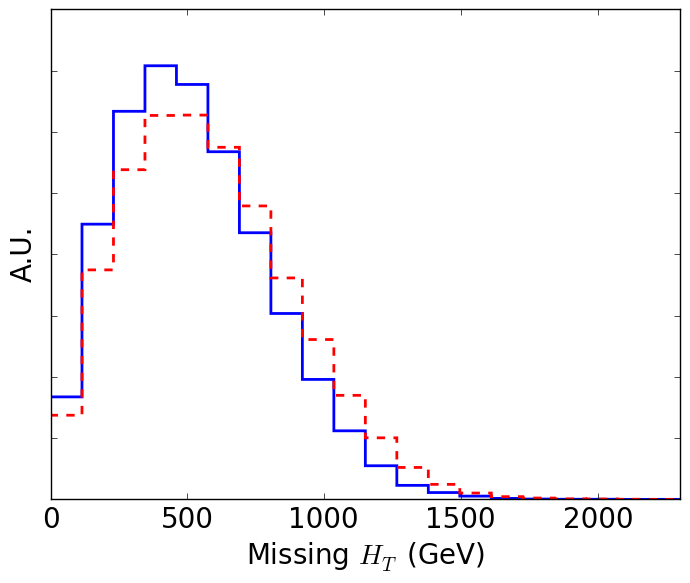}
	\includegraphics[width=0.305\linewidth]{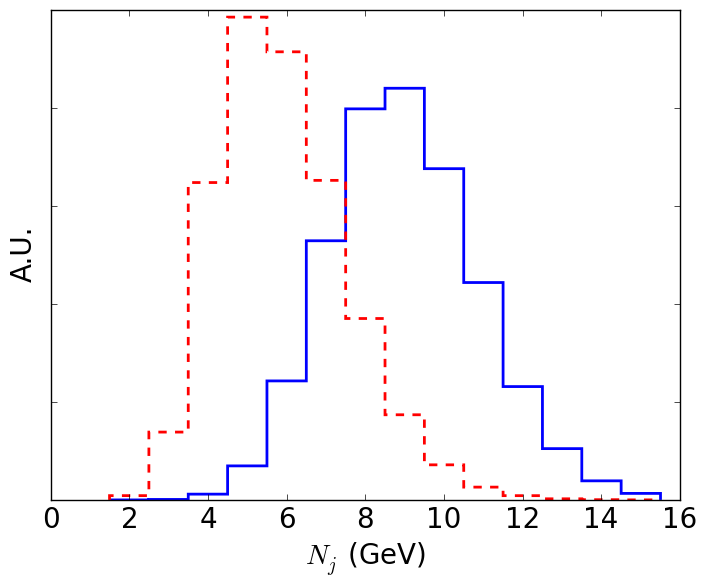}
	\caption{\label{Fig: tt} $H_T$ (top left), $\MHT$ ( top right) and $N_j$ (bottom) distributions for $\tilde{g}\rightarrow 2j+\tilde{\chi}$ (dashed red) and $\tilde{g}\rightarrow t + \bar t+\tilde{\chi}$ (solid blue). We have considered only hadronic top decays.}
\end{figure}

\begin{figure}[h!]
	\centering
	\includegraphics[width=0.3\linewidth]{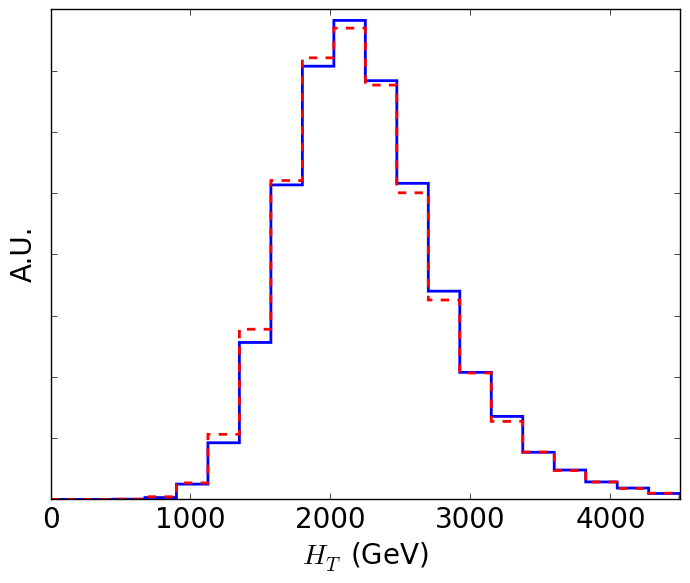}
	\includegraphics[width=0.3\linewidth]{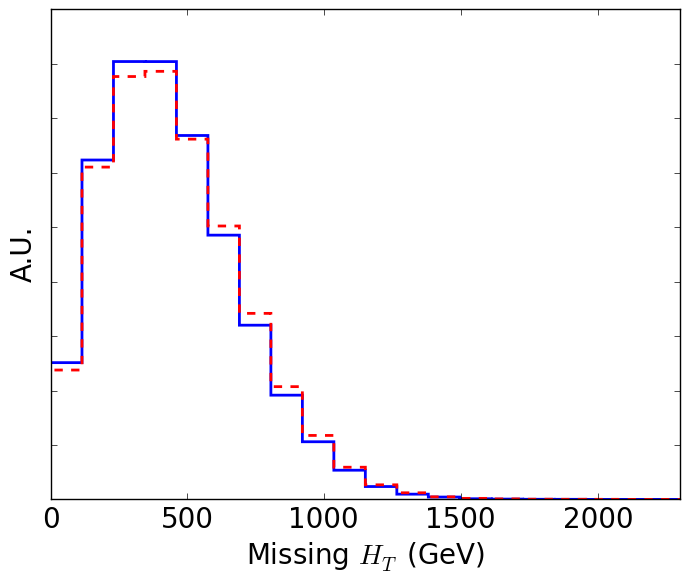}
	\includegraphics[width=0.305\linewidth]{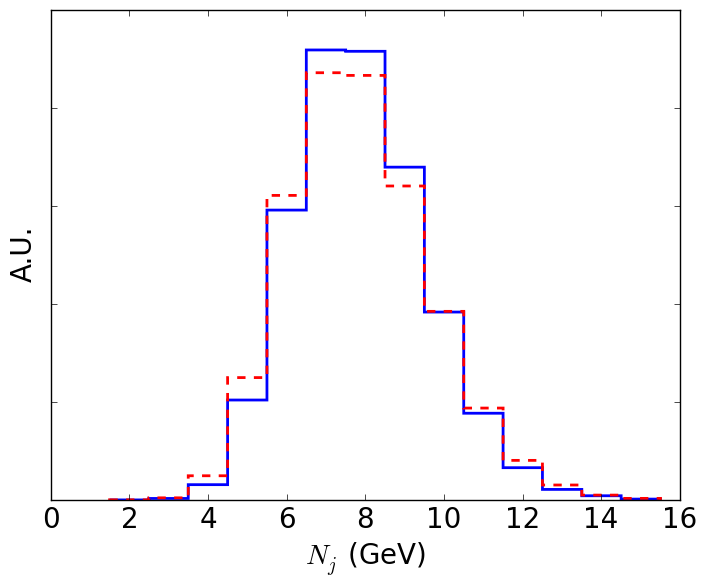}
	\caption{\label{Fig: qqZ} $H_T$ (top left), $\MHT$ ( top right) and $N_j$ (bottom) distributions for $\tilde{g}\rightarrow 3j+\tilde{\chi}$ (dashed red) and $\tilde{g}\rightarrow q + \bar q + Z + \tilde{\chi}$ (solid blue). We have considered only hadronic $Z$ decays.}
\end{figure}

A similar reasoning applies to more complicated processes such as $\widetilde{g}\rightarrow t\,\overline t\,\Nlino$ or $\widetilde{g}\rightarrow q\, \overline{q}\, Z\,\Nlino$. For a TeV-scale gluino, the masses of the top quark and the $Z$ can be neglected to good approximation and the two processes can be compared to $\widetilde{g}\rightarrow j \,j\,\Nlino$ and to $\widetilde{g}\rightarrow j\, j\, j\,\Nlino$ respectively. The comparisons are shown in Fig.~\ref{Fig: tt} for $\widetilde{g}\rightarrow t\,\overline{t}\,\Nlino$ and in Fig.~\ref{Fig: qqZ} for $\widetilde{g}\rightarrow q\,\overline{q}\, Z\,\Nlino$. In both cases, the $\MHT$ and $H_T$ distributions for the on-shell processes look similar to the ones corresponding to the appropriate choice of $n$-body operators. 

This demonstrates that although some of the higher-dimensional operators considered in this study will not lead to prompt gluino decays, most of our results can still be applied to models involving on-shell intermediate states. In particular, as shown in this section, a given cascade decay scenario can be mapped to a given effective decay operator with similar kinematic distributions for the ``missing energy'' and ``energy scale''-type variables. This mapping, however, cannot be solely determined by the number of final state objects and will depend on the mass hierarchy between the parent particle and the intermediate states. It is also important to note that, as shown in Figs~\ref{Fig: stop}, \ref{Fig: tt} and \ref{Fig: qqZ}, ``energy structure'' variables such as $N_j$ will remain by definition highly sensitive to the number of final states and to the quark or gluon nature of the jets, see  Appendix~\ref{App: quarkgluon}.  However,  this difference will not impact the conclusions drawn in the main body of the text.  

\pagebreak
\section{Comparing Quark And Gluon Final States}
\label{App: quarkgluon}
\begin{figure}[h!]
	\centering
	\includegraphics[width=0.35\linewidth]{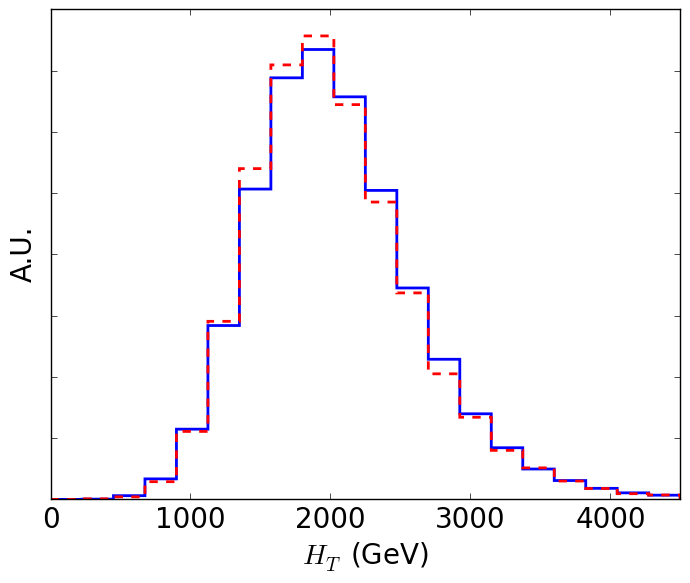}
	\includegraphics[width=0.35\linewidth]{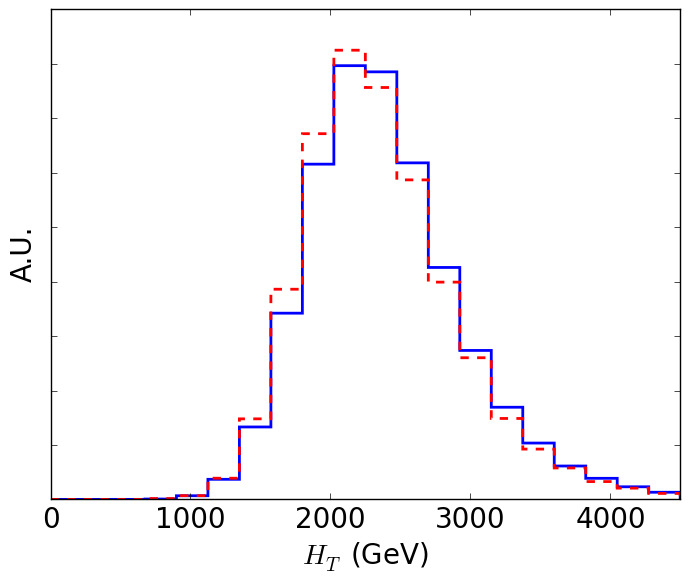}
	\includegraphics[width=0.35\linewidth]{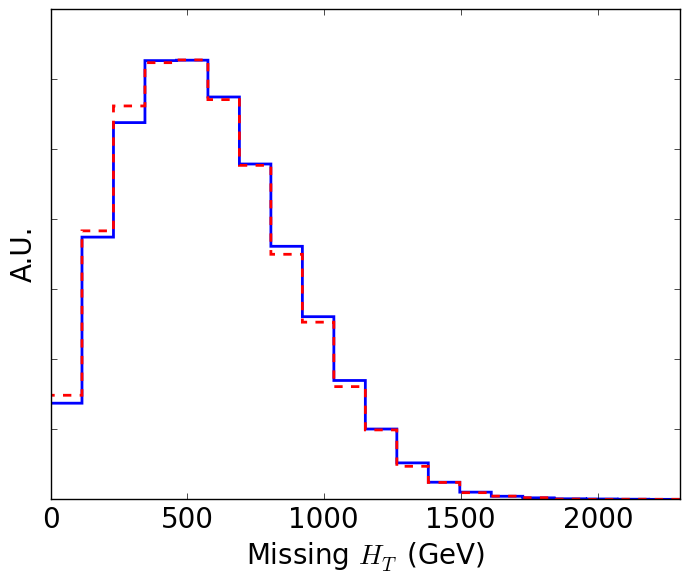}
	\includegraphics[width=0.365\linewidth]{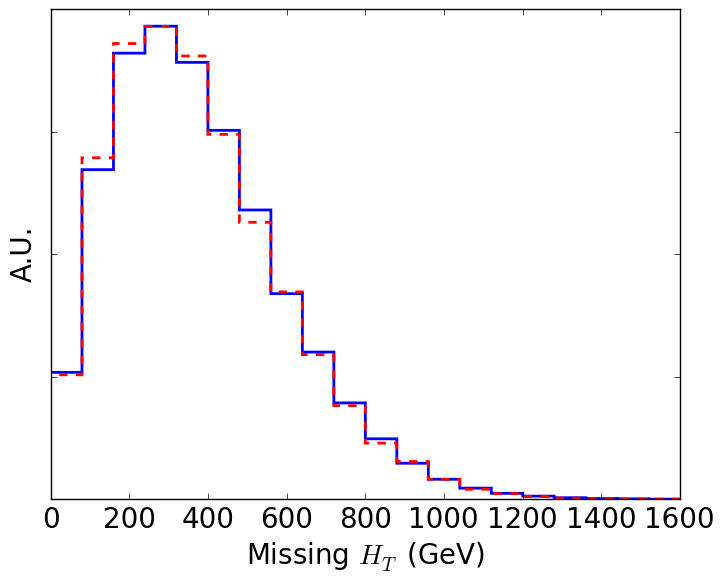}
	\includegraphics[width=0.35\linewidth]{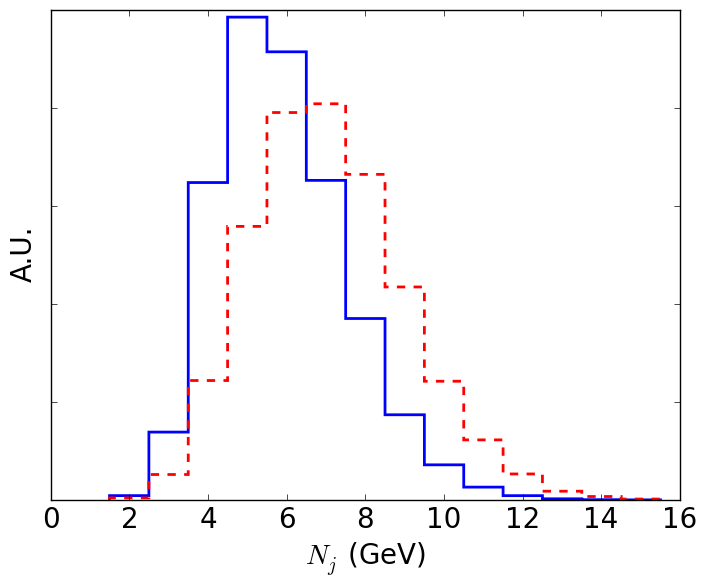}
	\includegraphics[width=0.35\linewidth]{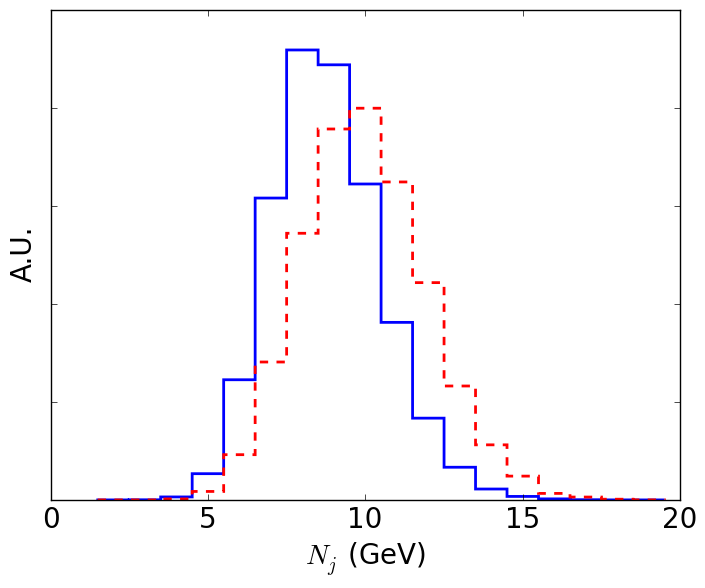}
	\caption{\label{Fig: gluon} $H_T$ (top), $\MHT$ (middle) and $N_j$ (bottom) distributions for gluino three-body (left) and five-body (right) decays. Here, we compare decays with only quarks in the final state (solid blue) to decays with only gluons in the final state (dashed red).}
\end{figure}

This appendix addresses the differences from having quarks versus gluons as the final state decay partons.  Figure~\ref{Fig: gluon} shows the $H_T$, $\MHT$ and $N_j$ distributions for gluino three- and five-body decays with either the mix of quarks and gluons used in the $n$-body operators given in Eqs.~(\ref{Eq:l1})-(\ref{Eq:l5}) or final states with only gluons.  As can be seen in this figure, the only variable that discriminates between quarks and gluons is the number of jets in the event, since gluons are associated with higher rates for hard splittings due to the larger Casimir. Since the angles between these jets due to showering and the initial parton tend to be collinear, the discrepancy between the $N_j$ distributions should decrease for larger jet radius. The other kinematic variables that we consider only minimally distinguish between quarks and gluons.

\pagebreak
\section{Comparing Shapes of Various Signals}
\label{App:signalShapes}

\begin{figure}[h!]
	\centering
	\includegraphics[width=0.5\linewidth]{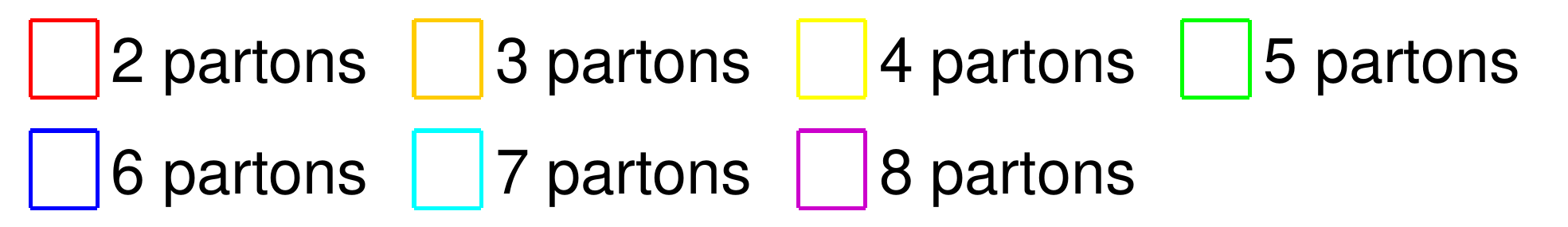}\\
	\includegraphics[width=0.32\linewidth]{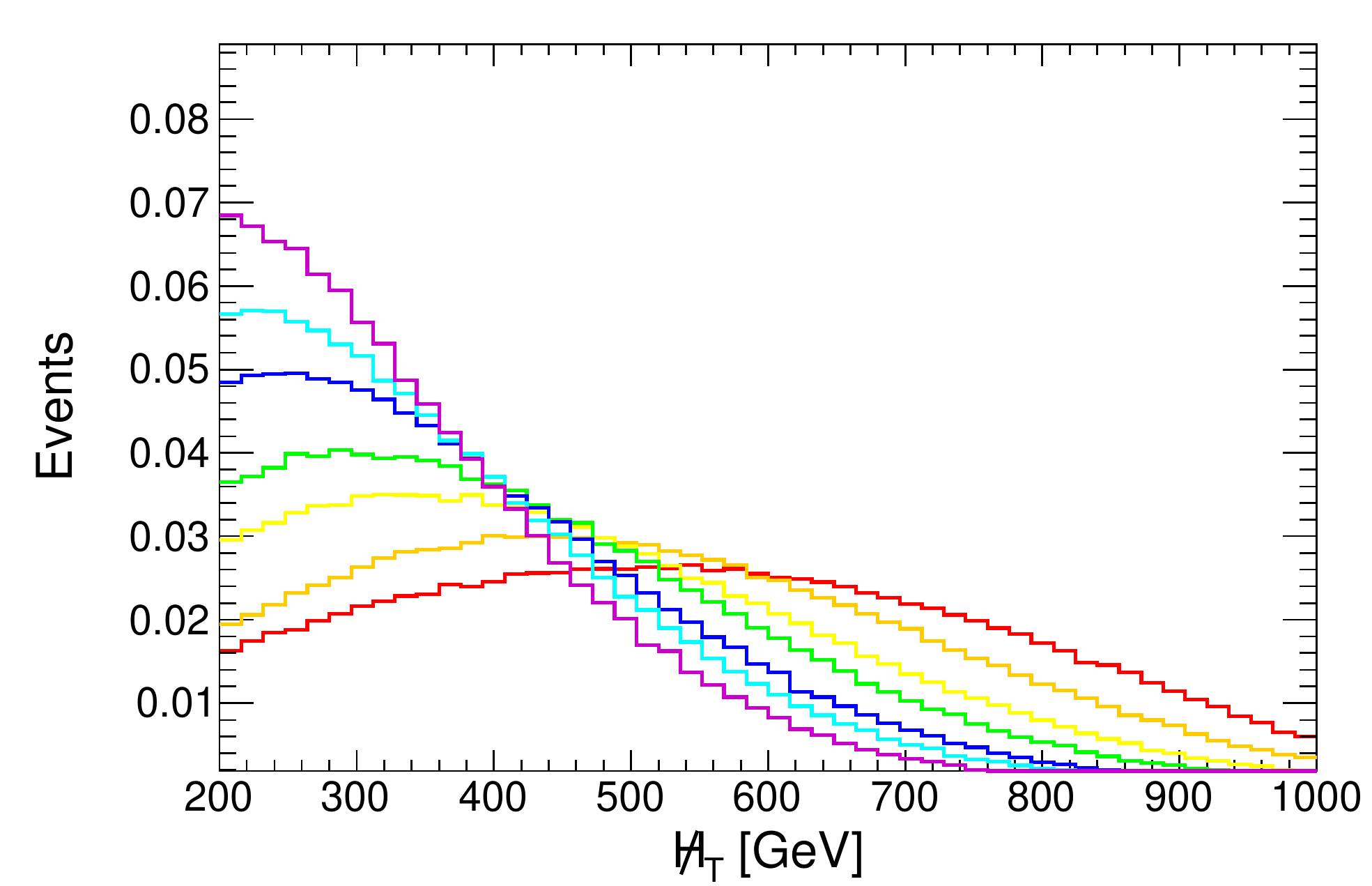}
	\includegraphics[width=0.32\linewidth]{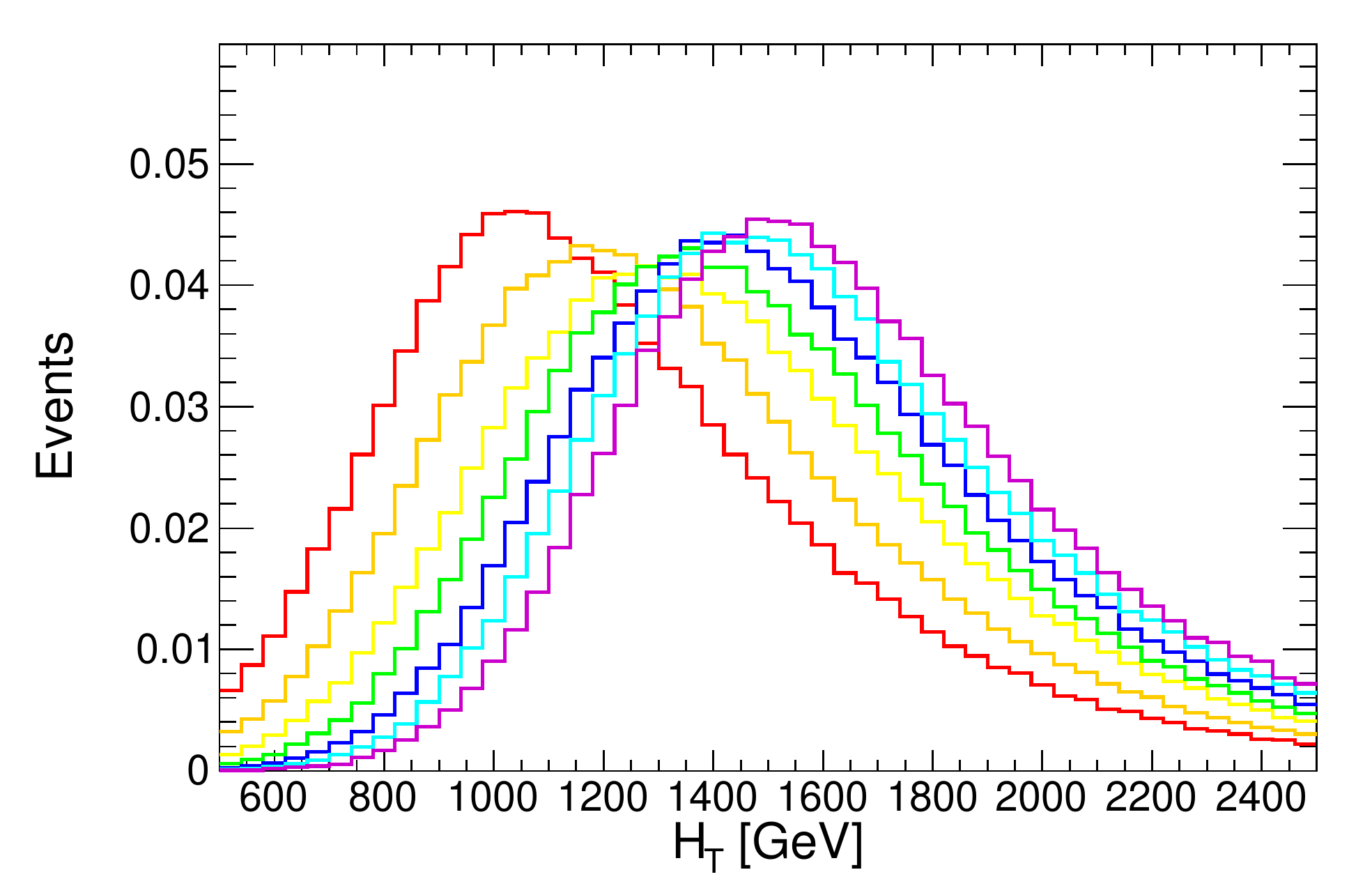}
	\includegraphics[width=0.32\linewidth]{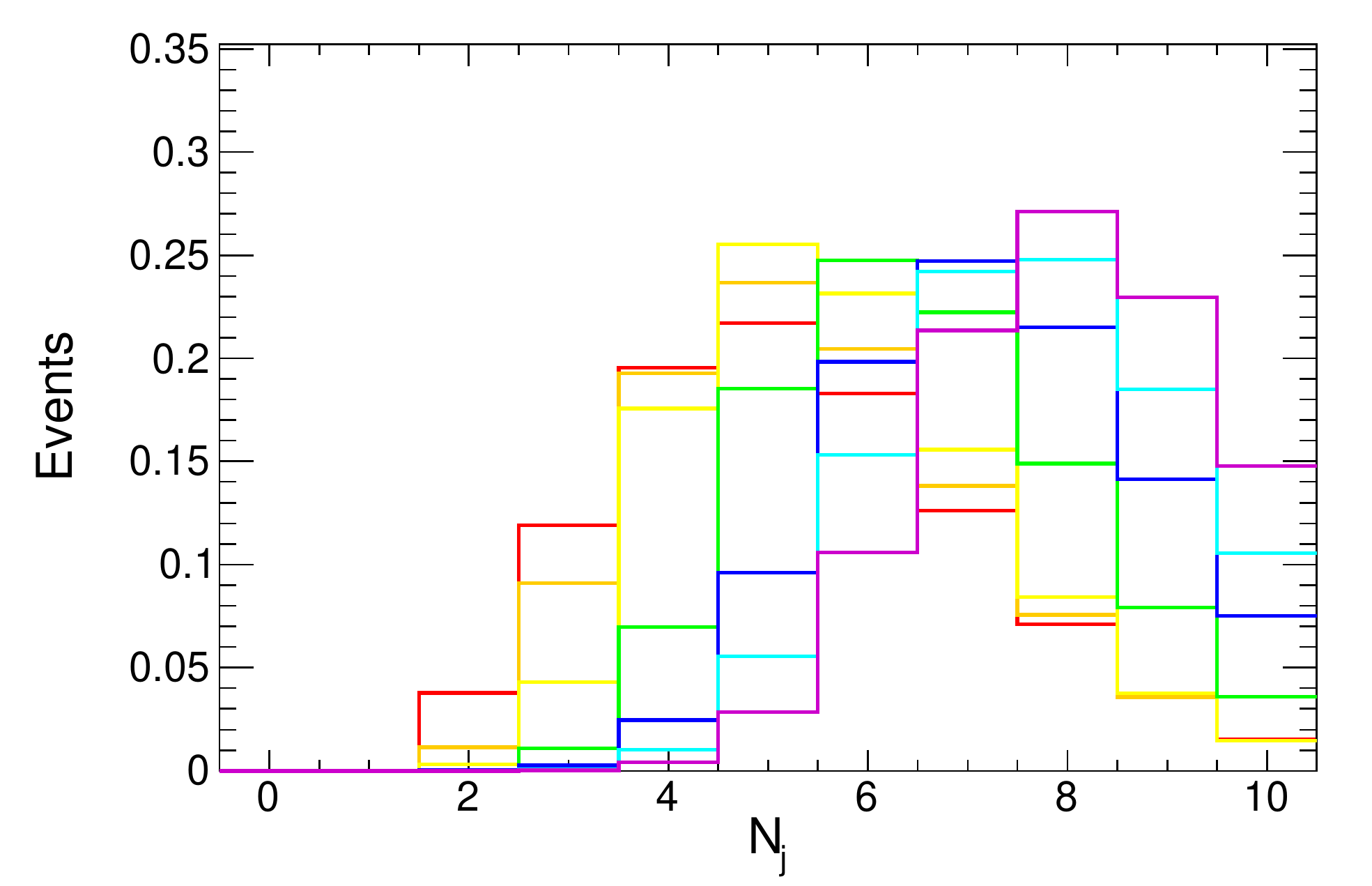}
	\caption{\label{Fig:compSigVsNpart}$\MHT$ (left), $H_T$ (middle), and $N_j$ (right) distributions for 1 TeV gluinos decaying to various n-parton final states and a massless neutralino.}
\end{figure}

\begin{figure}[h!]
	\centering
	\includegraphics[width=0.5\linewidth]{Figures/compareSignals/versusNpartons/compareSignalsLegend.pdf}\\
	\includegraphics[width=0.32\linewidth]{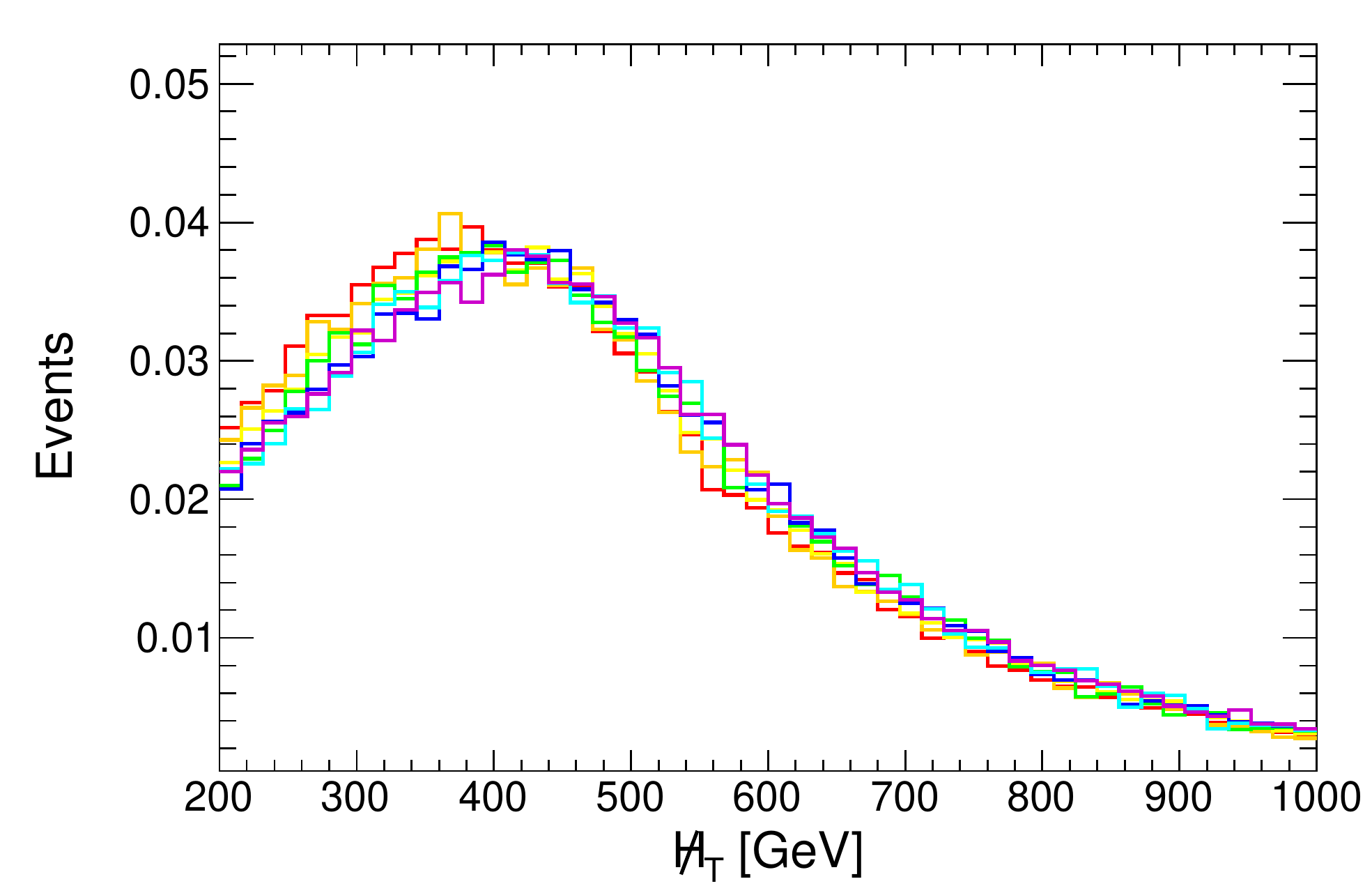}
	\includegraphics[width=0.32\linewidth]{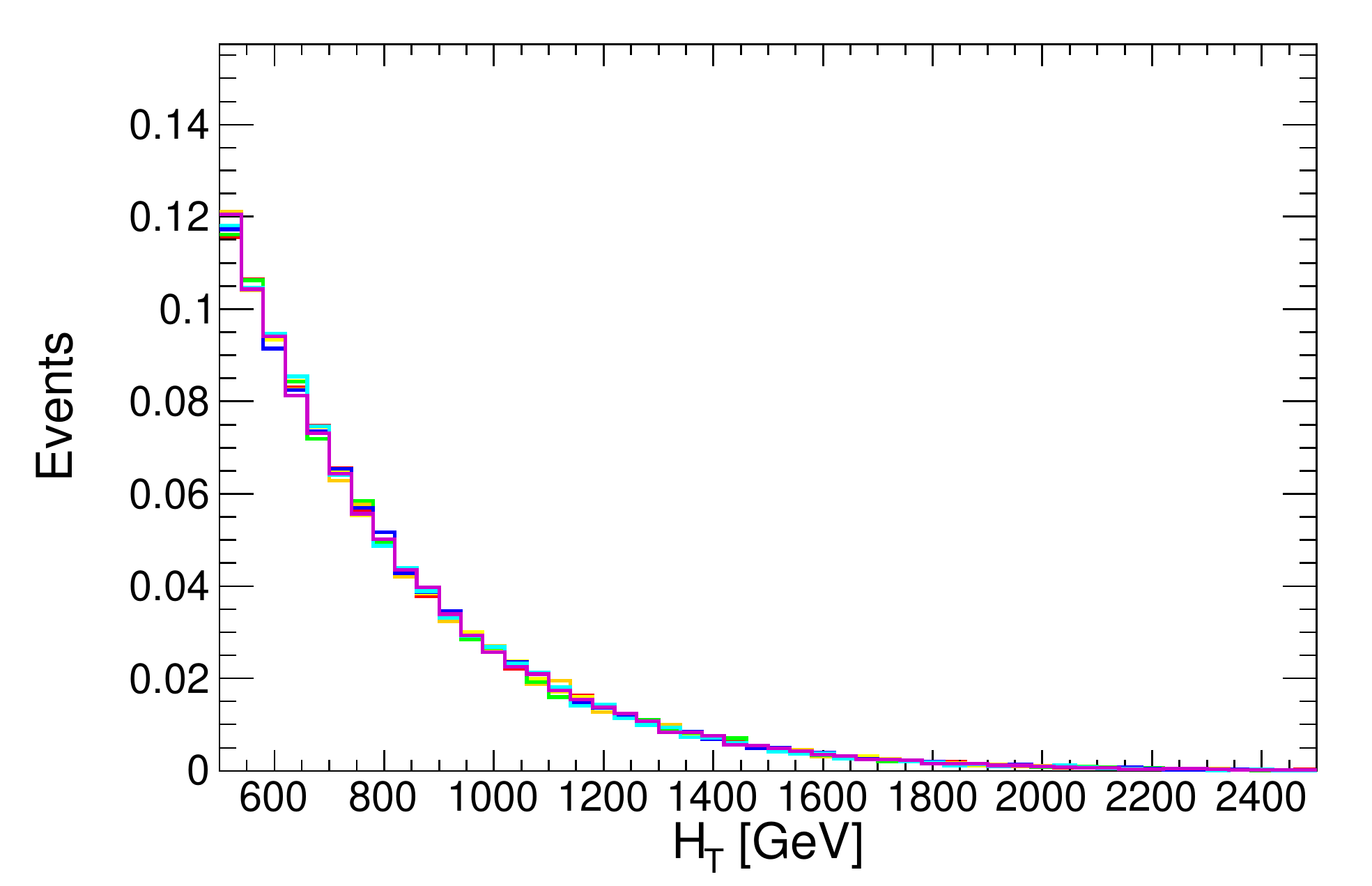}
	\includegraphics[width=0.32\linewidth]{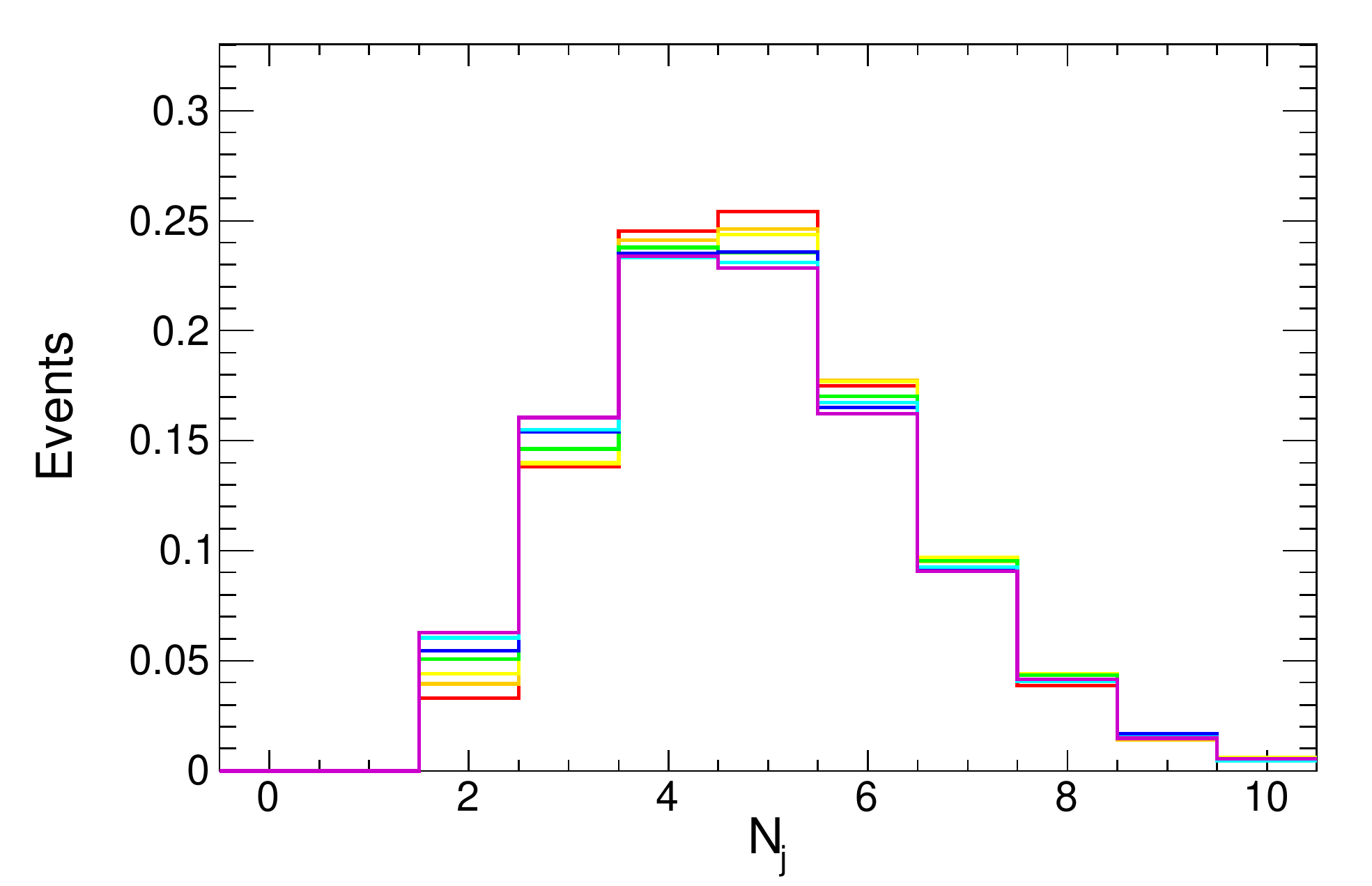}
	\caption{\label{Fig:compSigVsNpartComp}$\MHT$ (left), $H_T$ (middle), and $N_j$ (right) distributions for 1 TeV gluinos decaying to various n-parton final states and a massive (950 GeV) neturalino.}
\end{figure}

\begin{figure}[h!]
	\centering
	\includegraphics[width=0.32\linewidth]{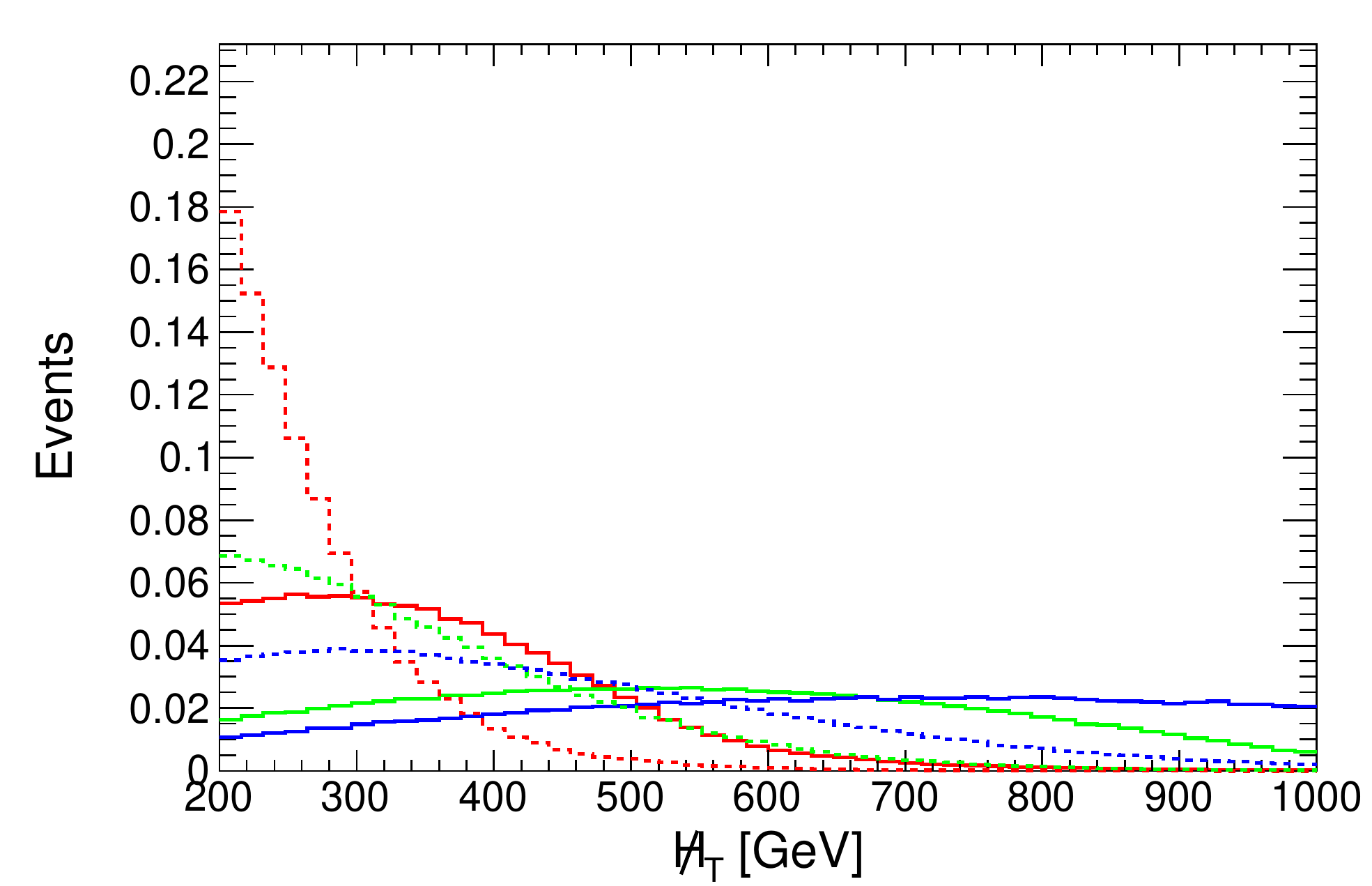}
	\includegraphics[width=0.32\linewidth]{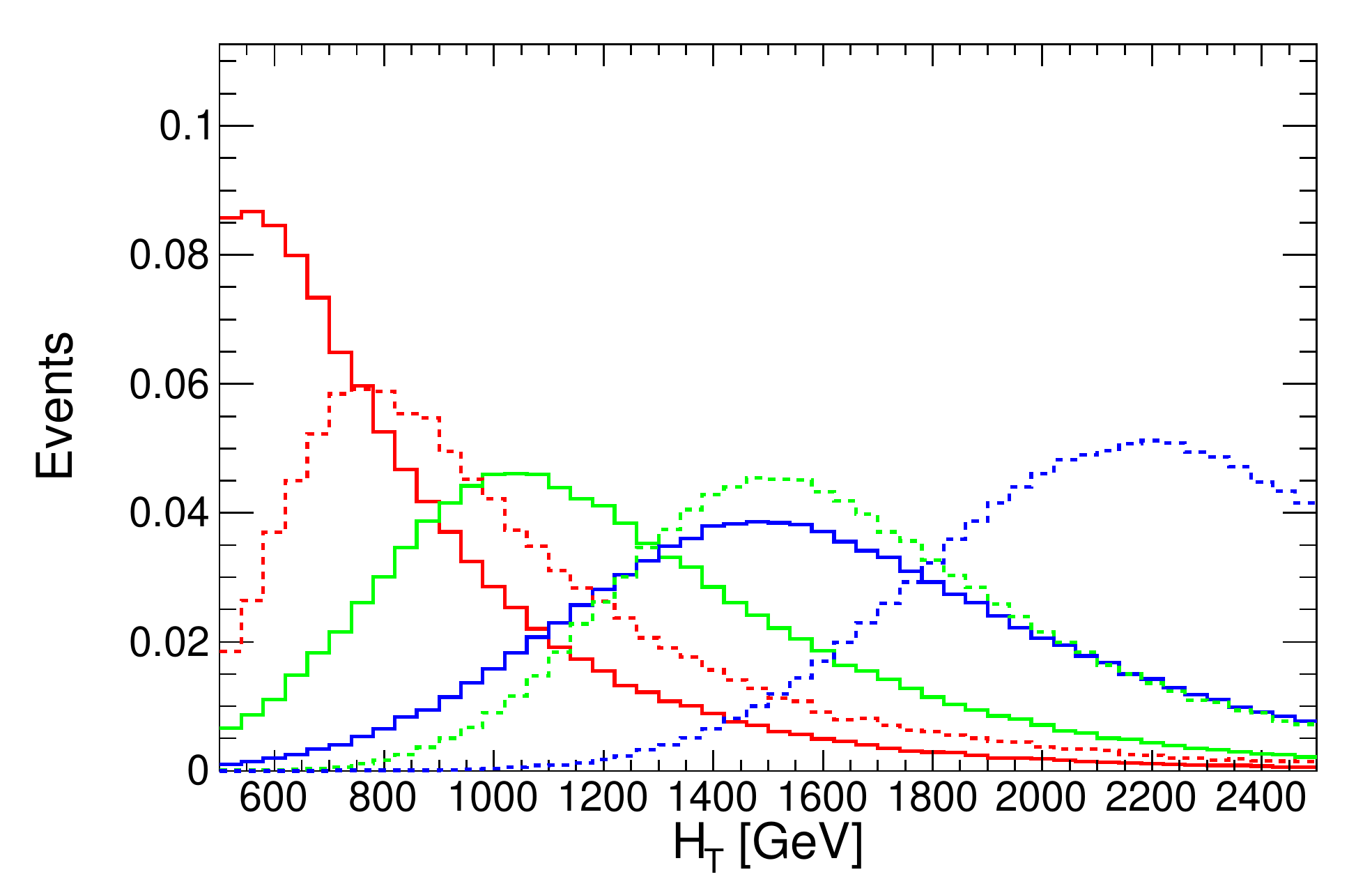}
	\includegraphics[width=0.32\linewidth]{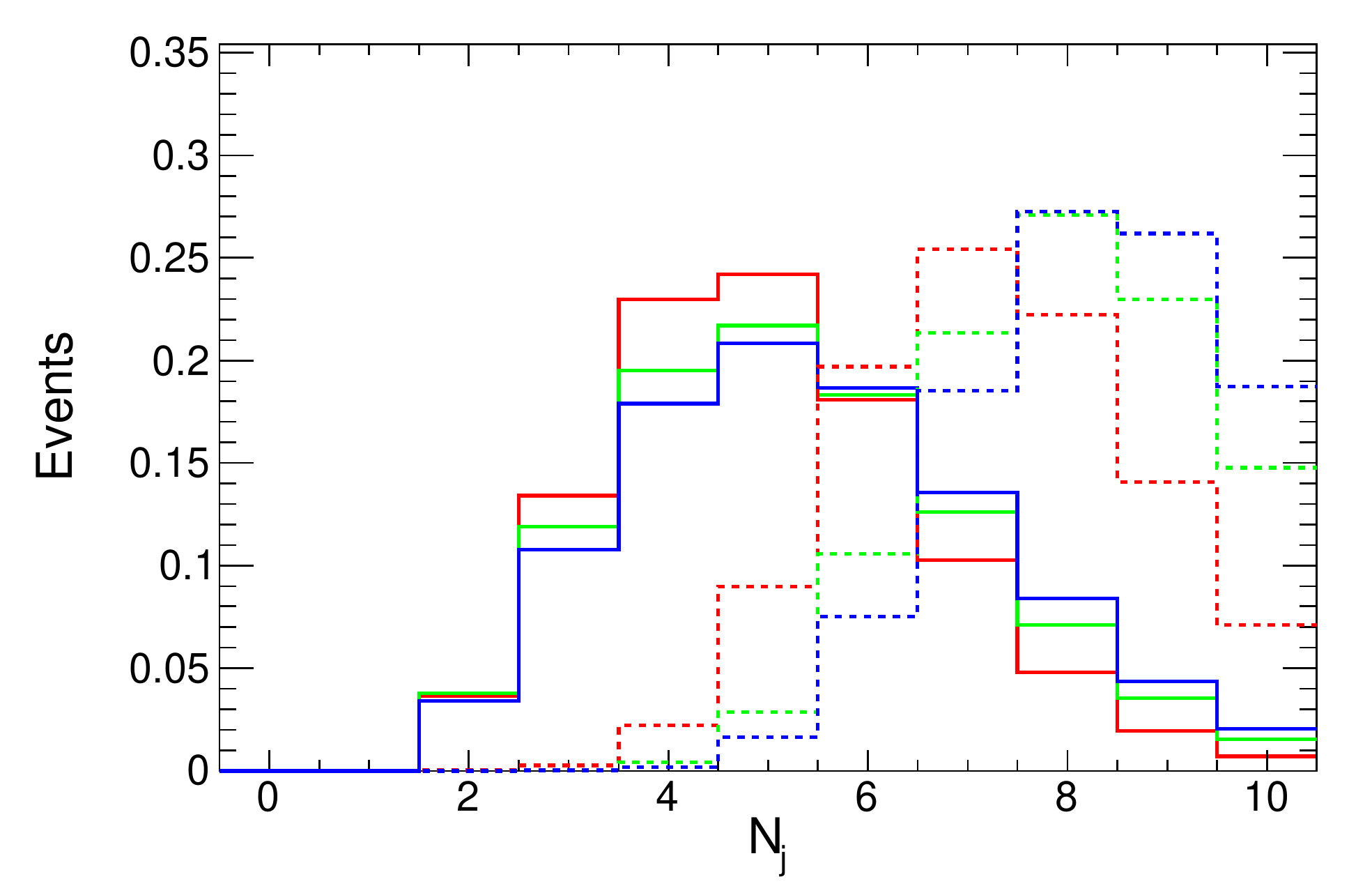}
	\caption{\label{Fig:compSigVsMass}$\MHT$ (left), $H_T$ (middle), and $N_j$ (right) distributions for various 
                       gluino mass hypotheses.  The gluinos are always restricted to decay to either one parton and a massless
		     neutralino (solid) or four partons and a	 massless neutralino (dashed).  Three different mass hypotheses
	   	     are shown: $m_{\tilde{g}}$=500 GeV (red), 1000 GeV (green), and 1500 GeV (blue).}
\end{figure}

In this appendix we compare kinematics of the $n$-body Simplified Models to show trends as a function of $n$-partons and the gluino mass.  We show only three variables, one of each type presented in Sec.~\ref{subsubsec:oneVar}, $H_T$, $\MHT$, and $N_j$.  We have checked that the trends for variables within the same classification are qualitatively similar.  Figures~\ref{Fig:compSigVsNpart} and~\ref{Fig:compSigVsNpartComp} compare kinematic distribution for a 1 TeV gluino decaying to different numbers of partons for an uncompressed and compressed mass spectrum respectively.  We find that the similarity observed in the signal distributions for the compressed mass spectrum, see in Figure~\ref{Fig:compSigVsNpartComp}, is a generic feature of compressed signals.  Figure~\ref{Fig:compSigVsMass} compares kinematic distributions for various gluino mass hypotheses with the gluino restricted to decay to either one parton and a massless neutralino or four partons and a massless neutralino.  

\end{spacing}
\begin{spacing}{1.1}
\bibliography{DissectingJetsMET}
\bibliographystyle{utphys}
\end{spacing}
\end{document}